\documentclass[aps,showpacs,floatfix,twocolumn,byrevtex,superscriptaddress,longbibliography]{revtex4-1}
\usepackage{amsmath}
\usepackage{amssymb}
\usepackage{amstext}
\usepackage{amsopn}
\usepackage{amsfonts}
\usepackage{amsxtra}
\usepackage[english]{babel}
\usepackage{graphicx}
\usepackage{float}
\usepackage{bm}
\usepackage{multirow}
\usepackage{dcolumn}
\usepackage{hyperref}
\usepackage{CJK}

\begin{document}

\title{Optical and photoemission investigation of structural and magnetic transitions in the
iron-based superconductor Sr$_\mathbf{0.67}$Na$_\mathbf{0.33}$Fe$_\mathbf{2}$As$_\mathbf{2}$}
\author{R. Yang}
\altaffiliation{Present address: Laboratorium f\"{u}r Festk\"{o}rperphysik, ETH Z\"{u}rich, CH-8093 Z\"{u}ich, Switzerland}
\affiliation{Condensed Matter Physics and Materials Science Division, Brookhaven National Laboratory,
  Upton, New York 11973, USA}
\affiliation{Beijing National Laboratory for Condensed Matter Physics, Institute of Physics,
  Chinese Academy of Sciences, Beijing 100190, China}
\affiliation{School of Physical Sciences, University of Chinese Academy of Sciences, Beijing 100049, China}
\author{J.~W.~Huang}
\affiliation{Beijing National Laboratory for Condensed Matter Physics, Institute of Physics,
  Chinese Academy of Sciences, Beijing 100190, China}
\affiliation{School of Physical Sciences, University of Chinese Academy of Sciences, Beijing 100049, China}
\author{N. Zaki}
\affiliation{Condensed Matter Physics and Materials Science Division, Brookhaven National Laboratory,
  Upton, New York 11973, USA}
\author{I. Pletikosi\'{c}}
\affiliation{Condensed Matter Physics and Materials Science Division, Brookhaven National Laboratory,
  Upton, New York 11973, USA}
\affiliation{Department of Physics, Princeton University, Princeton, NJ 08544, USA}
\author{Y.~M.~Dai}
\affiliation{Center for Superconducting Physics and Materials, National Laboratory of
  Solid State Microstructures and Department of Physics, Nanjing University, Nanjing 210093, China}
\author{H. Xiao}
\affiliation{Center for High Pressure Science and Technology Advanced Research, Beijing 100094, China}
\author{T.~Valla}
\author{P.~D.~Johnson}
\affiliation{Condensed Matter Physics and Materials Science Division, Brookhaven National Laboratory,
  Upton, New York 11973, USA}
\author{X.~J.~Zhou}
\affiliation{Beijing National Laboratory for Condensed Matter Physics, Institute of Physics,
  Chinese Academy of Sciences, Beijing 100190, China}
\affiliation{School of Physical Sciences, University of Chinese Academy of Sciences, Beijing 100049, China}
\affiliation{Songshan Lake Materials Laboratory, Dongguan 523808, China}
\affiliation{Beijing Academy of Quantum Information Science, Beijing 100193, China}
\author{X.~G.~Qiu}
\email[]{xgqiu@iphy.ac.cn}
\affiliation{Beijing National Laboratory for Condensed Matter Physics, Institute of Physics,
  Chinese Academy of Sciences, Beijing 100190, China}
\affiliation{School of Physical Sciences, University of Chinese Academy of Sciences, Beijing 100049, China}
\affiliation{Songshan Lake Materials Laboratory, Dongguan 523808, China}
\author{C.~C.~Homes}
\email[]{homes@bnl.gov}
\affiliation{Condensed Matter Physics and Materials Science Division, Brookhaven National Laboratory,
  Upton, New York 11973, USA}
\date{\today }
%%%%%%%%%%%%%%%%%%%%%%%%%%%%%%%%%%%%
%
% Abstract
%
\begin{abstract}
We report the temperature dependent optical conductivity and angle-resolved
photoemission spectroscopy (ARPES) studies of the multiband iron-based superconductor
Sr$_{0.67}$Na$_{0.33}$Fe$_2$As$_2$.  Measurements were made in the high-temperature
tetragonal paramagnetic phase; below the structural and magnetic transitions
at $T_{\rm N}\simeq 125$~K in the orthorhombic spin-density-wave (SDW)-like phase, and
$T_r\simeq 42$~K in the reentrant tetragonal double-$\mathbf{Q}$ magnetic phase where
both charge and SDW order exist; and below the superconducting transition at
$T_c\simeq 10$~K.
The free-carrier component in the optical conductivity is described by two Drude
contributions; one strong and broad, the other weak and narrow.  The broad Drude
component decreases dramatically below $T_{\rm N}$ and $T_r$, with much of its strength being
transferred to a bound excitation in the mid-infrared, while the narrow Drude component
shows no anomalies at either of the transitions, actually increasing in strength at low
temperature while narrowing dramatically.  The behavior of an infrared-active mode suggests
zone-folding below $T_r$.  Below $T_c$ the dramatic decrease in the low-frequency optical
conductivity signals the formation of a superconducting energy gap.
ARPES reveals hole-like bands at the center of the Brillouin zone (BZ), with both electron-
and hole-like bands at the corners.  Below $T_{\rm N}$, the hole pockets at the center of the BZ
decrease in size, consistent with the behavior of the broad Drude component; while
below $T_r$ the electron-like bands shift and split, giving rise to a low-energy excitation
in the optical conductivity at $\simeq 20$~meV.
The $C_2$ and $C_4$ magnetic states, with resulting spin-density-wave and charge-SDW
order, respectively, lead to a significant reconstruction of the Fermi surface that has profound
implications for the transport originating from the electron and hole pockets, but appears
to have relatively little impact on the superconductivity in this material.
\end{abstract}

%  72.15.-v  Electronic conduction in metals and alloys
%  74.70.-b  SC: Superconducting materials other than cuprates
%  78.20.-e  Optical properties of bulk materials and thin films
%  78.30.-j  Infrared and Raman spectra

\maketitle

%%%%%%%%%%%%%%%%%%%%%%%%%%%%%%%%%%%%%%%%%%%%%%%%%%%%%%%%%%%%%%%%%%%%%%%%%%%%%%%
%
% Introduction
%
\section{Introduction}
The discovery of iron-based superconductors prompted an intensive investigation in
the hope of identifying new compounds with high superconducting critical temperatures
($T_c$'s) \cite{Johnston2010,Paglione2010,Canfield2010,Si2016}.  In many of the iron-based
materials, superconductivity emerges with the suppression of antiferromagnetic (AFM) order,
suggesting that the pairing mechanism is related to the magnetism. Indeed, the iron-based
materials display a variety of magnetically-ordered ground states \cite{Dean2012,Dai2015,
Moroni2017,Kreyssig2018,Meier2018} that may either compete with or foster the emergence
of superconductivity.

%
% The 122 materials
%
One class of materials, \emph{Ae}Fe$_2$As$_2$, where \emph{Ae}$\,=\,$Ba, Ca or Sr
(the so-called ``122'' materials), is particularly useful as superconductivity
may be induced through a variety of chemical substitutions \cite{Rotter2008,Sefat2008,
Ni2008,Sasmal2008,Chen2008,Chu2009,Goko2009,Saha2009,Jiang2009,Shi2010,Cortes2011},
as well as through the application of pressure \cite{Ishikawa2009,Alireza2009,Colombier2009,
Kitagawa2009}.
%
% SrFe2As2
%
The phase diagram of Sr$_{1-x}$Na$_x$Fe$_2$As$_2$ has a number of interesting
features.  At room temperature, the parent compound SrFe$_2$As$_2$ is a paramagnetic
metal with a tetragonal ($I4/mmm$) structure.  The resistivity in the iron-arsenic
planes decreases with temperature until it drops anomalously as the material undergoes
a magnetic transition at $T_{\rm N} \simeq 195$~K to a spin-density-wave (SDW)-like AFM
ground state that is also accompanied by a structural transition to an orthorhombic
($Fmmm$) phase \cite{Tegel2008,Yan2008,Zhao2008,Hu2008,Hancock2010,Blomberg2011}.
The crystals are heavily twinned in the orthorhombic phase; however, the application
of uniaxial stress along the $(110)$ direction of the tetragonal unit cell results
in a nearly twin-free sample \cite{Tanatar2009,Fisher2011}.
The magnetic order may be described as AFM stripes, where the iron spins are aligned
antiferromagnetically along the \emph{a} axis and ferromagnetically along the \emph{b}
axis \cite{Goldman2008,Kofu2009}; this is also referred to as the magnetic $C_2$ phase
due to its twofold rotation symmetry.
As the sodium content increases, the magnetic and structural transition temperatures
decrease until both disappear at $x\simeq 0.48$; superconductivity appears well before
this point at $x\simeq 0.2$, and reaches a maximum of $T_c\simeq 37$~K for
$x \simeq 0.5 - 0.6$.
Between $0.29 < x < 0.42$, an additional magnetic and structural transition occurs
below $T_{\rm N}$ at $T_r$; the tetragonal ($I4/mmm$) phase reemerges, forming
a dome which lies completely within the AFM region.  This phase appears to be
a common element in the hole-doped 122 materials \cite{Kim2010,Hassinger2012,Bohmer2015,
Wang2016,Taddei2016,Wang2019,Hassinger2016,Taddei2017,Yi2018,Avci2014,Allred2016}; however, in
Sr$_{1-x}$Na$_x$Fe$_2$As$_2$ the dome is more robust and occurs over a wider doping
range at temperatures up to $T_r \simeq 65$~K \cite{Taddei2016,Wang2019}, which is higher
than has been observed in other compounds.
This magnetic order is described as the collinear superposition of two itinerant SDW's
with nesting wavevector $\mathbf{Q}$, leading to a double-$\mathbf{Q}$ SDW
\cite{Avci2014,Allred2016} in which half the iron sites are nonmagnetic, and half
have twice the moment measured in the orthorhombic AFM phase, oriented along the
\emph{c} axis \cite{Wasser2015,Mallett2015a}; this is referred to as the magnetic $C_4$
phase because of its fourfold rotational invariance.  This magnetic state is accompanied
by a charge-density wave (CDW) with the charge coupling to the square of the
magnetization, resulting in a charge-SDW (CSDW) \cite{Hoyer2016}.
%
%the underlying mechanism of the CSDW and its relation to the $C_2$ SDW remains
%uncertain.
%

%
%%%%%%%%%%%%%%%%%%%%%%%%%%%%%%%%%%%%%%%%%%%%%%%%%%%%%%%%%%%%%%%%%%%%%%%%%%%%%%%
%
% Figure 1
%
\begin{figure}[tb]
\includegraphics[width=3.4in]{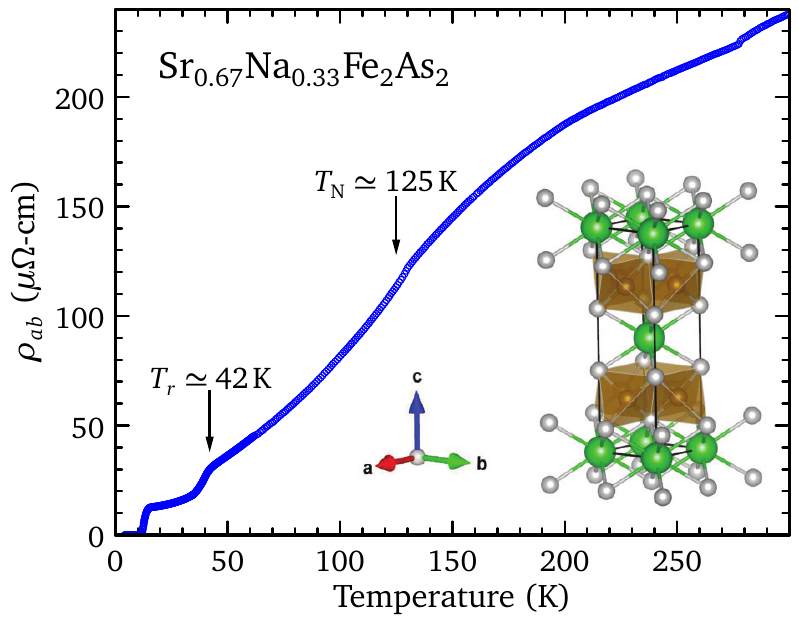}
\caption{The temperature dependence of the in-plane resistivity for
Sr$_{0.67}$Na$_{0.33}$Fe$_2$As$_2$ with inflection points at $T_{\rm N}
\simeq 125$~K and $T_r\simeq 42$~K; the resistivity at room temperature
has been adjusted to match the optical conductivity in the zero-frequency
limit.
Inset: The generic unit cell in the high-temperature tetragonal phase for
the 122 materials.}
\label{fig:resis}
\end{figure}

%
% In this work...
%
In this work, the complex optical properties and angle-resolved photoemission
spectroscopy (ARPES), of Sr$_{0.67}$Na$_{0.33}$Fe$_2$As$_2$ have been investigated
in the high-temperature tetragonal phase, as well as the magnetic $C_2$ and
$C_4$ phases.  The value of $x\simeq 0.33$ used in the current study is slightly
below the optimal value of $x\simeq 0.37$ that bisects the $C_4$ dome in the
Sr$_{1-x}$Na$_x$Fe$_2$As$_2$ phase diagram \cite{Taddei2016}.  Based on transport
studies, $T_{\rm N}\simeq 125$~K, $T_r \simeq 42$~K, and $T_c\simeq 10$K.
%
%The complex conductivity is described by two free-carrier (Drude) components
%with a strength related to the plasma frequency (proportional to the number
%of carriers in a band), and a scattering rate; in addition, there are a number
%of bound excitations from interband transitions.
%
In the high temperature tetragonal paramagnetic state, the optical response of the
free-carriers is described by two Drude terms (Sec.~IIIA); one strong and broad
(large scattering rate), and the other weak and narrower (smaller scattering rate);
as the temperature is reduced, the strength of the Drude terms show relatively
little temperature dependence, while the scattering rates slowly decrease.
Below $T_{\rm N}$, the Fermi surface reconstruction driven by the structural and magnetic
transitions causes both the strength and the scattering rate for the broad Drude
term to decrease dramatically; the missing spectral weight (the area under
the conductivity curve) associated with the free carriers is transferred
to a peak that emerges in the mid-infrared.  The narrow Drude term actually
increases slightly in strength below $T_{\rm N}$ while narrowing.
Below $T_r$, in the magnetic $C_4$ phase, the broad Drude term again narrows and
decreases in strength; while the strength of the narrow term does not appear to
change, its scattering rate decreases dramatically.  Based on the behavior of an
infrared-active lattice mode, the presence of CSDW order likely results in the
formation of a supercell resulting in zone folding, leading to a further reconstruction
of the Fermi surface; while spectral weight is again transferred from the broad Drude
to the midinfrared peak, a new low-energy peak emerges at $\simeq 20$~meV.  Below
$T_c$, there is a dramatic decrease in the low-frequency conductivity,
signalling the formation of a superconducting energy gap.
ARPES reveals several large hole pockets at the center of the Brillouin zone above
$T_{\rm N}$, one of which shifts below the Fermi level below $T_{\rm N}$ in the $C_2$ magnetic
phase, a trend which continues below $T_r$, suggesting that these bands may be related
to the broad Drude response.  At the corners of the Brillouin zone, there are both hole-
and electron-like bands.  Below $T_{\rm N}$ and $T_r$, several of these bands appear to
split and shift, but it is not clear if there are any significant changes to the
size of the associated Fermi surfaces, suggesting that some of these carriers may
be related to the narrow Drude term; below $T_r$ the band splitting is likely responsible
for the emergence of the low-energy peak.
The structural and magnetic transitions from which the $C_2$ (SDW) and $C_4$
(double-$\mathbf{Q}$ SDW) phases emerge result in a Fermi surface reconstruction
that has profound effects on the optical conductivity and electronic structure;
however, the superfluid stiffness appears to be more or less unaffected by the
CSDW order.

%
%%%%%%%%%%%%%%%%%%%%%%%%%%%%%%%%%%%%%%%%%%%%%%%%%%%%%%%%%%%%%%%%%%%%%%%%%%%%%%%
%
% Figure 2
%
\begin{figure*}[t]
\includegraphics[width=6.25in]{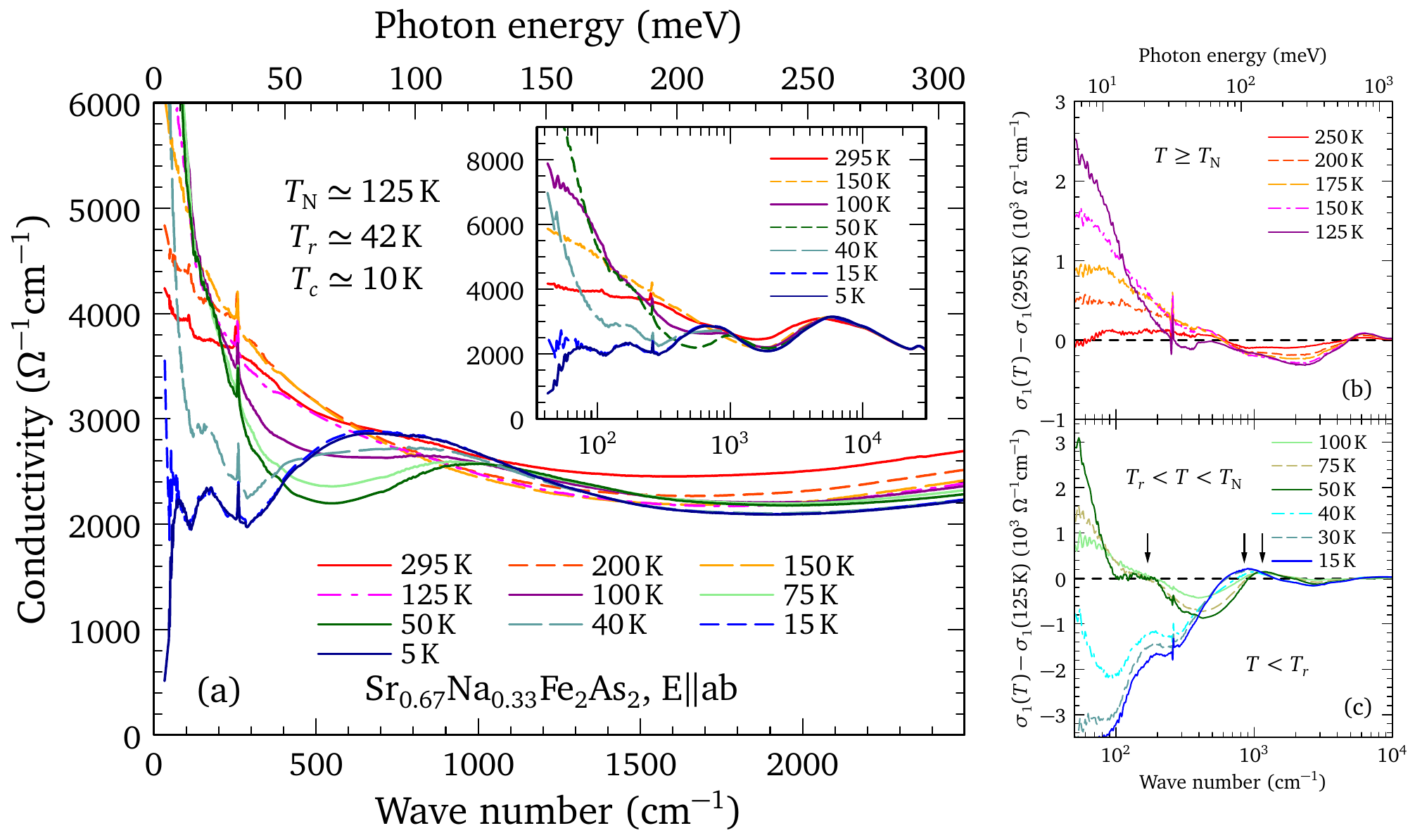}
\caption{(a) The temperature dependence of the real part of the optical conductivity of
Sr$_{0.67}$Na$_{0.33}$Fe$_2$As$_2$ in the infrared region for light polarized in the
Fe--As planes. Inset: the conductivity over a wide spectral range at several
temperatures.
(b) The $\sigma_1(\omega, T)-\sigma_1(\omega, 295\,{\rm K})$ difference plot for $T\geq T_{\rm N}$
over a wide spectral range showing the narrowing of the free-carrier response and the
transfer of spectral weight from high to low frequency.
(c) The $\sigma_1(\omega, T)-\sigma_1(\omega, 125\,{\rm K})$ difference plot.  Int the
$T_r < T < T_{\rm N}$ region the free-carrier response continues to narrow and a peak
emerges in the mid-infrared region; for $T<T_r$, the low-frequency conductivity is further
suppressed, the mid-infrared peak shifts to low energy, and a prominent peak is observed
at $\simeq 170$~cm$^{-1}$ (arrows).}
\label{fig:sigma}
\end{figure*}

%%%%%%%%%%%%%%%%%%%%%%%%%%%%%%%%%%%%%%%%%%%%%%%%%%%%%%%%%%%%%%%%%%%%%%%%%%%%%%%
%
% Experiment
%
\section{Experiment}
High-quality single crystals of Sr$_{0.67}$Na$_{0.33}$Fe$_2$As$_2$ with good
cleavage planes (001) were synthesized using a self-flux technique \cite{Taddei2016,Guo2019}.
The temperature dependence of the in-plane resistivity, shown in Fig.~\ref{fig:resis},
was measured using a standard four-probe configuration using a Quantum Design physical
property measurement system; the unit cell for the high-temperature tetragonal phase
is shown in the inset.  The resistivity decreases gradually with temperature, showing
a weak inflection point at $T_{\rm N} \simeq 125$~K with a more pronounced decrease
in the resistivity at $T_r \simeq 42$~K; the resistivity goes to zero below the
superconducting transition at $T_c \simeq 10$~K.
The reflectance from freshly-cleaved surfaces has been measured at a near-normal angle of
incidence over a wide temperature ($\simeq 5$ to 300~K) and frequency range ($\simeq 2$~meV
to about 5~eV) with Bruker IFS 113v and Vertex 80v Fourier transform spectrometers for
light polarized in the \emph{a-b} planes using an \emph{in situ} evaporation technique
\cite{Homes1993}.  The complex optical properties have been determined from a
Kramers-Kronig analysis of the reflectivity. The reflectivity is shown in supplementary
Fig.~S1; the details of the Kramers-Kronig analysis are described in the Supplementary
Material \cite{Suplmt}.
Temperature dependent ARPES measurements have been performed to track the evolution
of the electron and hole pockets in the various phases.  Measurements at BNL, which
focused on the electronic structure near the center of the Brillouin zone, were
performed using 21.2~eV light from a monochromator-filtered He~I source (Omicron VUV5k)
and a Scienta SES-R4000 electron spectrometer; emitted electrons were collected along
the direction perpendicular to the light-surface mirror plane.  Samples were cleaved at
low temperature and measured in an ultrahigh vacuum with a base pressure better than
$5\times 10^{-10}$~mbar.
Measurements at the National Laboratory for Superconductivity, Institute of Physics,
Chinese Academy of Sciences, were performed using a 21.2~eV helium discharge lamp and
a Scienta DA30L electron spectrometer.  The latter's overall energy resolution was
10~meV for Fermi surface mapping and 4~meV for the cuts; the angular resolution was
$\sim 0.1^\circ$. All the samples were cleaved at low temperature and measured in an
ultrahigh vacuum with a base pressure better than $5\times 10^{-11}$~mbar.
Note that because uniaxial strain is not applied to the samples below $T_{\rm N}$,
they will be heavily twinned, thus the optical and ARPES results represent
an average of the \emph{a} and \emph{b} axis response in the magnetic $C_2$ phase.

%
%%%%%%%%%%%%%%%%%%%%%%%%%%%%%%%%%%%%%%%%%%%%%%%%%%%%%%%%%%%%%%%%%%%%%%%%%%%%%%%
%
% Figure 3 - fits for the parent compound
%
\begin{figure*}[t]

\includegraphics[width=2.3in]{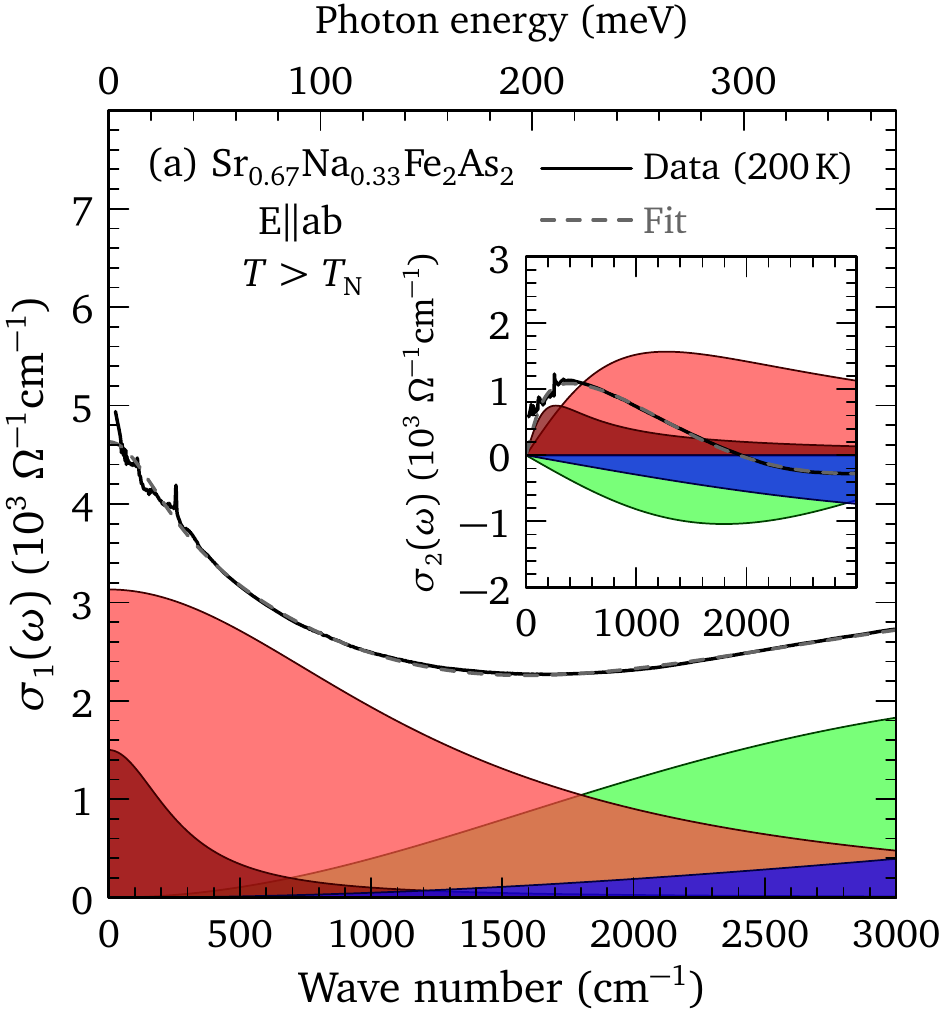}\
\includegraphics[width=2.3in]{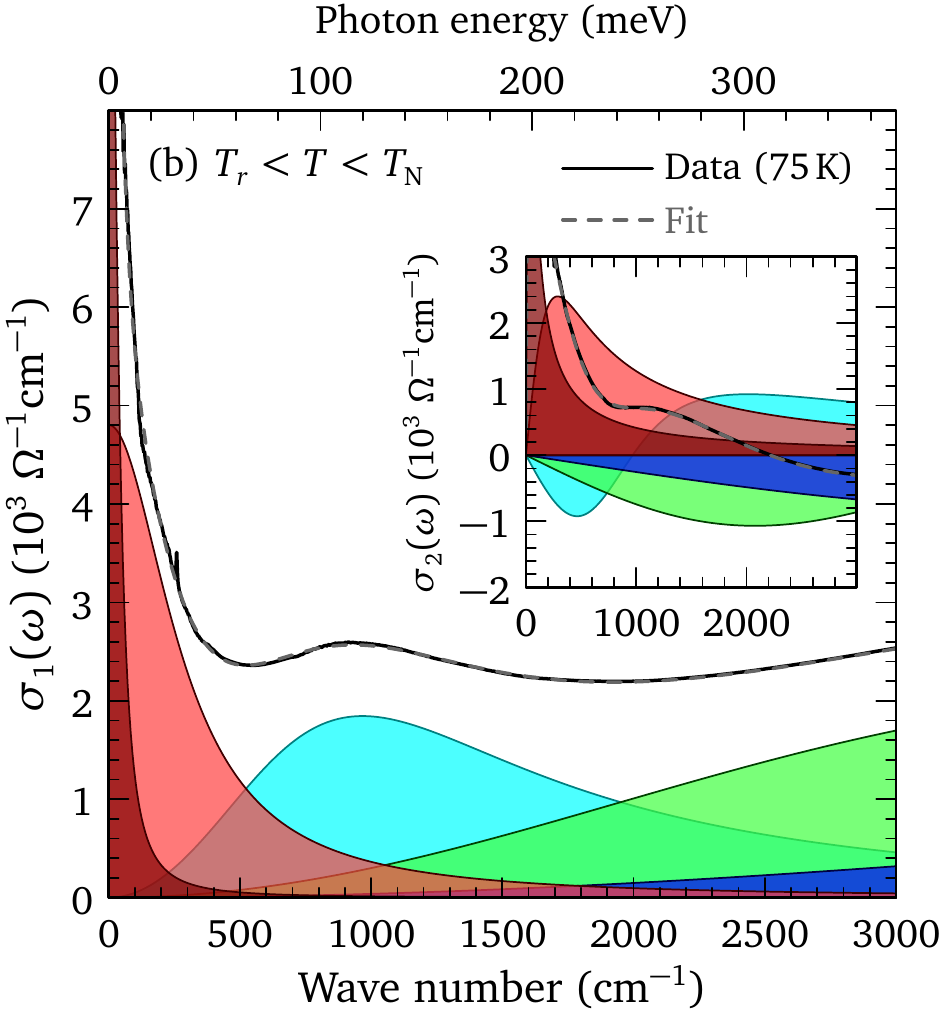}\
\includegraphics[width=2.3in]{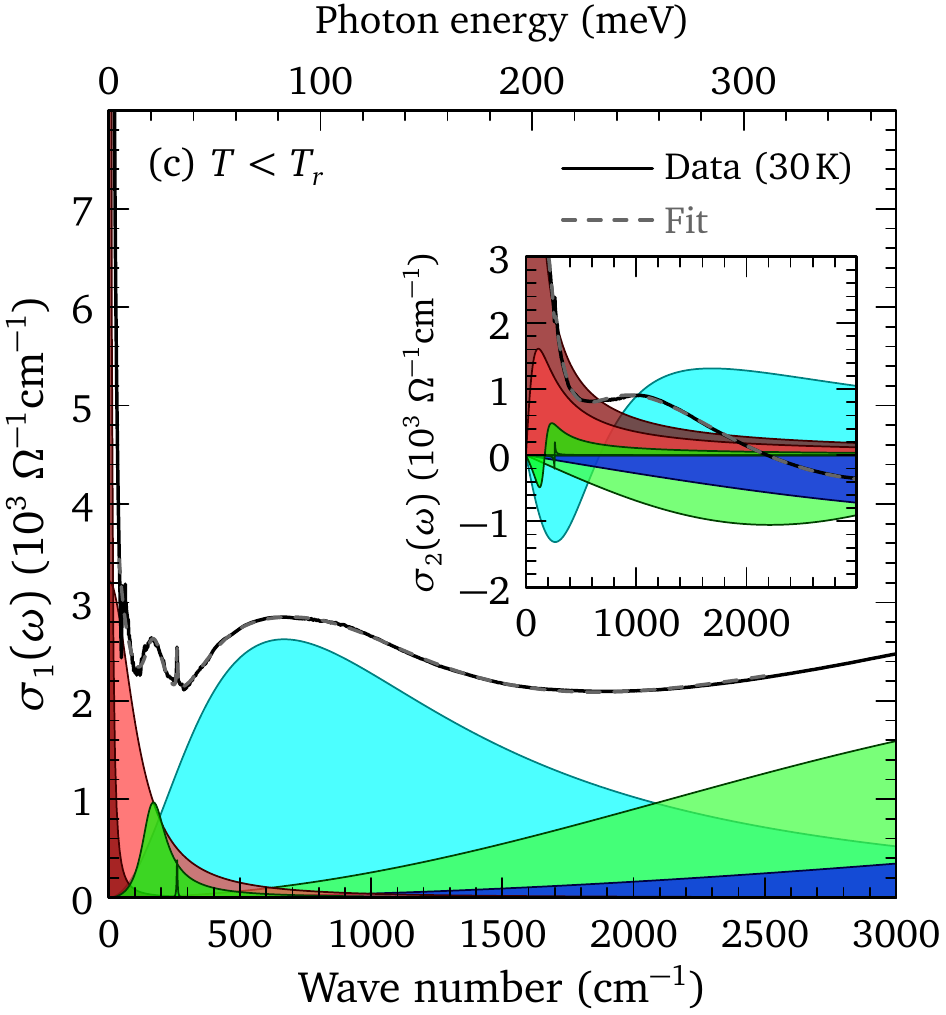}\
\caption{The Drude-Lorentz model fits to the real and imaginary (inset) parts of the in-plane optical conductivity
of Sr$_{0.67}$Na$_{0.33}$Fe$_2$As$_2$ decomposed into the narrow (D1) and broad (D2) Drude components, as well as
several bound excitations (a) above $T_{\rm N}$ at 200~K, (b) below $T_{\rm N}$ at 75~K showing the narrowing
of the Drude features and the emergence of a peak at $\simeq 950$cm$^{-1}$, and (c) below $T_r$ at 30~K, showing
further narrowing and peaks at $\simeq 170$ and 700~cm$^{-1}$.
}
\label{fig:fits}
\end{figure*}
%

%%%%%%%%%%%%%%%%%%%%%%%%%%%%%%%%%%%%%%%%%%%%%%%%%%%%%%%%%%%%%%%%%%%%%%%%%%%%%%%
%
% Experiment and results
%
\section{Results and Discussion}
\subsection{Optical properties}
%
% Optical conductivity results: The real part of the optical conductivity determined
% from the Kramers-Kronig analysis of the reflectance is shown in Fig.~\ref{fig:s2}.
%
The temperature dependence of the real part of the in-plane optical conductivity [$\sigma_{1}(\omega)$]
of Sr$_{0.67}$Na$_{0.33}$Fe$_2$As$_2$ is shown in the infrared region in Fig.~\ref{fig:sigma}(a)
(an additional plot of the optical conductivity is shown in supplementary Fig.~S2).  The character of
the conductivity changes dramatically through the structural and magnetic transitions, which can be
characterized by four distinct regions: (i) $T > T_{\rm N}$; (ii) $T_r < T < T_{\rm N}$; (iii)
$T < T_r$, and below the superconducting transition (iv) $T < T_c$.  The changes to the nature of the
conductivity are shown as the difference plots $\sigma_1(\omega, T) - \sigma_1(\omega,295\,{\rm K})$,
and $\sigma_1(\omega, T)-\sigma_1(\omega, 125\,{\rm K})$, shown in Figs.~\ref{fig:sigma}(b) and
\ref{fig:sigma}(c), respectively.

At room temperature, the free-carrier response appears Drude-like (a Lorentzian centered
at zero frequency with a scattering rate defined as the full width at half maximum), giving way
to a flat response at higher frequencies, until the first interband transitions are encountered at
about 1~eV.  As the temperature is reduced, the scattering rate decreases and there is a
slight reduction of the conductivity in the mid-infrared region as spectral weight
is transferred from high to low frequency, which leads to an increase
at low frequency and a decrease at high frequency in the difference spectra in Fig.~\ref{fig:sigma}(b).
%
% Below T_N
%
Below $T_{\rm N}$ in the $C_2$ phase, the free-carrier response narrows dramatically and a
peak-like structure emerges at about 950~cm$^{-1}$, somewhat lower than a similar feature
that was observed below $T_{\rm N}$ at $\simeq 1400$~cm$^{-1}$ in the parent compound SrFe$_2$As$_2$
\cite{Homes2016}.  This is illustrated by the upper three curves in Fig.~\ref{fig:sigma}(c)
that show the continuing increase in the low-frequency conductivity, as well as the emergence
of a peak in the mid-infrared region.  Interestingly, below $\simeq 75$~K, a low-energy peak at
$\simeq 170$~cm$^{-1}$ begins to emerge.
%
% Below T_r
%
This behavior continues until $T \leq T_r$, at which point the Drude-like response becomes
extremely narrow in the $C_4$ phase, illustrated by the dramatic suppression of the low-frequency
conductivity in the difference plot in Fig.~\ref{fig:sigma}(c), leaving clearly identifiable peaks
at $\simeq 170$ and 700~cm$^{-1}$.  Below $T_c\simeq 10$~K, there is a depletion of the
low-frequency conductivity with the emergence of a shoulder-like structure around 70 cm$^{-1}$
that signals the formation of a superconducting energy gap (supplementary Fig.~S2).

%
% Phonon
%
The sharp feature observed in the conductivity at $\simeq 260$~cm$^{-1}$ is attributed to a
normally infrared-active lattice vibration in the iron-arsenic planes; while this mode increases
in frequency with decreasing temperature, it does not display the anomalous increase in
oscillator strength below $T_{\rm N}$ that was observed in the parent compound \cite{Homes2018}.
However, below $T_r$ there is evidence for a new satellite mode appearing at
$\simeq 282$~cm$^{-1}$ (supplementary Fig.~S3); a similar feature has also been observed
in the $C_4$ phase of Ba$_{1-x}$K$_x$Fe$_2$As$_2$ and is attributed to Brillouin-zone
folding due to the formation of a supercell in the CSDW phase \cite{Mallett2015b}.

\begin{figure}[tb]
\includegraphics[width=2.85in]{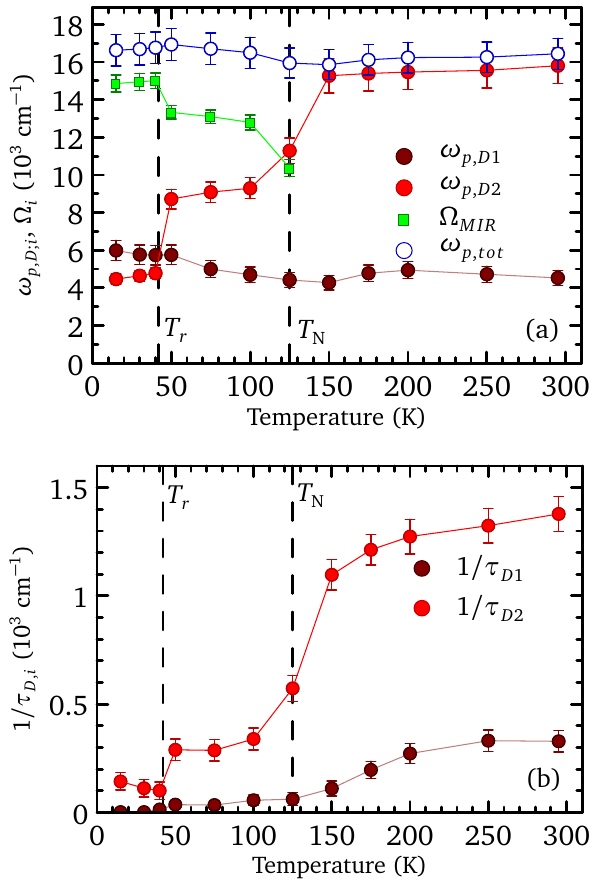}
\caption{(a) The temperature dependence of the plasma frequencies of the narrow (D1) and broad (D2)
Drude components, the oscillator strength of the mid-infrared peak ($\Omega_{\rm MIR}$, and
the total when these three components are added in quadrature ($\omega_{p, tot}$), for
Sr$_{0.67}$Na$_{0.33}$Fe$_2$As$_2$.
(b) The temperature dependence of the scattering rates of the narrow and broad Drude components.
}
\label{fig:results}
\end{figure}

%
% Fits to the complex conductivity
%
Previous optical studies of the iron-arsenic materials recognized that these are multiband materials with
hole and electron pockets at the center and corners of the Brillouin zone \cite{Singh2008,Fink2009}; a
minimal description consists of two electronic subsystems using the so-called two-Drude
model \cite{Wu2010}.  The complex dielectric function $\tilde\epsilon=\epsilon_1+i\epsilon_2$ can
be written as,
\begin{equation}
  \tilde\epsilon(\omega) = \epsilon_\infty - \sum_{j=1}^2 {{\omega_{p,D;j}^2}\over{\omega^2+i\omega/\tau_{D,j}}}
    + \sum_k {{\Omega_k^2}\over{\omega_k^2 - \omega^2 - i\omega\gamma_k}},
  \label{eq:eps}
\end{equation}
where $\epsilon_\infty$ is the real part at high frequency.  In the first sum,
$\omega_{p,D;j}^2 = 4\pi n_je^2/m^\ast_j$ and $1/\tau_{D,j}$ are the square of the
plasma frequency and scattering rate for the delocalized (Drude) carriers in the $j$th
band, respectively, and $n_j$ and $m^\ast_j$ are the carrier concentration and effective mass.
In the second summation, $\omega_k$, $\gamma_k$ and $\Omega_k$ are the position, width, and
strength of the $k$th vibration or bound excitation.  The complex conductivity is
$\tilde\sigma(\omega) = \sigma_1 +i\sigma_2 = -2\pi i \omega [\tilde\epsilon(\omega) -
\epsilon_\infty ]/Z_0$ (in units of $\Omega^{-1}$cm$^{-1}$); $Z_0\simeq 377$~$\Omega$ is
the impedance of free space.  The model is fit to the real and imaginary parts of the
optical conductivity simultaneously using a non-linear least-squares technique.
The results of the fits are shown in Figs.~\ref{fig:fits}(a), \ref{fig:fits}(b),
and \ref{fig:fits}(c) at 200~K ($T>T_{\rm N}$), 75~K ($T_r < T < T_{\rm N}$),
and 30~K ($T<T_r$), respectively; the combined response has been decomposed into
individual Drude and Lorentz components.  In agreement with previous studies on the
iron-based materials, the complex conductivity can be described by two Drude terms,
one weak and narrow (D1), the other strong and broad (D2), as well as several
Lorentzian oscillators.
The temperature dependence of the plasma frequencies, the D1 and D2 components, as
well as the strength of the mid-infrared (MIR) peak, are shown in Fig.~\ref{fig:results}(a);
the temperature dependence of the scattering rates for the two Drude components is shown
in Fig.~\ref{fig:results}(b).

%
%%%%%%%%%%%%%%%%%%%%%%%%%%%%%%%%%%%%%%%%%%%%%%%%%%%%%%%%%%%%%%%%%%%%%%%%%%%%%%%
%
% Figure 4 - fits for the parent compound
%
%

%
% T > T_N
%
\subsubsection{$T>T_{\rm N}$}
At room temperature, the plasma frequencies for the narrow and broad Drude terms,
$\omega_{p,D1}\simeq 4400$~cm$^{-1}$ and $\omega_{p,D2} \simeq 15\,800$~cm$^{-1}$,
respectively, are slightly less than those of the undoped parent compound SrFe$_2$As$_2$
($\omega_{p,D1}\simeq 5200$~cm$^{-1}$ and $\omega_{p,D2}\simeq 17\,700$~cm$^{-1}$); however,
the scattering rates of $1/\tau_{D1}\simeq 330$~cm$^{-1}$ and $1/\tau_{D2}\simeq 1400$~cm$^{-1}$
are noticeably lower than the values of $1/\tau_{D1}\simeq 470$~cm$^{-1}$ and
$1/\tau_{D2}\simeq2330$~cm$^{-1}$ observed in the undoped material \cite{Homes2016}.
This is somewhat surprising considering that in this material the layers
in between the Fe--As sheets are disordered.  While the plasma frequencies show
little temperature dependence between room temperature and $T_{\rm N}$, the scattering rates
for both Drude components decrease with temperature, with the narrow Drude decreasing from
about $1/\tau_{D1}\simeq 330$ to about 60~cm$^{-1}$, and the broad Drude decreasing from
$1/\tau_{D2}\simeq 1400$~cm$^{-1}$ to about 1100~cm$^{-1}$ just above $T_{\rm N}$.

%
% T_r < T < T_N
%
\subsubsection{$T_r < T < T_{\rm N}$}
Below $T_{\rm N}$ in the magnetic $C_2$ phase, the plasma frequency for the narrow
Drude increases slightly from $\omega_{p,D1}\simeq 4400$ to $\simeq 6000$~cm$^{-1}$,
while the scattering rate continues to decrease to $1/\tau_{D1}\simeq 40$~cm$^{-1}$
just above $T_r$.
The broad Drude displays much larger changes, with the plasma frequency decreasing from
$\omega_{p,D2} \simeq 15\,800$ to $9\,000$~cm$^{-1}$, which corresponds to a decrease in carrier
concentration of nearly 65\% ($\omega_p^2 \propto n/m^\ast$); the scattering rate also drops
dramatically from $1/\tau_{D2}\simeq 1100$~cm$^{-1}$ just above $T_{\rm N}$ to 300~cm$^{-1}$
in the $T_r < T < T_{\rm N}$ region.   The dramatic loss of spectral weight of the
broad Drude term is accompanied by the emergence of a new peak in the MIR region
with position $\omega_{\rm MIR}\simeq 950$~cm$^{-1}$, width $\gamma_{\rm MIR}\simeq 1550$~cm$^{-1}$,
and strength $\Omega_{\rm MIR}\simeq 13\,000$~cm$^{-1}$ [Fig.~\ref{fig:fits}(b)]; the
missing weight from the free carriers is transferred into this bound excitation, and
accordingly the total spectral weight is defined as $\omega_{p, tot}^2 =
\omega_{p,D1}^2 + \omega_{p,D2}^2 + \Omega_{\rm MIR}^2$, is constant, as shown in
Fig.~\ref{fig:results}(a).  This behavior is similar to what was previously observed in
the parent compound, and has been explained as the partial gapping of the pocket responsible
for the broad Drude term due and the appearance of a low-energy interband transition
\cite{Yin2011,Homes2016}.

%
% Figure 5: ARPES
%
\begin{figure*}[t]
\includegraphics[width=6.50in]{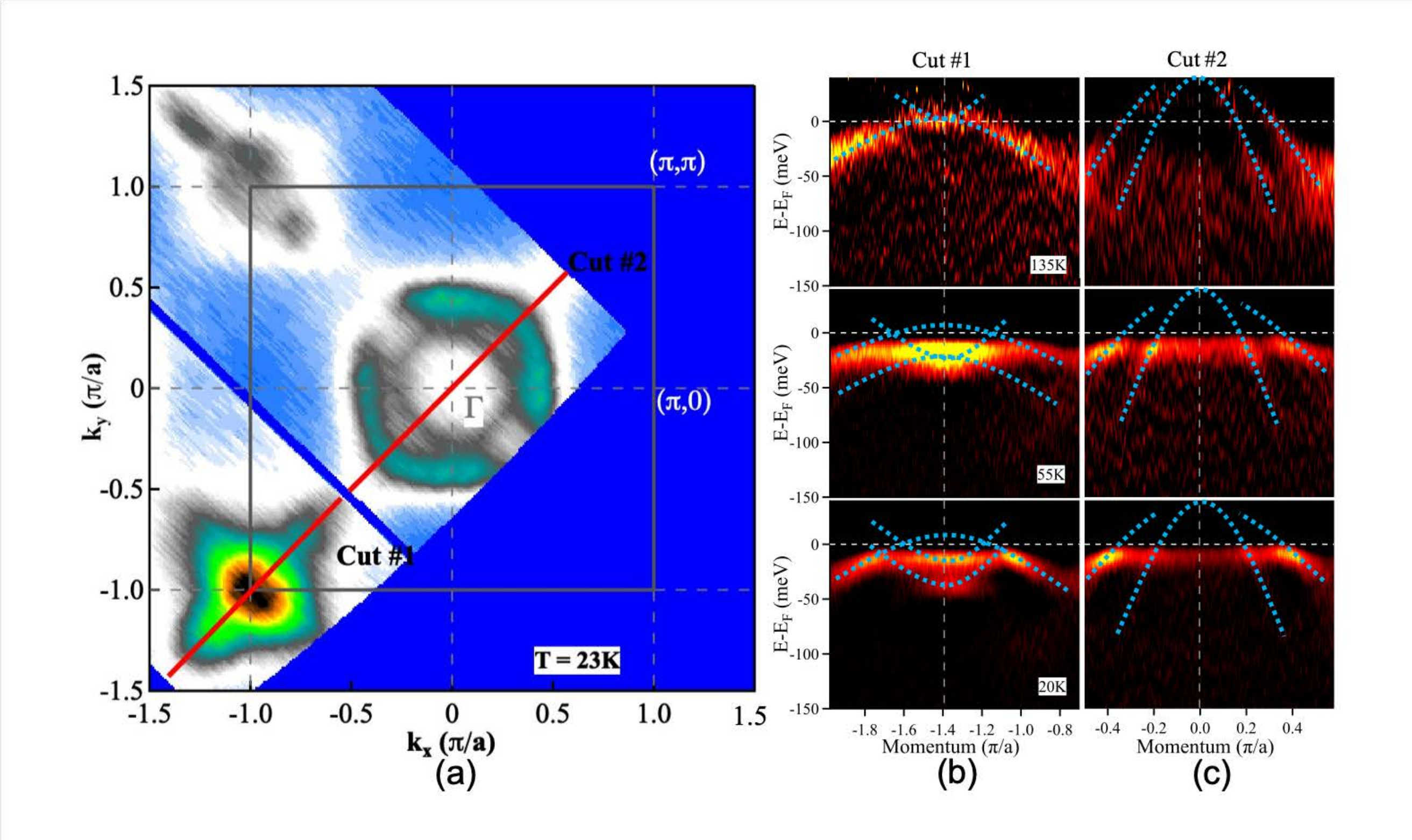}
\caption{(a) Fermi surface mapping of Sr$_{0.67}$Na$_{0.33}$Fe$_2$As$_2$ in the $C_4$ magnetic
phase at 23~K with the spectral weight integrated within a $\pm 10$~meV energy window with
respect to the Fermi level, showing the hole-like pockets at the center ($\Gamma$), and the
electron-like pockets at the corner (M)  of the Brillouin zone. Several different cuts are
shown along
the
$\Gamma\rightarrow {\rm M}$ path focus on the evolution of the hole and electron pockets.
(b) The temperature dependence of the second derivative of the energy bands measured along the
first cut around the M point at $(-\pi,-\pi)$ at 135~K ($T>T_{\rm N}$), 55~K ($T_r < T < T_{\rm N}$),
and 20~K ($T_c<T<T_r$).
(c) The temperature dependence of the second derivative of the energy bands measured along
the second cut around the $\Gamma$ point at 135, 55, and 20~K.  The dotted lines are drawn as
a guide to the eye.
}
\label{fig:arpes}
\end{figure*}

%
% T < T_r
%
\subsubsection{$T < T_r$}
As the temperature is reduced the system undergoes a further magnetic and structural
transition at $T_r\simeq  42$~K and enters the magnetic $C_4$ phase.  Below $T_r$ the plasma
frequency for the narrow Drude term appears to actually increase slightly; however, this
is accompanied by a dramatic collapse of $1/\tau_{D1}\simeq 40$~cm$^{-1}$ just above
$T_r$ to a value of $\simeq 2$~cm$^{-1}$ at 15~K; this is nearly an order of magnitude
smaller than what is observed in the parent compound \cite{Homes2016}.  Consequently,
the narrow Drude is no longer observable in $\sigma_1(\omega)$, leaving a relatively flat
optical conductivity due to the broad Drude term and Lorentzian components; instead,
its effects are determined from $\sigma_2(\omega)$ [shown in the inset of Fig.~\ref{fig:fits}(c)].
The plasma frequency of the broad Drude term continues to decrease from $\omega_{p,D2}
\simeq 9000$ to about $4200$~cm$^{-1}$ at 15~K, a further 80\% reduction in the carrier
concentration associated with this pocket, and over 90\% from the room temperature value;
this is comparable to what was observed in the parent compound for $T \ll T_{\rm N}$
\cite{Homes2016}.
In addition, the scattering rate decreases from $1/\tau_{D2}\simeq 300$~cm$^{-1}$ at
$T_r$ to $\simeq 120$~cm$^{-1}$ at 15~K.  At the same time, the peak at $\omega_{\rm MIR}
\simeq 950$~cm$^{-1}$ shifts down to about $\simeq 650$~cm$^{-1}$; while the width decreases
slightly to $\gamma_{\rm MIR}\simeq 1480$~cm$^{-1}$, the strength of this feature
increases to $\Omega_{\rm MIR}\simeq 15\,400$~cm$^{-1}$.  However, $\omega_{p,tot}$
continues to be conserved, indicating that the loss of spectral weight associated with
the free carriers in the broad Drude term has been transferred to this peak.

%
%%%%%%%%%%%%%%%%%%%%%%%%%%%%%%%%%%%%%%%%%%%%%%%%%%%%%%%%%%%%%%%%%%%%%%%%%%%%%%%
%
% Figure 6 - fits for the hole pocket
%
\begin{figure*}[t]
\includegraphics[width=2.3in]{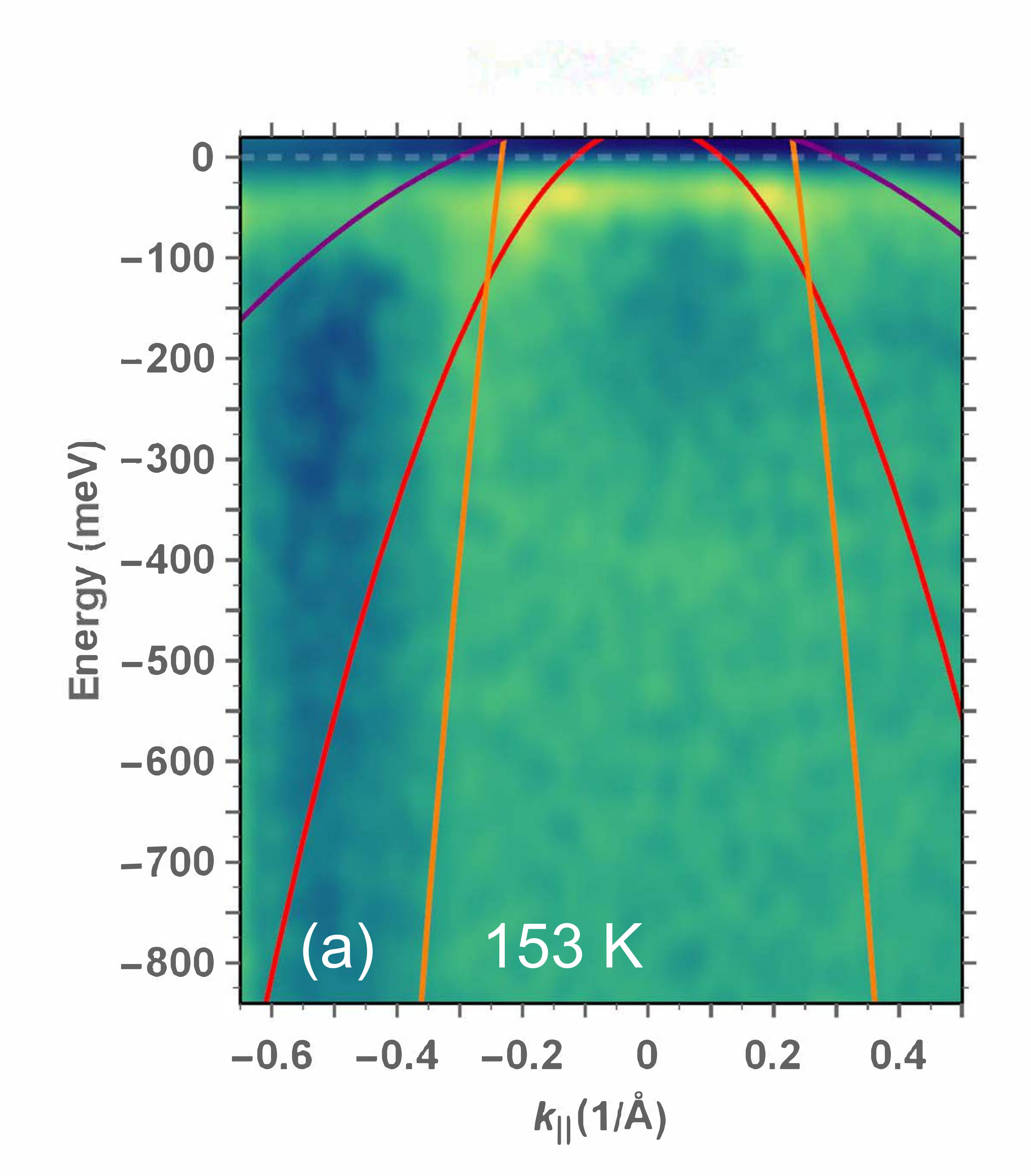}\
\includegraphics[width=2.3in]{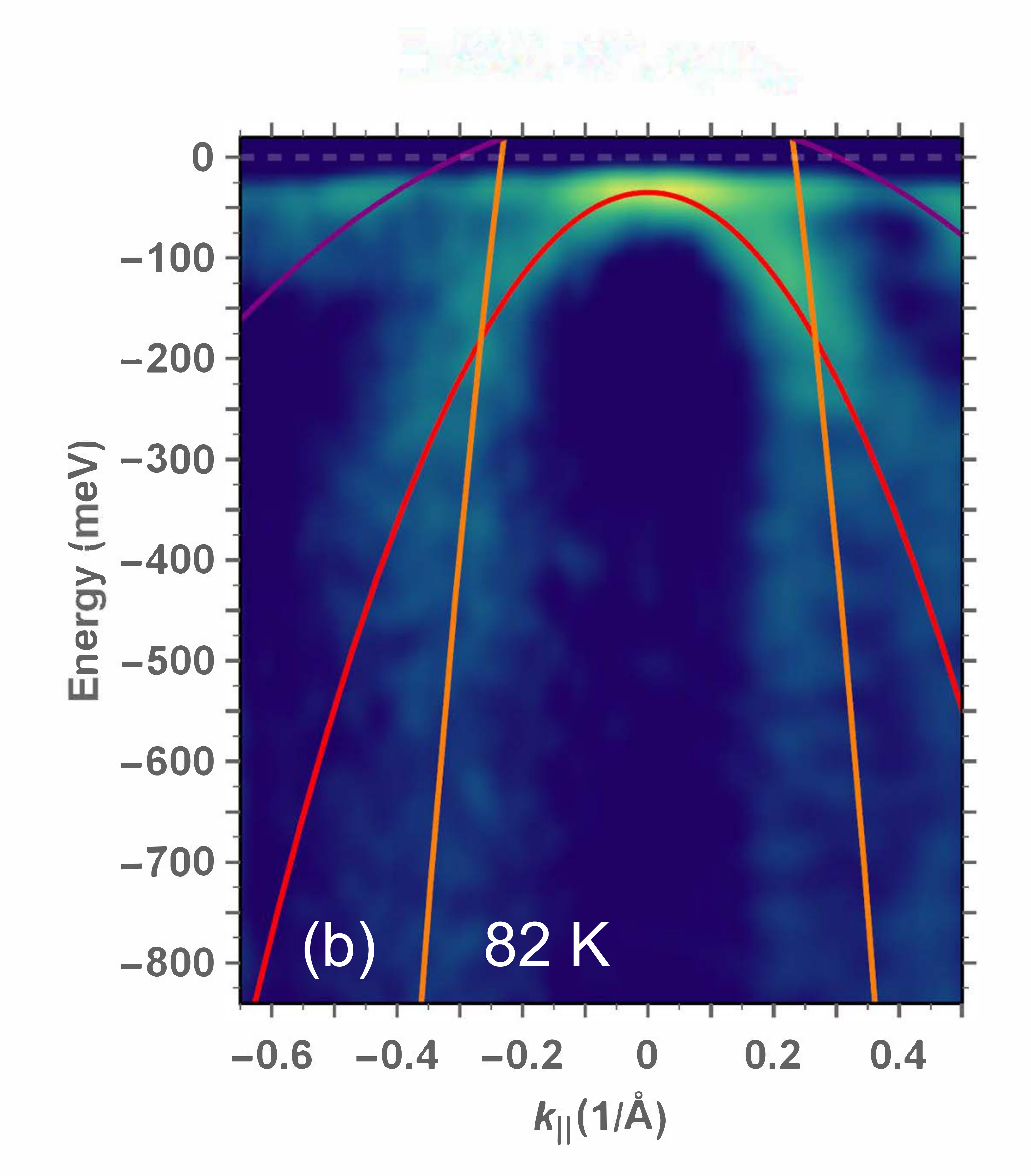}\
\includegraphics[width=2.3in]{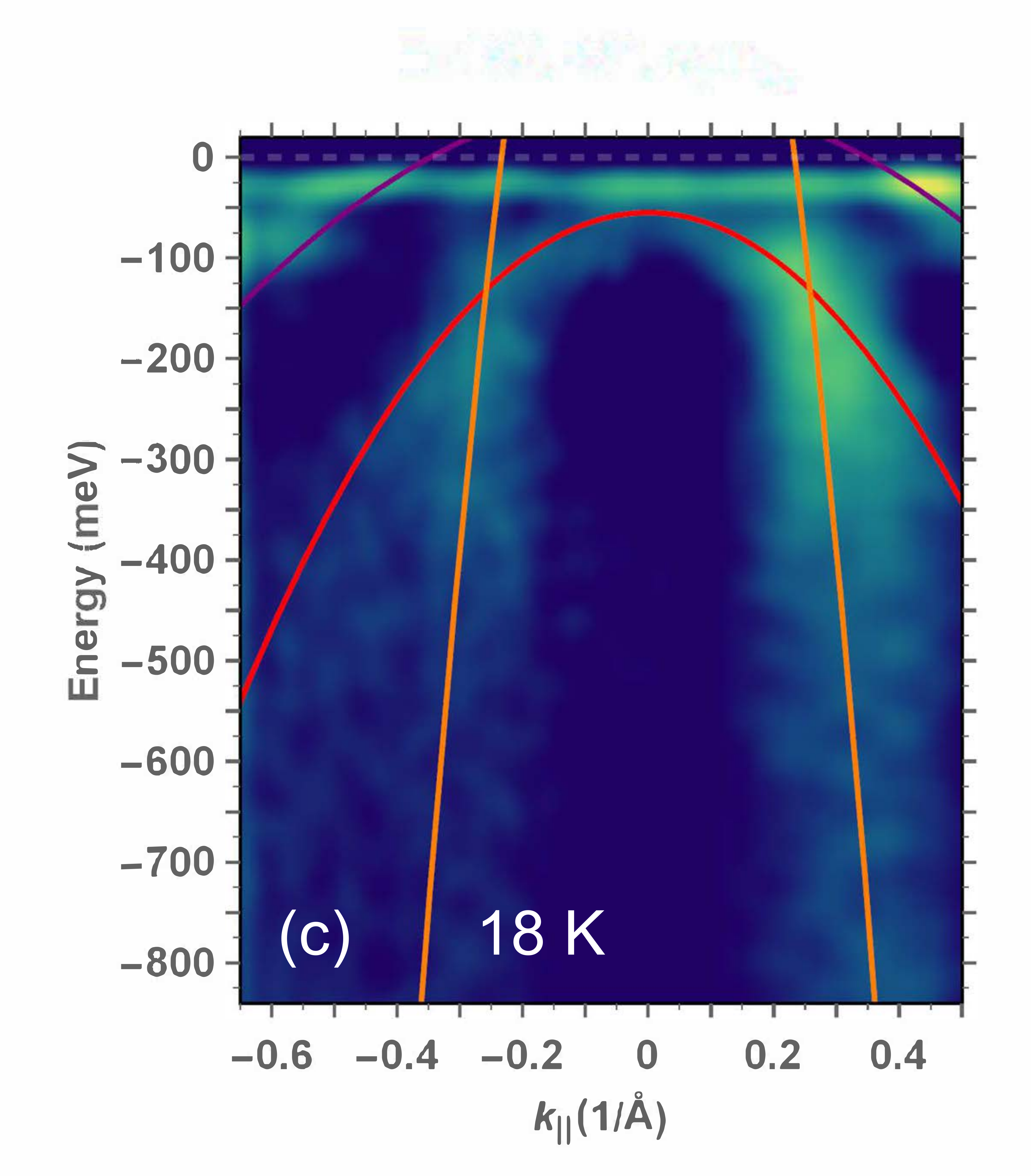}\
\caption{The temperature dependence of the second-derivative of the hole-like bands
of Sr$_{0.67}$Na$_{0.33}$Fe$_2$As$_2$ around the $\Gamma$ point along the $\Gamma
\rightarrow{\rm M}$ cut at: (a) above $T_{\rm N}$ at 153~K, (b) for $T_r < T < T_{\rm N}$
at 82~K, and (c) below $T_r$ at 18~K.  At high temperature three hole-like bands may
be resolved that cross $\epsilon_F$.  Below $T_{\rm N}$ one of these bands shift to
below the Fermi level; this trend continues below $T_r$ as the bands shift further
below $\epsilon_F$.  The lines are drawn as a guide to the eye.
}
\label{fig:holes}
\end{figure*}
%

%
% Below Tc
%
\subsubsection{$T<T_c$}
Below $T_c\simeq 10$~K there is a dramatic suppression of the low-frequency conductivity,
signalling the formation of a superconducting energy gap [Fig.~\ref{fig:sigma}(a) and
supplementary Fig.~S2].  Although the low-frequency data is somewhat limited, a comparison
of the optical conductivity for $T\gtrsim T_c$ and $T\ll T_c$ allows the superfluid
density, $\rho_s = \omega_{ps}^2$, where $\omega_{ps}$ is the superconducting plasma
frequency, to be determined from the missing spectral weight, calculated using the
Ferrell-Glover-Tinkham (FGT) sum rule \cite{Ferrell1958,Tinkham1959}.  The FGT sum rule
converges to $\omega_{ps}\simeq 5800\pm 500$~cm$^{-1}$, which corresponds to a
superconducting penetration depth of $\lambda\simeq 2700\pm 300$~\AA\ at 5~K,
comparable to the K-doped material \cite{Mallett2015a}; however, because
the lowest temperature obtained was only $\simeq T_c/2$, it is almost certain that
$\omega_{ps}$ is underestimated.  From Fig.~\ref{fig:sigma}(a) and supplementary
Fig.~S2, the characteristic energy scale for the superconducting energy gap is about
$2\Delta\simeq 50$~cm$^{-1}$.  In the narrow Drude band, $1/\tau_{D1}\ll 2\Delta$,
placing this material in the clean limit; as a result, most of the weight in the
condensate will come from this band.  In the broad Drude band, $1/\tau_{D2} > 2\Delta$,
placing this band in the dirty limit; consequently, only a small fraction of the
weight in this band will collapse into the condensate.  This is another example
of a multiband iron-based superconductor that is simultaneously in both the
clean and dirty limits \cite{Homes2015}.
One of the interesting properties of this material is its relatively low resistivity
just above $T_c$, $\rho_{ab}\simeq 20$~$\mu\Omega\,$cm, or $\sigma_{dc} \simeq
5\times10^4$~$\Omega^{-1}{\rm cm}^{-1}$ [Fig~\ref{fig:resis}].  These values place
this material just below the universal scaling line $\rho_s(T\ll T_c) \propto
\sigma_{dc}(T\gtrsim T_c)\,T_c$ \cite{Homes2004,Homes2005a,Homes2005b}, in close proximity
to other doped ``122'' superconductors, as well as many cuprate materials \cite{Tu2010}.

%
% Appearance of new peak at very low energy
%
\subsection{Low-energy peak}
The dramatic collapse of the scattering rate below $T_r$ of the narrow Drude allows a new
low-energy peak at $\omega_0\simeq 170$~cm$^{-1}$, with width $\gamma_0\simeq 110$~cm$^{-1}$ and
oscillator strength of $\Omega_0\simeq 2230$~cm$^{-1}$, to be observed [Figs.~\ref{fig:sigma}(a),
\ref{fig:fits}(c), and supplementary Fig.~S2].  This is close to where a peak was observed in
(CaFe$_{1-x}$Pt$_{x}$As)$_{10}$Pt$_{3}$As$_{8}$ for $x=0.1$ at $\simeq 120$~cm$^{-1}$ \cite{Yang2019};
that feature was attributed to a localization process due to impurity scattering described by a
classical generalization of the Drude model \cite{Smith2001},
\begin{equation}
  \tilde\sigma(\omega)= \left(\frac{2\pi}{Z_0}\right) \frac{\omega_p^2\tau}{(1-i\omega\tau)}
  \left[ 1+\frac{c}{(1-i\omega\tau)} \right],
  \label{eq:smith}
\end{equation}
where $c$ is the persistence of velocity that is retained for a single collision.
%
%This model has the interesting attribute that for $c=0$ a simple Drude is
%recovered, while for $c=-1$ the carriers are completely localized in the form of
%a Lorentzian oscillator with a peak at $\omega\tau=1$, width $2/\tau$, and an
%oscillator strength that is identical to the plasma frequency.
%
The scattering rate for the narrow Drude is far too small to yield a peak at the
experimentally-observed position, while the broad Drude predicts a localization
peak at $\simeq 120$~cm$^{-1}$, well below the experimentally-observed value
of $\omega_0 \simeq 170$~cm$^{-1}$ \footnote{Replacing the broad Drude term with
the expression in Eq.~(\ref{eq:smith}) and fitting to the real and imaginary
parts of the optical conductivity using a non-linear least-squares technique
yields $\omega_p \simeq 5350$~cm$^{-1}$, $1/\tau \simeq 146$~cm$^{-1}$, and
$c=-0.7$.  The plasma frequency is larger because it now describes both the
localized as well as free carriers; $\omega_{p}^2 \simeq \omega_{p,D2}^2 +
\Omega_0^2$.}.
Thus, it is likely that the low-energy peak originates from a further reconstruction
of the Fermi surface in the $C_4$ phase rather than any sort of localization process.
Indeed, a remarkably similar peak has also been observed to emerge at $\simeq 150$~cm$^{-1}$
in the optical conductivity of underdoped Ba$_{1-x}$K$_x$Fe$_2$As$_2$ at low temperature
\cite{Dai2012}; this feature may also be a related to the magnetic $C_4$ phase
observed in that compound.

%
% Angle resolved photoemission
%
\subsection{ARPES}
A simple density functional theory calculation of SrFe$_2$As$_2$ in the paramagnetic
high-temperature tetragonal phase reveals a familiar band structure consisting of
three hole-like pockets at the center of the Brillouin zone ($\Gamma$), and two
electron-like pockets at the corners (M); the orbital character is primarily Fe
$d_{xz}/d_{yz}$ in nature (shown in supplementary Fig.~S4, details of the
calculation are discussed in the Supplementary Material.)
The Fermi surface of Sr$_{0.67}$Na$_{0.33}$Fe$_2$As$_2$, with the spectral weight
integrated within a $\pm 10$~meV energy window with respect to the Fermi level, is
shown below $T_r$ in the $C_4$ magnetic phase at 23~K, in Fig.~\ref{fig:arpes}(a).
Two momentum cuts have been made along the $\Gamma \rightarrow {\rm M}$ path; the
first examines the temperature dependence of the anisotropic electron-like bands
around an M point, Fig.~\ref{fig:arpes}(b), and the second details the behavior
of the isotropic hole-like pockets around the $\Gamma$ point, shown in
Fig.~\ref{fig:arpes}(c).  This Fermi surface is qualitatively similar to what
was observed in Ba$_{1-x}$K$_x$Fe$_2$As$_2$ \cite{Zabolotnyy2009,Derondeau2017}

%
% electron pockets
%
At high temperature, the cut along the $\Gamma \rightarrow {\rm M}$ direction at the M
point there appears to be a hole-like band as well as a possible electron-like band
at 135~K, shown in the upper panel of Fig.~\ref{fig:arpes}(b).  In the simple picture for
the Fermi surface of SrFe$_2$As$_2$ (supplementary Fig.~S4) this result can be reproduced
by lowering the Fermi level $\epsilon_F$ by about 0.2~eV, which is consistent with the
removal of electrons due to sodium substitution (hole doping).  As the temperature is
lowered below $T_{\rm N}$ and enters the magnetic $C_2$ phase, the hole-like band may
split, while the electron-like band appears to shift below $\epsilon_F$.  Below $T_r$
in the $C_4$ magnetic phase, a single hole-like band is recovered, while the electron-like
band now appears to be split into two bands, with a separation of $\simeq 20$~meV, which
is comparable to the position of the low-energy peak (this behavior is explored further
in supplementary Fig.~S5).

%
% Hole pockets
%
The initial investigation into the temperature dependence of the energy bands
around the $\Gamma$ point in Fig.~\ref{fig:arpes}(c) revealed two large hole
pockets at the Fermi level, but relatively little temperature dependence.  This
prompted a more detailed investigation of the hole-like bands along the $\Gamma
\rightarrow {\rm M}$ path, shown in Fig.~\ref{fig:holes} (further detail is provided
in supplementary Figs.~S6 and S7).   Above $T_{\rm N}$ the bands are rather
broad, but at least three bands may be resolved, all of which cross the Fermi level,
resulting in several large hole-like Fermi surfaces, shown in the second-derivative
curves in Fig.~\ref{fig:holes}(a).  Below $T_{\rm N}$ the bands sharpen considerably in
the $C_2$ phase, and one of the bands is observed to shift to $\simeq 40$~meV below
the Fermi level, shown in Fig.~\ref{fig:holes}(b), leading to the removal of a hole-like Fermi
surface; this is consistent with the Fermi surface reconstruction below $T_{\rm N}$
observed in the parent compounds \cite{Yi2009,Yin2011}.  This trend continues in the
magnetic $C_4$ phase, with the band shifting to $\simeq 60$~meV below the Fermi
level, Fig.~\ref{fig:holes}(c).
%
% Discussion of optics and ARPES
%

\vspace*{-0.2cm}
\subsection{Discussion}
%
% hole pockets
%
Both the electron and hole pockets appear to undergo significant changes in
response to the Fermi surface reconstruction in the magnetic $C_2$ and $C_4$
phases that exhibit SDW and CSDW order, respectively.
In the case of the hole pockets, the fact that one of the bands shifts
below $\epsilon_F$ below $T_{\rm N}$ in the magnetic $C_2$ phase, shifting further
below $T_r$ in the magnetic $C_4$ phase, signals the decrease in the
size of the Fermi surface associated with the hole pockets.  It is possible that
this may be related to the dramatic decrease in the spectral weight of the broad
Drude component as described by the plasma frequency in Fig.~\ref{fig:results}(a);
from $\omega_{p,D2}^2\propto n/m^\ast$ we infer a significant decrease in the
carriers associated with the hole pockets at low temperature ($\simeq 90$\%
reduction of the room temperature value).
%
%It is estimated for $T<T_N, T_r$ in Figs.~\ref{fig:holes}(b) and \ref{fig:holes}(c),
%respectively, that the direct transition at the $\Gamma$ point from the bands just
%below $\epsilon_F$ to the band that lies just above $\epsilon_F$ is $\simeq 80 $~meV,
%which is similar to the position of the mid-infrared peak at low temperature; however,
%the uncertainty regarding the position of the upper band rules out definitive assignment.
%

%
% electron pockets
%
The evolution of the electron-like bands is more complicated, as the bands at the
M point have both electron- and hole-like character.  The initial splitting of the
hole-like band below $T_{\rm N}$ is consistent with the lifting of the degeneracy between
the $d_{xz}$ and $d_{yz}$ orbitals; however, the fact that one of the hole-like
bands lies completely below the Fermi level suggests no significant changes to the size
of the Fermi surfaces.  Below $T_r$ the orbital degeneracy is restored, but the presence of
CSDW order leads to the formation of a supercell; the electron-like bands are split
as a result of zone-folding, which may lead to an increase in the size of the Fermi
surface.  This is consistent with the slight increase in the plasma frequency of the
narrow Drude component at low temperature, shown in Fig.~\ref{fig:results}(a).
Furthermore, the splitting between the two electron-like bands of $\simeq 20$~meV,
is very close to the position of the low-energy peak.  This suggests that,
similar to the mid-infrared peak, the low-energy peak emerges in response to
the Fermi surface reconstruction driven by the $C_4$ magnetic phase and the CSDW
order at low temperature \cite{Yi2018}.

%
% Conclusions
%
\section{Summary}
The ARPES and complex optical properties of freshly-cleaved surfaces of the iron-based
superconductor Sr$_{0.67}$Na$_{0.33}$Fe$_2$As$_2$ have been determined for light polarized
in the iron-arsenic (\emph{a-b}) planes at a variety of temperatures for the room temperature
tetragonal paramagnetic phase, the orthorhombic $C_2$ SDW magnetic phase, the tetragonal $C_4$
double-$\mathbf{Q}$ SDW (CSDW) phase, as well as below $T_c$ in the superconducting state.
The free-carrier response is described by two Drude components, one broad and strong, the other
narrow and weak.  The strength of the narrow component shows little temperature dependence,
increasing slightly in strength at low temperature, while narrowing dramatically.  The broad
Drude component decreases dramatically in strength and narrows below $T_{\rm N}$ at the same
time a peak emerges in the mid-infrared; the decrease in the spectral weight associated with
the free carriers is transferred into the emergent peak.    Below $T_r$, this trend continues,
with the emergence of a new low-energy peak at $\simeq 20$~meV.  The appearance of a new
infrared-active mode in the Fe--As planes below $T_r$ is attributed to zone-folding due to
the formation of a supercell in response to the CSDW; this suggests that the low-energy peak
originates from a further Fermi surface reconstruction in the $C_4$ phase.
Below $T_c$ the low-frequency conductivity decreases dramatically, signalling the formation of
a superconducting energy gap.
ARPES reveals large hole-like Fermi surfaces at the $\Gamma$ point, one of which
is apparently removed below the structural and magnetic transitions, suggesting
that they may be related to the behavior of the broad Drude component.  The electron-
and hole-like bands at the corners of the Brillouin zone shift and split below $T_{\rm N}$
and $T_r$, but the Fermi surfaces do not appear to undergo any significant change in size,
suggesting they may be related to the narrow Drude component; the apparent splitting of
the electron-like bands in the  $C_4$ phase would appear to explain the emergence of the
low-energy peak at $\simeq 20$~meV in the optical conductivity.
While the $C_2$ and $C_4$ magnetic transitions, with resulting SDW and CSDW order,
respectively, lead to a significant reconstruction of the Fermi surface that has
profound implications for the transport originating from the electron- and hole-like
pockets, they appear to have relatively little impact on the superconductivity in
this material.

%
%%%%%%%%%%%%%%%%%%%%%%%%%%%%%%%%%%%%%%%%%%%%%%%%%%%%%%%%%%%%%%%%%%%%%%%%%%%%%%%
%
% Acknowledgment
%
%
% XJZ: NSFC add 11888101
% XJZ: MOST add 2016YFA0300300
%
\begin{acknowledgments}
Work at Chinese Academy of Science was supported by NSFC (Project Nos. 11774400,
11888101, and 11974412) and  MOST (Project Nos. 2015CB921102, 2016YFA0300300,
and 2017YFA0302903).  Work at HPSTAR was supported by NSAF, Grant No.~U1530402.
Work at Brookhaven National Laboratory was supported by the Office of Science, U.S.
Department of Energy under Contract No. DE-SC0012704.
\end{acknowledgments}

%%%%%%%%%%%%%%%%%%%%%%%%%%%%%%%%%%%%%%%%%%%%%%%%%%%%%%%%%%%%%%%%%%%%%%%%%%%%%%%
%
% The bibliography (BibTeX)
%
%\bibliography{references}

\begin{thebibliography}{77}%
\makeatletter
\providecommand \@ifxundefined [1]{%
 \@ifx{#1\undefined}
}%
\providecommand \@ifnum [1]{%
 \ifnum #1\expandafter \@firstoftwo
 \else \expandafter \@secondoftwo
 \fi
}%
\providecommand \@ifx [1]{%
 \ifx #1\expandafter \@firstoftwo
 \else \expandafter \@secondoftwo
 \fi
}%
\providecommand \natexlab [1]{#1}%
\providecommand \enquote  [1]{``#1''}%
\providecommand \bibnamefont  [1]{#1}%
\providecommand \bibfnamefont [1]{#1}%
\providecommand \citenamefont [1]{#1}%
\providecommand \href@noop [0]{\@secondoftwo}%
\providecommand \href [0]{\begingroup \@sanitize@url \@href}%
\providecommand \@href[1]{\@@startlink{#1}\@@href}%
\providecommand \@@href[1]{\endgroup#1\@@endlink}%
\providecommand \@sanitize@url [0]{\catcode `\\12\catcode `\$12\catcode
  `\&12\catcode `\#12\catcode `\^12\catcode `\_12\catcode `\%12\relax}%
\providecommand \@@startlink[1]{}%
\providecommand \@@endlink[0]{}%
\providecommand \url  [0]{\begingroup\@sanitize@url \@url }%
\providecommand \@url [1]{\endgroup\@href {#1}{\urlprefix }}%
\providecommand \urlprefix  [0]{URL }%
\providecommand \Eprint [0]{\href }%
\providecommand \doibase [0]{http://dx.doi.org/}%
\providecommand \selectlanguage [0]{\@gobble}%
\providecommand \bibinfo  [0]{\@secondoftwo}%
\providecommand \bibfield  [0]{\@secondoftwo}%
\providecommand \translation [1]{[#1]}%
\providecommand \BibitemOpen [0]{}%
\providecommand \bibitemStop [0]{}%
\providecommand \bibitemNoStop [0]{.\EOS\space}%
\providecommand \EOS [0]{\spacefactor3000\relax}%
\providecommand \BibitemShut  [1]{\csname bibitem#1\endcsname}%
\let\auto@bib@innerbib\@empty
%</preamble>
\bibitem [{\citenamefont {Johnston}(2010)}]{Johnston2010}%
  \BibitemOpen
  \bibfield  {author} {\bibinfo {author} {\bibfnamefont {David~C.}\
  \bibnamefont {Johnston}},\ }\bibfield  {title} {\enquote {\bibinfo {title}
  {The puzzle of high temperature superconductivity in layered iron pnictides
  and chalcogenides},}\ }\href {\doibase 10.1080/00018732.2010.513480}
  {\bibfield  {journal} {\bibinfo  {journal} {Adv. Phys.}\ }\textbf {\bibinfo
  {volume} {59}},\ \bibinfo {pages} {803--1061} (\bibinfo {year}
  {2010})}\BibitemShut {NoStop}%
\bibitem [{\citenamefont {Paglione}\ and\ \citenamefont
  {Greene}(2010)}]{Paglione2010}%
  \BibitemOpen
  \bibfield  {author} {\bibinfo {author} {\bibfnamefont {Johnpierre}\
  \bibnamefont {Paglione}}\ and\ \bibinfo {author} {\bibfnamefont {Richard~L.}\
  \bibnamefont {Greene}},\ }\bibfield  {title} {\enquote {\bibinfo {title}
  {High-temperature superconductivity in iron-based materials},}\ }\href
  {\doibase 10.1038/nphys1759} {\bibfield  {journal} {\bibinfo  {journal} {Nat.
  Phys.}\ }\textbf {\bibinfo {volume} {6}},\ \bibinfo {pages} {645--658}
  (\bibinfo {year} {2010})}\BibitemShut {NoStop}%
\bibitem [{\citenamefont {Canfield}\ and\ \citenamefont
  {Bud'ko}(2010)}]{Canfield2010}%
  \BibitemOpen
  \bibfield  {author} {\bibinfo {author} {\bibfnamefont {Paul~C.}\ \bibnamefont
  {Canfield}}\ and\ \bibinfo {author} {\bibfnamefont {Sergey~L.}\ \bibnamefont
  {Bud'ko}},\ }\bibfield  {title} {\enquote {\bibinfo {title} {{FeAs-Based
  Superconductivity: A Case Study of the Effects of Transition Metal Doping on
  BaFe$_2$As$_2$}},}\ }\href {\doibase
  10.1146/annurev-conmatphys-070909-104041} {\bibfield  {journal} {\bibinfo
  {journal} {Ann. Rev. Cond. Mat. Phys.}\ }\textbf {\bibinfo {volume} {1}},\
  \bibinfo {pages} {27--50} (\bibinfo {year} {2010})}\BibitemShut {NoStop}%
\bibitem [{\citenamefont {Si}\ \emph {et~al.}(2016)\citenamefont {Si},
  \citenamefont {Yu},\ and\ \citenamefont {Abrahams}}]{Si2016}%
  \BibitemOpen
  \bibfield  {author} {\bibinfo {author} {\bibfnamefont {Qimiao}\ \bibnamefont
  {Si}}, \bibinfo {author} {\bibfnamefont {Rong}\ \bibnamefont {Yu}}, \ and\
  \bibinfo {author} {\bibfnamefont {Elihu}\ \bibnamefont {Abrahams}},\
  }\bibfield  {title} {\enquote {\bibinfo {title} {High-temperature
  superconductivity in iron pnictides and chalcogenides},}\ }\href {\doibase
  10.1038/natrevmats.2016.17} {\bibfield  {journal} {\bibinfo  {journal} {Nat.
  Rev. Mater.}\ }\textbf {\bibinfo {volume} {1}},\ \bibinfo {pages} {16017}
  (\bibinfo {year} {2016})}\BibitemShut {NoStop}%
\bibitem [{\citenamefont {Dean}\ \emph {et~al.}(2012)\citenamefont {Dean},
  \citenamefont {Kim}, \citenamefont {Kreyssig}, \citenamefont {Kim},
  \citenamefont {Liu}, \citenamefont {Ryan}, \citenamefont {Thaler},
  \citenamefont {Bud'ko}, \citenamefont {Strassheim}, \citenamefont {Canfield},
  \citenamefont {Hill},\ and\ \citenamefont {Goldman}}]{Dean2012}%
  \BibitemOpen
  \bibfield  {author} {\bibinfo {author} {\bibfnamefont {M.~P.~M.}\
  \bibnamefont {Dean}}, \bibinfo {author} {\bibfnamefont {M.~G.}\ \bibnamefont
  {Kim}}, \bibinfo {author} {\bibfnamefont {A.}~\bibnamefont {Kreyssig}},
  \bibinfo {author} {\bibfnamefont {J.~W.}\ \bibnamefont {Kim}}, \bibinfo
  {author} {\bibfnamefont {X.}~\bibnamefont {Liu}}, \bibinfo {author}
  {\bibfnamefont {P.~J.}\ \bibnamefont {Ryan}}, \bibinfo {author}
  {\bibfnamefont {A.}~\bibnamefont {Thaler}}, \bibinfo {author} {\bibfnamefont
  {S.~L.}\ \bibnamefont {Bud'ko}}, \bibinfo {author} {\bibfnamefont
  {W.}~\bibnamefont {Strassheim}}, \bibinfo {author} {\bibfnamefont {P.~C.}\
  \bibnamefont {Canfield}}, \bibinfo {author} {\bibfnamefont {J.~P.}\
  \bibnamefont {Hill}}, \ and\ \bibinfo {author} {\bibfnamefont {A.~I.}\
  \bibnamefont {Goldman}},\ }\bibfield  {title} {\enquote {\bibinfo {title}
  {{Magnetically polarized Ir dopant atoms in superconducting
  Ba(Fe$_{1-x}$Ir$_{x}$)$_{2}$As$_{2}$}},}\ }\href {\doibase
  10.1103/PhysRevB.85.140514} {\bibfield  {journal} {\bibinfo  {journal} {Phys.
  Rev. B}\ }\textbf {\bibinfo {volume} {85}},\ \bibinfo {pages} {140514(R)}
  (\bibinfo {year} {2012})}\BibitemShut {NoStop}%
\bibitem [{\citenamefont {Dai}(2015)}]{Dai2015}%
  \BibitemOpen
  \bibfield  {author} {\bibinfo {author} {\bibfnamefont {Pengcheng}\
  \bibnamefont {Dai}},\ }\bibfield  {title} {\enquote {\bibinfo {title}
  {Antiferromagnetic order and spin dynamics in iron-based superconductors},}\
  }\href {\doibase 10.1103/RevModPhys.87.855} {\bibfield  {journal} {\bibinfo
  {journal} {Rev. Mod. Phys.}\ }\textbf {\bibinfo {volume} {87}},\ \bibinfo
  {pages} {855--896} (\bibinfo {year} {2015})}\BibitemShut {NoStop}%
\bibitem [{\citenamefont {Moroni}\ \emph {et~al.}(2017)\citenamefont {Moroni},
  \citenamefont {Carretta}, \citenamefont {Allodi}, \citenamefont {De~Renzi},
  \citenamefont {Gastiasoro}, \citenamefont {Andersen}, \citenamefont
  {Materne}, \citenamefont {Klauss}, \citenamefont {Kobayashi}, \citenamefont
  {Sato},\ and\ \citenamefont {Sanna}}]{Moroni2017}%
  \BibitemOpen
  \bibfield  {author} {\bibinfo {author} {\bibfnamefont {M.}~\bibnamefont
  {Moroni}}, \bibinfo {author} {\bibfnamefont {P.}~\bibnamefont {Carretta}},
  \bibinfo {author} {\bibfnamefont {G.}~\bibnamefont {Allodi}}, \bibinfo
  {author} {\bibfnamefont {R.}~\bibnamefont {De~Renzi}}, \bibinfo {author}
  {\bibfnamefont {M.~N.}\ \bibnamefont {Gastiasoro}}, \bibinfo {author}
  {\bibfnamefont {B.~M.}\ \bibnamefont {Andersen}}, \bibinfo {author}
  {\bibfnamefont {P.}~\bibnamefont {Materne}}, \bibinfo {author} {\bibfnamefont
  {H.-H.}\ \bibnamefont {Klauss}}, \bibinfo {author} {\bibfnamefont
  {Y.}~\bibnamefont {Kobayashi}}, \bibinfo {author} {\bibfnamefont
  {M.}~\bibnamefont {Sato}}, \ and\ \bibinfo {author} {\bibfnamefont
  {S.}~\bibnamefont {Sanna}},\ }\bibfield  {title} {\enquote {\bibinfo {title}
  {{Fast recovery of the stripe magnetic order by Mn/Fe substitution in F-doped
  LaFeAsO superconductors}},}\ }\href {\doibase 10.1103/PhysRevB.95.180501}
  {\bibfield  {journal} {\bibinfo  {journal} {Phys. Rev. B}\ }\textbf {\bibinfo
  {volume} {95}},\ \bibinfo {pages} {180501(R)} (\bibinfo {year}
  {2017})}\BibitemShut {NoStop}%
\bibitem [{\citenamefont {Kreyssig}\ \emph {et~al.}(2018)\citenamefont
  {Kreyssig}, \citenamefont {Wilde}, \citenamefont {B\"ohmer}, \citenamefont
  {Tian}, \citenamefont {Meier}, \citenamefont {Li}, \citenamefont {Ueland},
  \citenamefont {Xu}, \citenamefont {Bud'ko}, \citenamefont {Canfield},
  \citenamefont {McQueeney},\ and\ \citenamefont {Goldman}}]{Kreyssig2018}%
  \BibitemOpen
  \bibfield  {author} {\bibinfo {author} {\bibfnamefont {A.}~\bibnamefont
  {Kreyssig}}, \bibinfo {author} {\bibfnamefont {J.~M.}\ \bibnamefont {Wilde}},
  \bibinfo {author} {\bibfnamefont {A.~E.}\ \bibnamefont {B\"ohmer}}, \bibinfo
  {author} {\bibfnamefont {W.}~\bibnamefont {Tian}}, \bibinfo {author}
  {\bibfnamefont {W.~R.}\ \bibnamefont {Meier}}, \bibinfo {author}
  {\bibfnamefont {Bing}\ \bibnamefont {Li}}, \bibinfo {author} {\bibfnamefont
  {B.~G.}\ \bibnamefont {Ueland}}, \bibinfo {author} {\bibfnamefont {Mingyu}\
  \bibnamefont {Xu}}, \bibinfo {author} {\bibfnamefont {S.~L.}\ \bibnamefont
  {Bud'ko}}, \bibinfo {author} {\bibfnamefont {P.~C.}\ \bibnamefont
  {Canfield}}, \bibinfo {author} {\bibfnamefont {R.~J.}\ \bibnamefont
  {McQueeney}}, \ and\ \bibinfo {author} {\bibfnamefont {A.~I.}\ \bibnamefont
  {Goldman}},\ }\bibfield  {title} {\enquote {\bibinfo {title}
  {Antiferromagnetic order in {CaK}({Fe}$_{1-x}${Ni}$_{x}$)$_{4}${As}$_{4}$ and
  its interplay with superconductivity},}\ }\href {\doibase
  10.1103/PhysRevB.97.224521} {\bibfield  {journal} {\bibinfo  {journal} {Phys.
  Rev. B}\ }\textbf {\bibinfo {volume} {97}},\ \bibinfo {pages} {224521}
  (\bibinfo {year} {2018})}\BibitemShut {NoStop}%
\bibitem [{\citenamefont {Meier}\ \emph {et~al.}(2018)\citenamefont {Meier},
  \citenamefont {Ding}, \citenamefont {Kreyssig}, \citenamefont {Bud'ko},
  \citenamefont {Sapkota}, \citenamefont {Kothapalli}, \citenamefont {Borisov},
  \citenamefont {Valent\'{i}}, \citenamefont {Batista}, \citenamefont {Orth},
  \citenamefont {Fernandes}, \citenamefont {Goldman}, \citenamefont {Furukawa},
  \citenamefont {B\"{o}hmer},\ and\ \citenamefont {Canfield}}]{Meier2018}%
  \BibitemOpen
  \bibfield  {author} {\bibinfo {author} {\bibfnamefont {William~R.}\
  \bibnamefont {Meier}}, \bibinfo {author} {\bibfnamefont {Qing-Ping}\
  \bibnamefont {Ding}}, \bibinfo {author} {\bibfnamefont {Andreas}\
  \bibnamefont {Kreyssig}}, \bibinfo {author} {\bibfnamefont {Sergey~L.}\
  \bibnamefont {Bud'ko}}, \bibinfo {author} {\bibfnamefont {Aashish}\
  \bibnamefont {Sapkota}}, \bibinfo {author} {\bibfnamefont {Karunakar}\
  \bibnamefont {Kothapalli}}, \bibinfo {author} {\bibfnamefont {Vladislav}\
  \bibnamefont {Borisov}}, \bibinfo {author} {\bibfnamefont {Roser}\
  \bibnamefont {Valent\'{i}}}, \bibinfo {author} {\bibfnamefont {Cristian~D.}\
  \bibnamefont {Batista}}, \bibinfo {author} {\bibfnamefont {Peter~P.}\
  \bibnamefont {Orth}}, \bibinfo {author} {\bibfnamefont {Rafael~M.}\
  \bibnamefont {Fernandes}}, \bibinfo {author} {\bibfnamefont {Alan~I.}\
  \bibnamefont {Goldman}}, \bibinfo {author} {\bibfnamefont {Yuji}\
  \bibnamefont {Furukawa}}, \bibinfo {author} {\bibfnamefont {Anna~E.}\
  \bibnamefont {B\"{o}hmer}}, \ and\ \bibinfo {author} {\bibfnamefont
  {Paul~C.}\ \bibnamefont {Canfield}},\ }\bibfield  {title} {\enquote {\bibinfo
  {title} {Hedgehog spin-vortex crystal stabilized in a hole-doped iron-based
  superconductor},}\ }\href {\doibase 10.1038/s41535-017-0076-x} {\bibfield
  {journal} {\bibinfo  {journal} {npj Quantum Materials}\ }\textbf {\bibinfo
  {volume} {3}},\ \bibinfo {pages} {5} (\bibinfo {year} {2018})}\BibitemShut
  {NoStop}%
\bibitem [{\citenamefont {Rotter}\ \emph {et~al.}(2008)\citenamefont {Rotter},
  \citenamefont {Tegel},\ and\ \citenamefont {Johrendt}}]{Rotter2008}%
  \BibitemOpen
  \bibfield  {author} {\bibinfo {author} {\bibfnamefont {Marianne}\
  \bibnamefont {Rotter}}, \bibinfo {author} {\bibfnamefont {Marcus}\
  \bibnamefont {Tegel}}, \ and\ \bibinfo {author} {\bibfnamefont {Dirk}\
  \bibnamefont {Johrendt}},\ }\bibfield  {title} {\enquote {\bibinfo {title}
  {{Superconductivity at 38~K in the Iron Arsenide
  (Ba$_{1-x}$K$_x$)Fe$_2$As$_2$}},}\ }\href {\doibase
  10.1103/PhysRevLett.101.107006} {\bibfield  {journal} {\bibinfo  {journal}
  {Phys. Rev. Lett.}\ }\textbf {\bibinfo {volume} {101}},\ \bibinfo {pages}
  {107006} (\bibinfo {year} {2008})}\BibitemShut {NoStop}%
\bibitem [{\citenamefont {Sefat}\ \emph {et~al.}(2008)\citenamefont {Sefat},
  \citenamefont {Jin}, \citenamefont {McGuire}, \citenamefont {Sales},
  \citenamefont {Singh},\ and\ \citenamefont {Mandrus}}]{Sefat2008}%
  \BibitemOpen
  \bibfield  {author} {\bibinfo {author} {\bibfnamefont {Athena~S.}\
  \bibnamefont {Sefat}}, \bibinfo {author} {\bibfnamefont {Rongying}\
  \bibnamefont {Jin}}, \bibinfo {author} {\bibfnamefont {Michael~A.}\
  \bibnamefont {McGuire}}, \bibinfo {author} {\bibfnamefont {Brian~C.}\
  \bibnamefont {Sales}}, \bibinfo {author} {\bibfnamefont {David~J.}\
  \bibnamefont {Singh}}, \ and\ \bibinfo {author} {\bibfnamefont {David}\
  \bibnamefont {Mandrus}},\ }\bibfield  {title} {\enquote {\bibinfo {title}
  {{Superconductivity at 22~K in Co-Doped BaFe$_{2}$As$_{2}$ Crystals}},}\
  }\href {\doibase 10.1103/PhysRevLett.101.117004} {\bibfield  {journal}
  {\bibinfo  {journal} {Phys. Rev. Lett.}\ }\textbf {\bibinfo {volume} {101}},\
  \bibinfo {pages} {117004} (\bibinfo {year} {2008})}\BibitemShut {NoStop}%
\bibitem [{\citenamefont {Ni}\ \emph {et~al.}(2008)\citenamefont {Ni},
  \citenamefont {Tillman}, \citenamefont {Yan}, \citenamefont {Kracher},
  \citenamefont {Hannahs}, \citenamefont {Bud'ko},\ and\ \citenamefont
  {Canfield}}]{Ni2008}%
  \BibitemOpen
  \bibfield  {author} {\bibinfo {author} {\bibfnamefont {N.}~\bibnamefont
  {Ni}}, \bibinfo {author} {\bibfnamefont {M.~E.}\ \bibnamefont {Tillman}},
  \bibinfo {author} {\bibfnamefont {J.-Q.}\ \bibnamefont {Yan}}, \bibinfo
  {author} {\bibfnamefont {A.}~\bibnamefont {Kracher}}, \bibinfo {author}
  {\bibfnamefont {S.~T.}\ \bibnamefont {Hannahs}}, \bibinfo {author}
  {\bibfnamefont {S.~L.}\ \bibnamefont {Bud'ko}}, \ and\ \bibinfo {author}
  {\bibfnamefont {P.~C.}\ \bibnamefont {Canfield}},\ }\bibfield  {title}
  {\enquote {\bibinfo {title} {Effects of {Co} substitution on thermodynamic
  and transport properties and anisotropic ${H}_{c2}$ in
  {Ba}({Fe}$_{1-x}${Co}$_{x}$)$_{2}${As}$_{2}$ single crystals},}\ }\href
  {\doibase 10.1103/PhysRevB.78.214515} {\bibfield  {journal} {\bibinfo
  {journal} {Phys. Rev. B}\ }\textbf {\bibinfo {volume} {78}},\ \bibinfo
  {pages} {214515} (\bibinfo {year} {2008})}\BibitemShut {NoStop}%
\bibitem [{\citenamefont {Sasmal}\ \emph {et~al.}(2008)\citenamefont {Sasmal},
  \citenamefont {Lv}, \citenamefont {Lorenz}, \citenamefont {Guloy},
  \citenamefont {Chen}, \citenamefont {Xue},\ and\ \citenamefont
  {Chu}}]{Sasmal2008}%
  \BibitemOpen
  \bibfield  {author} {\bibinfo {author} {\bibfnamefont {Kalyan}\ \bibnamefont
  {Sasmal}}, \bibinfo {author} {\bibfnamefont {Bing}\ \bibnamefont {Lv}},
  \bibinfo {author} {\bibfnamefont {Bernd}\ \bibnamefont {Lorenz}}, \bibinfo
  {author} {\bibfnamefont {Arnold~M.}\ \bibnamefont {Guloy}}, \bibinfo {author}
  {\bibfnamefont {Feng}\ \bibnamefont {Chen}}, \bibinfo {author} {\bibfnamefont
  {Yu-Yi}\ \bibnamefont {Xue}}, \ and\ \bibinfo {author} {\bibfnamefont
  {Ching-Wu}\ \bibnamefont {Chu}},\ }\bibfield  {title} {\enquote {\bibinfo
  {title} {{Superconducting Fe-Based Compounds ($A_{1-x}$Sr$_x$)Fe$_2$As$_2$
  with $A=\,$K and Cs with Transition Temperatures up to 37~K}},}\ }\href
  {\doibase 10.1103/PhysRevLett.101.107007} {\bibfield  {journal} {\bibinfo
  {journal} {Phys. Rev. Lett.}\ }\textbf {\bibinfo {volume} {101}},\ \bibinfo
  {pages} {107007} (\bibinfo {year} {2008})}\BibitemShut {NoStop}%
\bibitem [{\citenamefont {Chen}\ \emph {et~al.}(2008)\citenamefont {Chen},
  \citenamefont {Li}, \citenamefont {Li}, \citenamefont {Hu}, \citenamefont
  {Dong}, \citenamefont {Jun~Zhou}, \citenamefont {Zheng}, \citenamefont
  {Wang},\ and\ \citenamefont {Luo}}]{Chen2008}%
  \BibitemOpen
  \bibfield  {author} {\bibinfo {author} {\bibfnamefont {Gen-Fu}\ \bibnamefont
  {Chen}}, \bibinfo {author} {\bibfnamefont {Zheng}\ \bibnamefont {Li}},
  \bibinfo {author} {\bibfnamefont {Gang}\ \bibnamefont {Li}}, \bibinfo
  {author} {\bibfnamefont {Wan-Zheng}\ \bibnamefont {Hu}}, \bibinfo {author}
  {\bibfnamefont {Jing}\ \bibnamefont {Dong}}, \bibinfo {author} {\bibfnamefont
  {Xiao-Dong~Zhang}\ \bibnamefont {Jun~Zhou}}, \bibinfo {author} {\bibfnamefont
  {Ping}\ \bibnamefont {Zheng}}, \bibinfo {author} {\bibfnamefont {Nan-Lin}\
  \bibnamefont {Wang}}, \ and\ \bibinfo {author} {\bibfnamefont {Jian-Lin}\
  \bibnamefont {Luo}},\ }\bibfield  {title} {\enquote {\bibinfo {title}
  {{Superconductivity in Hole-Doped (Sr$_{1-x}$K$_x$)Fe$_2$As$_2$}},}\ }\href
  {\doibase 10.1088/0256-307X/25/9/083} {\bibfield  {journal} {\bibinfo
  {journal} {Chin. Phys. Lett.}\ }\textbf {\bibinfo {volume} {25}},\ \bibinfo
  {pages} {3403} (\bibinfo {year} {2008})}\BibitemShut {NoStop}%
\bibitem [{\citenamefont {Chu}\ \emph {et~al.}(2009)\citenamefont {Chu},
  \citenamefont {Analytis}, \citenamefont {Kucharczyk},\ and\ \citenamefont
  {Fisher}}]{Chu2009}%
  \BibitemOpen
  \bibfield  {author} {\bibinfo {author} {\bibfnamefont {Jiun-Haw}\
  \bibnamefont {Chu}}, \bibinfo {author} {\bibfnamefont {James~G.}\
  \bibnamefont {Analytis}}, \bibinfo {author} {\bibfnamefont {Chris}\
  \bibnamefont {Kucharczyk}}, \ and\ \bibinfo {author} {\bibfnamefont {Ian~R.}\
  \bibnamefont {Fisher}},\ }\bibfield  {title} {\enquote {\bibinfo {title}
  {{Determination of the phase diagram of the electron-doped superconductor
  {Ba}({Fe}$_{1-x}${Co}$_{x}$)$_{2}${As}$_{2}$}},}\ }\href {\doibase
  10.1103/PhysRevB.79.014506} {\bibfield  {journal} {\bibinfo  {journal} {Phys.
  Rev. B}\ }\textbf {\bibinfo {volume} {79}},\ \bibinfo {pages} {014506}
  (\bibinfo {year} {2009})}\BibitemShut {NoStop}%
\bibitem [{\citenamefont {Goko}\ \emph {et~al.}(2009)\citenamefont {Goko},
  \citenamefont {Aczel}, \citenamefont {Baggio-Saitovitch}, \citenamefont
  {Bud'ko}, \citenamefont {Canfield}, \citenamefont {Carlo}, \citenamefont
  {Chen}, \citenamefont {Dai}, \citenamefont {Hamann}, \citenamefont {Hu},
  \citenamefont {Kageyama}, \citenamefont {Luke}, \citenamefont {Luo},
  \citenamefont {Nachumi}, \citenamefont {Ni}, \citenamefont {Reznik},
  \citenamefont {Sanchez-Candela}, \citenamefont {Savici}, \citenamefont
  {Sikes}, \citenamefont {Wang}, \citenamefont {Wiebe}, \citenamefont
  {Williams}, \citenamefont {Yamamoto}, \citenamefont {Yu},\ and\ \citenamefont
  {Uemura}}]{Goko2009}%
  \BibitemOpen
  \bibfield  {author} {\bibinfo {author} {\bibfnamefont {T.}~\bibnamefont
  {Goko}}, \bibinfo {author} {\bibfnamefont {A.~A.}\ \bibnamefont {Aczel}},
  \bibinfo {author} {\bibfnamefont {E.}~\bibnamefont {Baggio-Saitovitch}},
  \bibinfo {author} {\bibfnamefont {S.~L.}\ \bibnamefont {Bud'ko}}, \bibinfo
  {author} {\bibfnamefont {P.~C.}\ \bibnamefont {Canfield}}, \bibinfo {author}
  {\bibfnamefont {J.~P.}\ \bibnamefont {Carlo}}, \bibinfo {author}
  {\bibfnamefont {G.~F.}\ \bibnamefont {Chen}}, \bibinfo {author}
  {\bibfnamefont {Pengcheng}\ \bibnamefont {Dai}}, \bibinfo {author}
  {\bibfnamefont {A.~C.}\ \bibnamefont {Hamann}}, \bibinfo {author}
  {\bibfnamefont {W.~Z.}\ \bibnamefont {Hu}}, \bibinfo {author} {\bibfnamefont
  {H.}~\bibnamefont {Kageyama}}, \bibinfo {author} {\bibfnamefont {G.~M.}\
  \bibnamefont {Luke}}, \bibinfo {author} {\bibfnamefont {J.~L.}\ \bibnamefont
  {Luo}}, \bibinfo {author} {\bibfnamefont {B.}~\bibnamefont {Nachumi}},
  \bibinfo {author} {\bibfnamefont {N.}~\bibnamefont {Ni}}, \bibinfo {author}
  {\bibfnamefont {D.}~\bibnamefont {Reznik}}, \bibinfo {author} {\bibfnamefont
  {D.~R.}\ \bibnamefont {Sanchez-Candela}}, \bibinfo {author} {\bibfnamefont
  {A.~T.}\ \bibnamefont {Savici}}, \bibinfo {author} {\bibfnamefont {K.~J.}\
  \bibnamefont {Sikes}}, \bibinfo {author} {\bibfnamefont {N.~L.}\ \bibnamefont
  {Wang}}, \bibinfo {author} {\bibfnamefont {C.~R.}\ \bibnamefont {Wiebe}},
  \bibinfo {author} {\bibfnamefont {T.~J.}\ \bibnamefont {Williams}}, \bibinfo
  {author} {\bibfnamefont {T.}~\bibnamefont {Yamamoto}}, \bibinfo {author}
  {\bibfnamefont {W.}~\bibnamefont {Yu}}, \ and\ \bibinfo {author}
  {\bibfnamefont {Y.~J.}\ \bibnamefont {Uemura}},\ }\bibfield  {title}
  {\enquote {\bibinfo {title} {{Superconducting state coexisting with a
  phase-separated static magnetic order in ({Ba},{K}){Fe}$_{2}${As}$_{2}$,
  ({Sr},{Na}){Fe}$_{2}${As}$_{2}$, and {CaFe}$_{2}${As}$_{2}$}},}\ }\href
  {\doibase 10.1103/PhysRevB.80.024508} {\bibfield  {journal} {\bibinfo
  {journal} {Phys. Rev. B}\ }\textbf {\bibinfo {volume} {80}},\ \bibinfo
  {pages} {024508} (\bibinfo {year} {2009})}\BibitemShut {NoStop}%
\bibitem [{\citenamefont {Saha}\ \emph {et~al.}(2009)\citenamefont {Saha},
  \citenamefont {Butch}, \citenamefont {Kirshenbaum},\ and\ \citenamefont
  {Paglione}}]{Saha2009}%
  \BibitemOpen
  \bibfield  {author} {\bibinfo {author} {\bibfnamefont {S.~R.}\ \bibnamefont
  {Saha}}, \bibinfo {author} {\bibfnamefont {N.~P.}\ \bibnamefont {Butch}},
  \bibinfo {author} {\bibfnamefont {K.}~\bibnamefont {Kirshenbaum}}, \ and\
  \bibinfo {author} {\bibfnamefont {Johnpierre}\ \bibnamefont {Paglione}},\
  }\bibfield  {title} {\enquote {\bibinfo {title} {{Evolution of bulk
  superconductivity in {SrFe}$_{2}${As}$_{2}$ with Ni substitution}},}\ }\href
  {\doibase 10.1103/PhysRevB.79.224519} {\bibfield  {journal} {\bibinfo
  {journal} {Phys. Rev. B}\ }\textbf {\bibinfo {volume} {79}},\ \bibinfo
  {pages} {224519} (\bibinfo {year} {2009})}\BibitemShut {NoStop}%
\bibitem [{\citenamefont {Jiang}\ \emph {et~al.}(2009)\citenamefont {Jiang},
  \citenamefont {Xing}, \citenamefont {Xuan}, \citenamefont {Wang},
  \citenamefont {Ren}, \citenamefont {Feng}, \citenamefont {Dai}, \citenamefont
  {Xu},\ and\ \citenamefont {Cao}}]{Jiang2009}%
  \BibitemOpen
  \bibfield  {author} {\bibinfo {author} {\bibfnamefont {Shuai}\ \bibnamefont
  {Jiang}}, \bibinfo {author} {\bibfnamefont {Hui}\ \bibnamefont {Xing}},
  \bibinfo {author} {\bibfnamefont {Guofang}\ \bibnamefont {Xuan}}, \bibinfo
  {author} {\bibfnamefont {Cao}\ \bibnamefont {Wang}}, \bibinfo {author}
  {\bibfnamefont {Zhi}\ \bibnamefont {Ren}}, \bibinfo {author} {\bibfnamefont
  {Chunmu}\ \bibnamefont {Feng}}, \bibinfo {author} {\bibfnamefont {Jianhui}\
  \bibnamefont {Dai}}, \bibinfo {author} {\bibfnamefont {Zhu'an}\ \bibnamefont
  {Xu}}, \ and\ \bibinfo {author} {\bibfnamefont {Guanghan}\ \bibnamefont
  {Cao}},\ }\bibfield  {title} {\enquote {\bibinfo {title} {Superconductivity
  up to 30~{K} in the vicinity of the quantum critical point in
  {BaFe}$_2$({As}$_{1-x}${P}$_x$)$_2$},}\ }\href {\doibase
  10.1088/0953-8984/21/38/382203} {\bibfield  {journal} {\bibinfo  {journal}
  {J. Phys.: Condens. Matter}\ }\textbf {\bibinfo {volume} {21}},\ \bibinfo
  {pages} {382203} (\bibinfo {year} {2009})}\BibitemShut {NoStop}%
\bibitem [{\citenamefont {Shi}\ \emph {et~al.}(2010)\citenamefont {Shi},
  \citenamefont {Yang}, \citenamefont {Tian}, \citenamefont {Lu}, \citenamefont
  {Wang}, \citenamefont {Qin}, \citenamefont {Song},\ and\ \citenamefont
  {Li}}]{Shi2010}%
  \BibitemOpen
  \bibfield  {author} {\bibinfo {author} {\bibfnamefont {H.~L.}\ \bibnamefont
  {Shi}}, \bibinfo {author} {\bibfnamefont {H.~X.}\ \bibnamefont {Yang}},
  \bibinfo {author} {\bibfnamefont {H.~F.}\ \bibnamefont {Tian}}, \bibinfo
  {author} {\bibfnamefont {J.~B.}\ \bibnamefont {Lu}}, \bibinfo {author}
  {\bibfnamefont {Z.~W.}\ \bibnamefont {Wang}}, \bibinfo {author}
  {\bibfnamefont {Y.~B.}\ \bibnamefont {Qin}}, \bibinfo {author} {\bibfnamefont
  {Y.~J.}\ \bibnamefont {Song}}, \ and\ \bibinfo {author} {\bibfnamefont
  {J.~Q.}\ \bibnamefont {Li}},\ }\bibfield  {title} {\enquote {\bibinfo {title}
  {{Structural properties and superconductivity of SrFe$_2$As$_{2-x}$P$_x$ and
  ($0.0\leq x \leq 1.0$) and CaFe$_2$As$_{2-y}$P$_y$ ($0.0 \leq y \leq
  0.3$)}},}\ }\href {http://stacks.iop.org/0953-8984/22/i=12/a=125702}
  {\bibfield  {journal} {\bibinfo  {journal} {J. Phys.: Condens. Matter}\
  }\textbf {\bibinfo {volume} {22}},\ \bibinfo {pages} {125702} (\bibinfo
  {year} {2010})}\BibitemShut {NoStop}%
\bibitem [{\citenamefont {Cortes-Gil}\ and\ \citenamefont
  {Clarke}(2011)}]{Cortes2011}%
  \BibitemOpen
  \bibfield  {author} {\bibinfo {author} {\bibfnamefont {Raquel}\ \bibnamefont
  {Cortes-Gil}}\ and\ \bibinfo {author} {\bibfnamefont {Simon~J.}\ \bibnamefont
  {Clarke}},\ }\bibfield  {title} {\enquote {\bibinfo {title} {{Structure,
  Magnetism, and Superconductivity of the Layered Iron Arsenides
  Sr$_{1-x}$Na$_x$Fe$_2$As$_2$}},}\ }\href {\doibase 10.1021/cm1028244}
  {\bibfield  {journal} {\bibinfo  {journal} {Chem. Mater.}\ }\textbf {\bibinfo
  {volume} {23}},\ \bibinfo {pages} {1009--1016} (\bibinfo {year}
  {2011})}\BibitemShut {NoStop}%
\bibitem [{\citenamefont {Ishikawa}\ \emph {et~al.}(2009)\citenamefont
  {Ishikawa}, \citenamefont {Eguchi}, \citenamefont {Kodama}, \citenamefont
  {Fujimaki}, \citenamefont {Einaga}, \citenamefont {Ohmura}, \citenamefont
  {Nakayama}, \citenamefont {Mitsuda},\ and\ \citenamefont
  {Yamada}}]{Ishikawa2009}%
  \BibitemOpen
  \bibfield  {author} {\bibinfo {author} {\bibfnamefont {Fumihiro}\
  \bibnamefont {Ishikawa}}, \bibinfo {author} {\bibfnamefont {Naoya}\
  \bibnamefont {Eguchi}}, \bibinfo {author} {\bibfnamefont {Michihiro}\
  \bibnamefont {Kodama}}, \bibinfo {author} {\bibfnamefont {Koji}\ \bibnamefont
  {Fujimaki}}, \bibinfo {author} {\bibfnamefont {Mari}\ \bibnamefont {Einaga}},
  \bibinfo {author} {\bibfnamefont {Ayako}\ \bibnamefont {Ohmura}}, \bibinfo
  {author} {\bibfnamefont {Atsuko}\ \bibnamefont {Nakayama}}, \bibinfo {author}
  {\bibfnamefont {Akihiro}\ \bibnamefont {Mitsuda}}, \ and\ \bibinfo {author}
  {\bibfnamefont {Yuh}\ \bibnamefont {Yamada}},\ }\bibfield  {title} {\enquote
  {\bibinfo {title} {Zero-resistance superconducting phase in
  {BaFe}$_{2}${As}$_{2}$ under high pressure},}\ }\href {\doibase
  10.1103/PhysRevB.79.172506} {\bibfield  {journal} {\bibinfo  {journal} {Phys.
  Rev. B}\ }\textbf {\bibinfo {volume} {79}},\ \bibinfo {pages} {172506}
  (\bibinfo {year} {2009})}\BibitemShut {NoStop}%
\bibitem [{\citenamefont {Alireza}\ \emph {et~al.}(2009)\citenamefont
  {Alireza}, \citenamefont {Ko}, \citenamefont {Gillett}, \citenamefont
  {Petrone}, \citenamefont {Cole}, \citenamefont {Sebastian},\ and\
  \citenamefont {Lonzarich}}]{Alireza2009}%
  \BibitemOpen
  \bibfield  {author} {\bibinfo {author} {\bibfnamefont {Patricia~L.}\
  \bibnamefont {Alireza}}, \bibinfo {author} {\bibfnamefont {Y.~T.~Chris}\
  \bibnamefont {Ko}}, \bibinfo {author} {\bibfnamefont {Jack}\ \bibnamefont
  {Gillett}}, \bibinfo {author} {\bibfnamefont {Chiara~M.}\ \bibnamefont
  {Petrone}}, \bibinfo {author} {\bibfnamefont {Jacqui~M.}\ \bibnamefont
  {Cole}}, \bibinfo {author} {\bibfnamefont {Suchitra~E.}\ \bibnamefont
  {Sebastian}}, \ and\ \bibinfo {author} {\bibfnamefont {Gilbert~G.}\
  \bibnamefont {Lonzarich}},\ }\bibfield  {title} {\enquote {\bibinfo {title}
  {{Superconductivity up to 29~K in SrFe$_2$As$_2$ and BaFe$_2$As$_2$ at high
  pressures}},}\ }\href {\doibase 10.1088/0953-8984/21/1/012208} {\bibfield
  {journal} {\bibinfo  {journal} {J. Phys: Cond. Matter}\ }\textbf {\bibinfo
  {volume} {21}},\ \bibinfo {pages} {012208} (\bibinfo {year}
  {2009})}\BibitemShut {NoStop}%
\bibitem [{\citenamefont {Colombier}\ \emph {et~al.}(2009)\citenamefont
  {Colombier}, \citenamefont {Bud'ko}, \citenamefont {Ni},\ and\ \citenamefont
  {Canfield}}]{Colombier2009}%
  \BibitemOpen
  \bibfield  {author} {\bibinfo {author} {\bibfnamefont {E.}~\bibnamefont
  {Colombier}}, \bibinfo {author} {\bibfnamefont {S.~L.}\ \bibnamefont
  {Bud'ko}}, \bibinfo {author} {\bibfnamefont {N.}~\bibnamefont {Ni}}, \ and\
  \bibinfo {author} {\bibfnamefont {P.~C.}\ \bibnamefont {Canfield}},\
  }\bibfield  {title} {\enquote {\bibinfo {title} {Complete pressure-dependent
  phase diagrams for {SrFe}$_{2}${As}$_{2}$ and {BaFe}$_{2}${As}$_{2}$},}\
  }\href {\doibase 10.1103/PhysRevB.79.224518} {\bibfield  {journal} {\bibinfo
  {journal} {Phys. Rev. B}\ }\textbf {\bibinfo {volume} {79}},\ \bibinfo
  {pages} {224518} (\bibinfo {year} {2009})}\BibitemShut {NoStop}%
\bibitem [{\citenamefont {Kitagawa}\ \emph {et~al.}(2009)\citenamefont
  {Kitagawa}, \citenamefont {Katayama}, \citenamefont {Gotou}, \citenamefont
  {Yagi}, \citenamefont {Ohgushi}, \citenamefont {Matsumoto}, \citenamefont
  {Uwatoko},\ and\ \citenamefont {Takigawa}}]{Kitagawa2009}%
  \BibitemOpen
  \bibfield  {author} {\bibinfo {author} {\bibfnamefont {K.}~\bibnamefont
  {Kitagawa}}, \bibinfo {author} {\bibfnamefont {N.}~\bibnamefont {Katayama}},
  \bibinfo {author} {\bibfnamefont {H.}~\bibnamefont {Gotou}}, \bibinfo
  {author} {\bibfnamefont {T.}~\bibnamefont {Yagi}}, \bibinfo {author}
  {\bibfnamefont {K.}~\bibnamefont {Ohgushi}}, \bibinfo {author} {\bibfnamefont
  {T.}~\bibnamefont {Matsumoto}}, \bibinfo {author} {\bibfnamefont
  {Y.}~\bibnamefont {Uwatoko}}, \ and\ \bibinfo {author} {\bibfnamefont
  {M.}~\bibnamefont {Takigawa}},\ }\bibfield  {title} {\enquote {\bibinfo
  {title} {{Spontaneous Formation of a Superconducting and Antiferromagnetic
  Hybrid State in {SrFe}$_{2}${As}$_{2}$ under High Pressure}},}\ }\href
  {\doibase 10.1103/PhysRevLett.103.257002} {\bibfield  {journal} {\bibinfo
  {journal} {Phys. Rev. Lett.}\ }\textbf {\bibinfo {volume} {103}},\ \bibinfo
  {pages} {257002} (\bibinfo {year} {2009})}\BibitemShut {NoStop}%
\bibitem [{\citenamefont {Tegel}\ \emph {et~al.}(2008)\citenamefont {Tegel},
  \citenamefont {Rotter}, \citenamefont {Wei$\beta$}, \citenamefont
  {Schappacher}, \citenamefont {P\"{o}ttgen},\ and\ \citenamefont
  {Johrendt}}]{Tegel2008}%
  \BibitemOpen
  \bibfield  {author} {\bibinfo {author} {\bibfnamefont {Marcus}\ \bibnamefont
  {Tegel}}, \bibinfo {author} {\bibfnamefont {Marianne}\ \bibnamefont
  {Rotter}}, \bibinfo {author} {\bibfnamefont {Veronika}\ \bibnamefont
  {Wei$\beta$}}, \bibinfo {author} {\bibfnamefont {Falko~M}\ \bibnamefont
  {Schappacher}}, \bibinfo {author} {\bibfnamefont {Rainer}\ \bibnamefont
  {P\"{o}ttgen}}, \ and\ \bibinfo {author} {\bibfnamefont {Dirk}\ \bibnamefont
  {Johrendt}},\ }\bibfield  {title} {\enquote {\bibinfo {title} {Structural and
  magnetic phase transitions in the ternary iron arsenides {SrFe}$_2${As}$_2$
  and {EuFe}$_2${As}$_2$},}\ }\href {\doibase 10.1088/0953-8984/20/45/452201}
  {\bibfield  {journal} {\bibinfo  {journal} {J. Phys.: Condens. Matter}\
  }\textbf {\bibinfo {volume} {20}},\ \bibinfo {pages} {452201} (\bibinfo
  {year} {2008})}\BibitemShut {NoStop}%
\bibitem [{\citenamefont {Yan}\ \emph {et~al.}(2008)\citenamefont {Yan},
  \citenamefont {Kreyssig}, \citenamefont {Nandi}, \citenamefont {Ni},
  \citenamefont {Bud'ko}, \citenamefont {Kracher}, \citenamefont {McQueeney},
  \citenamefont {McCallum}, \citenamefont {Lograsso}, \citenamefont {Goldman},\
  and\ \citenamefont {Canfield}}]{Yan2008}%
  \BibitemOpen
  \bibfield  {author} {\bibinfo {author} {\bibfnamefont {J.-Q.}\ \bibnamefont
  {Yan}}, \bibinfo {author} {\bibfnamefont {A.}~\bibnamefont {Kreyssig}},
  \bibinfo {author} {\bibfnamefont {S.}~\bibnamefont {Nandi}}, \bibinfo
  {author} {\bibfnamefont {N.}~\bibnamefont {Ni}}, \bibinfo {author}
  {\bibfnamefont {S.~L.}\ \bibnamefont {Bud'ko}}, \bibinfo {author}
  {\bibfnamefont {A.}~\bibnamefont {Kracher}}, \bibinfo {author} {\bibfnamefont
  {R.~J.}\ \bibnamefont {McQueeney}}, \bibinfo {author} {\bibfnamefont {R.~W.}\
  \bibnamefont {McCallum}}, \bibinfo {author} {\bibfnamefont {T.~A.}\
  \bibnamefont {Lograsso}}, \bibinfo {author} {\bibfnamefont {A.~I.}\
  \bibnamefont {Goldman}}, \ and\ \bibinfo {author} {\bibfnamefont {P.~C.}\
  \bibnamefont {Canfield}},\ }\bibfield  {title} {\enquote {\bibinfo {title}
  {Structural transition and anisotropic properties of single-crystalline
  {SrFe}$_{2}${As}$_{2}$},}\ }\href {\doibase 10.1103/PhysRevB.78.024516}
  {\bibfield  {journal} {\bibinfo  {journal} {Phys. Rev. B}\ }\textbf {\bibinfo
  {volume} {78}},\ \bibinfo {pages} {024516} (\bibinfo {year}
  {2008})}\BibitemShut {NoStop}%
\bibitem [{\citenamefont {Zhao}\ \emph {et~al.}(2008)\citenamefont {Zhao},
  \citenamefont {Ratcliff}, \citenamefont {Lynn}, \citenamefont {Chen},
  \citenamefont {Luo}, \citenamefont {Wang}, \citenamefont {Hu},\ and\
  \citenamefont {Dai}}]{Zhao2008}%
  \BibitemOpen
  \bibfield  {author} {\bibinfo {author} {\bibfnamefont {Jun}\ \bibnamefont
  {Zhao}}, \bibinfo {author} {\bibfnamefont {W.}~\bibnamefont {Ratcliff}},
  \bibinfo {author} {\bibfnamefont {J.~W.}\ \bibnamefont {Lynn}}, \bibinfo
  {author} {\bibfnamefont {G.~F.}\ \bibnamefont {Chen}}, \bibinfo {author}
  {\bibfnamefont {J.~L.}\ \bibnamefont {Luo}}, \bibinfo {author} {\bibfnamefont
  {N.~L.}\ \bibnamefont {Wang}}, \bibinfo {author} {\bibfnamefont {Jiangping}\
  \bibnamefont {Hu}}, \ and\ \bibinfo {author} {\bibfnamefont {Pengcheng}\
  \bibnamefont {Dai}},\ }\bibfield  {title} {\enquote {\bibinfo {title} {Spin
  and lattice structures of single-crystalline {SrFe}$_{2}${As}$_{2}$},}\
  }\href {\doibase 10.1103/PhysRevB.78.140504} {\bibfield  {journal} {\bibinfo
  {journal} {Phys. Rev. B}\ }\textbf {\bibinfo {volume} {78}},\ \bibinfo
  {pages} {140504(R)} (\bibinfo {year} {2008})}\BibitemShut {NoStop}%
\bibitem [{\citenamefont {Hu}\ \emph {et~al.}(2008)\citenamefont {Hu},
  \citenamefont {Dong}, \citenamefont {Li}, \citenamefont {Li}, \citenamefont
  {Zheng}, \citenamefont {Chen}, \citenamefont {Luo},\ and\ \citenamefont
  {Wang}}]{Hu2008}%
  \BibitemOpen
  \bibfield  {author} {\bibinfo {author} {\bibfnamefont {W.~Z.}\ \bibnamefont
  {Hu}}, \bibinfo {author} {\bibfnamefont {J.}~\bibnamefont {Dong}}, \bibinfo
  {author} {\bibfnamefont {G.}~\bibnamefont {Li}}, \bibinfo {author}
  {\bibfnamefont {Z.}~\bibnamefont {Li}}, \bibinfo {author} {\bibfnamefont
  {P.}~\bibnamefont {Zheng}}, \bibinfo {author} {\bibfnamefont {G.~F.}\
  \bibnamefont {Chen}}, \bibinfo {author} {\bibfnamefont {J.~L.}\ \bibnamefont
  {Luo}}, \ and\ \bibinfo {author} {\bibfnamefont {N.~L.}\ \bibnamefont
  {Wang}},\ }\bibfield  {title} {\enquote {\bibinfo {title} {{Origin of the
  Spin Density Wave Instability in \emph{A}{Fe}$_{2}${As}$_{2}$
  (\emph{A}$=${Ba}, {Sr}) as Revealed by Optical Spectroscopy}},}\ }\href
  {\doibase 10.1103/PhysRevLett.101.257005} {\bibfield  {journal} {\bibinfo
  {journal} {Phys. Rev. Lett.}\ }\textbf {\bibinfo {volume} {101}},\ \bibinfo
  {pages} {257005} (\bibinfo {year} {2008})}\BibitemShut {NoStop}%
\bibitem [{\citenamefont {Hancock}\ \emph {et~al.}(2010)\citenamefont
  {Hancock}, \citenamefont {Mirzaei}, \citenamefont {Gillett}, \citenamefont
  {Sebastian}, \citenamefont {Teyssier}, \citenamefont {Viennois},
  \citenamefont {Giannini},\ and\ \citenamefont {van~der Marel}}]{Hancock2010}%
  \BibitemOpen
  \bibfield  {author} {\bibinfo {author} {\bibfnamefont {J.~N.}\ \bibnamefont
  {Hancock}}, \bibinfo {author} {\bibfnamefont {S.~I.}\ \bibnamefont
  {Mirzaei}}, \bibinfo {author} {\bibfnamefont {J.}~\bibnamefont {Gillett}},
  \bibinfo {author} {\bibfnamefont {S.~E.}\ \bibnamefont {Sebastian}}, \bibinfo
  {author} {\bibfnamefont {J.}~\bibnamefont {Teyssier}}, \bibinfo {author}
  {\bibfnamefont {R.}~\bibnamefont {Viennois}}, \bibinfo {author}
  {\bibfnamefont {E.}~\bibnamefont {Giannini}}, \ and\ \bibinfo {author}
  {\bibfnamefont {D.}~\bibnamefont {van~der Marel}},\ }\bibfield  {title}
  {\enquote {\bibinfo {title} {Strong coupling to magnetic fluctuations in the
  charge dynamics of iron-based superconductors},}\ }\href {\doibase
  10.1103/PhysRevB.82.014523} {\bibfield  {journal} {\bibinfo  {journal} {Phys.
  Rev. B}\ }\textbf {\bibinfo {volume} {82}},\ \bibinfo {pages} {014523}
  (\bibinfo {year} {2010})}\BibitemShut {NoStop}%
\bibitem [{\citenamefont {Blomberg}\ \emph {et~al.}(2011)\citenamefont
  {Blomberg}, \citenamefont {Tanatar}, \citenamefont {Kreyssig}, \citenamefont
  {Ni}, \citenamefont {Thaler}, \citenamefont {Hu}, \citenamefont {Bud'ko},
  \citenamefont {Canfield}, \citenamefont {Goldman},\ and\ \citenamefont
  {Prozorov}}]{Blomberg2011}%
  \BibitemOpen
  \bibfield  {author} {\bibinfo {author} {\bibfnamefont {E.~C.}\ \bibnamefont
  {Blomberg}}, \bibinfo {author} {\bibfnamefont {M.~A.}\ \bibnamefont
  {Tanatar}}, \bibinfo {author} {\bibfnamefont {A.}~\bibnamefont {Kreyssig}},
  \bibinfo {author} {\bibfnamefont {N.}~\bibnamefont {Ni}}, \bibinfo {author}
  {\bibfnamefont {A.}~\bibnamefont {Thaler}}, \bibinfo {author} {\bibfnamefont
  {Rongwei}\ \bibnamefont {Hu}}, \bibinfo {author} {\bibfnamefont {S.~L.}\
  \bibnamefont {Bud'ko}}, \bibinfo {author} {\bibfnamefont {P.~C.}\
  \bibnamefont {Canfield}}, \bibinfo {author} {\bibfnamefont {A.~I.}\
  \bibnamefont {Goldman}}, \ and\ \bibinfo {author} {\bibfnamefont
  {R.}~\bibnamefont {Prozorov}},\ }\bibfield  {title} {\enquote {\bibinfo
  {title} {In-plane anisotropy of electrical resistivity in strain-detwinned
  {SrFe}$_{2}${As}$_{2}$},}\ }\href {\doibase 10.1103/PhysRevB.83.134505}
  {\bibfield  {journal} {\bibinfo  {journal} {Phys. Rev. B}\ }\textbf {\bibinfo
  {volume} {83}},\ \bibinfo {pages} {134505} (\bibinfo {year}
  {2011})}\BibitemShut {NoStop}%
\bibitem [{\citenamefont {Tanatar}\ \emph {et~al.}(2009)\citenamefont
  {Tanatar}, \citenamefont {Kreyssig}, \citenamefont {Nandi}, \citenamefont
  {Ni}, \citenamefont {Bud'ko}, \citenamefont {Canfield}, \citenamefont
  {Goldman},\ and\ \citenamefont {Prozorov}}]{Tanatar2009}%
  \BibitemOpen
  \bibfield  {author} {\bibinfo {author} {\bibfnamefont {M.~A.}\ \bibnamefont
  {Tanatar}}, \bibinfo {author} {\bibfnamefont {A.}~\bibnamefont {Kreyssig}},
  \bibinfo {author} {\bibfnamefont {S.}~\bibnamefont {Nandi}}, \bibinfo
  {author} {\bibfnamefont {N.}~\bibnamefont {Ni}}, \bibinfo {author}
  {\bibfnamefont {S.~L.}\ \bibnamefont {Bud'ko}}, \bibinfo {author}
  {\bibfnamefont {P.~C.}\ \bibnamefont {Canfield}}, \bibinfo {author}
  {\bibfnamefont {A.~I.}\ \bibnamefont {Goldman}}, \ and\ \bibinfo {author}
  {\bibfnamefont {R.}~\bibnamefont {Prozorov}},\ }\bibfield  {title} {\enquote
  {\bibinfo {title} {{Direct imaging of the structural domains in the iron
  pnictides $A{\text{Fe}}_{2}{\text{As}}_{2}$
  $(A=\text{Ca},\text{Sr},\text{Ba})$}},}\ }\href {\doibase
  10.1103/PhysRevB.79.180508} {\bibfield  {journal} {\bibinfo  {journal} {Phys.
  Rev. B}\ }\textbf {\bibinfo {volume} {79}},\ \bibinfo {pages} {180508(R)}
  (\bibinfo {year} {2009})}\BibitemShut {NoStop}%
\bibitem [{\citenamefont {Fisher}\ \emph {et~al.}(2011)\citenamefont {Fisher},
  \citenamefont {Degiorgi},\ and\ \citenamefont {Shen}}]{Fisher2011}%
  \BibitemOpen
  \bibfield  {author} {\bibinfo {author} {\bibfnamefont {I.~R.}\ \bibnamefont
  {Fisher}}, \bibinfo {author} {\bibfnamefont {L.}~\bibnamefont {Degiorgi}}, \
  and\ \bibinfo {author} {\bibfnamefont {Z.~X.}\ \bibnamefont {Shen}},\
  }\bibfield  {title} {\enquote {\bibinfo {title} {{In-plane electronic
  anisotropy of underdoped `122' Fe-arsenide superconductors revealed by
  measurements of detwinned single crystals}},}\ }\href {\doibase
  10.1088/0034-4885/74/12/124506} {\bibfield  {journal} {\bibinfo  {journal}
  {Rep. Prog. Phys.}\ }\textbf {\bibinfo {volume} {74}},\ \bibinfo {pages}
  {124506} (\bibinfo {year} {2011})}\BibitemShut {NoStop}%
\bibitem [{\citenamefont {Goldman}\ \emph {et~al.}(2008)\citenamefont
  {Goldman}, \citenamefont {Argyriou}, \citenamefont {Ouladdiaf}, \citenamefont
  {Chatterji}, \citenamefont {Kreyssig}, \citenamefont {Nandi}, \citenamefont
  {Ni}, \citenamefont {Bud'ko}, \citenamefont {Canfield},\ and\ \citenamefont
  {McQueeney}}]{Goldman2008}%
  \BibitemOpen
  \bibfield  {author} {\bibinfo {author} {\bibfnamefont {A.~I.}\ \bibnamefont
  {Goldman}}, \bibinfo {author} {\bibfnamefont {D.~N.}\ \bibnamefont
  {Argyriou}}, \bibinfo {author} {\bibfnamefont {B.}~\bibnamefont {Ouladdiaf}},
  \bibinfo {author} {\bibfnamefont {T.}~\bibnamefont {Chatterji}}, \bibinfo
  {author} {\bibfnamefont {A.}~\bibnamefont {Kreyssig}}, \bibinfo {author}
  {\bibfnamefont {S.}~\bibnamefont {Nandi}}, \bibinfo {author} {\bibfnamefont
  {N.}~\bibnamefont {Ni}}, \bibinfo {author} {\bibfnamefont {S.~L.}\
  \bibnamefont {Bud'ko}}, \bibinfo {author} {\bibfnamefont {P.~C.}\
  \bibnamefont {Canfield}}, \ and\ \bibinfo {author} {\bibfnamefont {R.~J.}\
  \bibnamefont {McQueeney}},\ }\bibfield  {title} {\enquote {\bibinfo {title}
  {{Lattice and magnetic instabilities in {CaFe}$_{2}${As}$_{2}$: A
  single-crystal neutron diffraction study}},}\ }\href {\doibase
  10.1103/PhysRevB.78.100506} {\bibfield  {journal} {\bibinfo  {journal} {Phys.
  Rev. B}\ }\textbf {\bibinfo {volume} {78}},\ \bibinfo {pages} {100506(R)}
  (\bibinfo {year} {2008})}\BibitemShut {NoStop}%
\bibitem [{\citenamefont {Kofu}\ \emph {et~al.}(2009)\citenamefont {Kofu},
  \citenamefont {Qiu}, \citenamefont {Bao}, \citenamefont {Lee}, \citenamefont
  {Chang}, \citenamefont {Wu}, \citenamefont {Wu},\ and\ \citenamefont
  {Chen}}]{Kofu2009}%
  \BibitemOpen
  \bibfield  {author} {\bibinfo {author} {\bibfnamefont {M.}~\bibnamefont
  {Kofu}}, \bibinfo {author} {\bibfnamefont {Y.}~\bibnamefont {Qiu}}, \bibinfo
  {author} {\bibfnamefont {Wei}\ \bibnamefont {Bao}}, \bibinfo {author}
  {\bibfnamefont {S.-H.}\ \bibnamefont {Lee}}, \bibinfo {author} {\bibfnamefont
  {S.}~\bibnamefont {Chang}}, \bibinfo {author} {\bibfnamefont
  {T.}~\bibnamefont {Wu}}, \bibinfo {author} {\bibfnamefont {G.}~\bibnamefont
  {Wu}}, \ and\ \bibinfo {author} {\bibfnamefont {X.~H.}\ \bibnamefont
  {Chen}},\ }\bibfield  {title} {\enquote {\bibinfo {title} {{Neutron
  scattering investigation of the magnetic order in single crystalline
  BaFe$_2$As$_2$}},}\ }\href {\doibase 10.1088/1367-2630/11/5/055001}
  {\bibfield  {journal} {\bibinfo  {journal} {New J. Phys.}\ }\textbf {\bibinfo
  {volume} {11}},\ \bibinfo {pages} {055001} (\bibinfo {year}
  {2009})}\BibitemShut {NoStop}%
\bibitem [{\citenamefont {Kim}\ \emph {et~al.}(2010)\citenamefont {Kim},
  \citenamefont {Kreyssig}, \citenamefont {Thaler}, \citenamefont {Pratt},
  \citenamefont {Tian}, \citenamefont {Zarestky}, \citenamefont {Green},
  \citenamefont {Bud'ko}, \citenamefont {Canfield}, \citenamefont {McQueeney},\
  and\ \citenamefont {Goldman}}]{Kim2010}%
  \BibitemOpen
  \bibfield  {author} {\bibinfo {author} {\bibfnamefont {M.~G.}\ \bibnamefont
  {Kim}}, \bibinfo {author} {\bibfnamefont {A.}~\bibnamefont {Kreyssig}},
  \bibinfo {author} {\bibfnamefont {A.}~\bibnamefont {Thaler}}, \bibinfo
  {author} {\bibfnamefont {D.~K.}\ \bibnamefont {Pratt}}, \bibinfo {author}
  {\bibfnamefont {W.}~\bibnamefont {Tian}}, \bibinfo {author} {\bibfnamefont
  {J.~L.}\ \bibnamefont {Zarestky}}, \bibinfo {author} {\bibfnamefont {M.~A.}\
  \bibnamefont {Green}}, \bibinfo {author} {\bibfnamefont {S.~L.}\ \bibnamefont
  {Bud'ko}}, \bibinfo {author} {\bibfnamefont {P.~C.}\ \bibnamefont
  {Canfield}}, \bibinfo {author} {\bibfnamefont {R.~J.}\ \bibnamefont
  {McQueeney}}, \ and\ \bibinfo {author} {\bibfnamefont {A.~I.}\ \bibnamefont
  {Goldman}},\ }\bibfield  {title} {\enquote {\bibinfo {title}
  {Antiferromagnetic ordering in the absence of structural distortion in
  {Ba}({Fe}$_{1-x}${Mn}$_x$)$_2${As}$_2$},}\ }\href {\doibase
  10.1103/PhysRevB.82.220503} {\bibfield  {journal} {\bibinfo  {journal} {Phys.
  Rev. B}\ }\textbf {\bibinfo {volume} {82}},\ \bibinfo {pages} {220503(R)}
  (\bibinfo {year} {2010})}\BibitemShut {NoStop}%
\bibitem [{\citenamefont {Hassinger}\ \emph {et~al.}(2012)\citenamefont
  {Hassinger}, \citenamefont {Gredat}, \citenamefont {Valade}, \citenamefont
  {de~Cotret}, \citenamefont {Juneau-Fecteau}, \citenamefont {Reid},
  \citenamefont {Kim}, \citenamefont {Tanatar}, \citenamefont {Prozorov},
  \citenamefont {Shen}, \citenamefont {Wen}, \citenamefont {Doiron-Leyraud},\
  and\ \citenamefont {Taillefer}}]{Hassinger2012}%
  \BibitemOpen
  \bibfield  {author} {\bibinfo {author} {\bibfnamefont {E.}~\bibnamefont
  {Hassinger}}, \bibinfo {author} {\bibfnamefont {G.}~\bibnamefont {Gredat}},
  \bibinfo {author} {\bibfnamefont {F.}~\bibnamefont {Valade}}, \bibinfo
  {author} {\bibfnamefont {S.~Ren\'e}\ \bibnamefont {de~Cotret}}, \bibinfo
  {author} {\bibfnamefont {A.}~\bibnamefont {Juneau-Fecteau}}, \bibinfo
  {author} {\bibfnamefont {J.-Ph.}\ \bibnamefont {Reid}}, \bibinfo {author}
  {\bibfnamefont {H.}~\bibnamefont {Kim}}, \bibinfo {author} {\bibfnamefont
  {M.~A.}\ \bibnamefont {Tanatar}}, \bibinfo {author} {\bibfnamefont
  {R.}~\bibnamefont {Prozorov}}, \bibinfo {author} {\bibfnamefont
  {B.}~\bibnamefont {Shen}}, \bibinfo {author} {\bibfnamefont {H.-H.}\
  \bibnamefont {Wen}}, \bibinfo {author} {\bibfnamefont {N.}~\bibnamefont
  {Doiron-Leyraud}}, \ and\ \bibinfo {author} {\bibfnamefont {Louis}\
  \bibnamefont {Taillefer}},\ }\bibfield  {title} {\enquote {\bibinfo {title}
  {{Pressure-induced Fermi-surface reconstruction in the iron-arsenide
  superconductor Ba$_{1-x}$K$_{x}$Fe$_{2}$As$_{2}$: Evidence of a phase
  transition inside the antiferromagnetic phase}},}\ }\href {\doibase
  10.1103/PhysRevB.86.140502} {\bibfield  {journal} {\bibinfo  {journal} {Phys.
  Rev. B}\ }\textbf {\bibinfo {volume} {86}},\ \bibinfo {pages} {140502(R)}
  (\bibinfo {year} {2012})}\BibitemShut {NoStop}%
\bibitem [{\citenamefont {B\"{o}hmer}\ \emph {et~al.}(2015)\citenamefont
  {B\"{o}hmer}, \citenamefont {Hardy}, \citenamefont {Wang}, \citenamefont
  {Wolf}, \citenamefont {Schweiss},\ and\ \citenamefont
  {Meingast}}]{Bohmer2015}%
  \BibitemOpen
  \bibfield  {author} {\bibinfo {author} {\bibfnamefont {A.~E.}\ \bibnamefont
  {B\"{o}hmer}}, \bibinfo {author} {\bibfnamefont {F.}~\bibnamefont {Hardy}},
  \bibinfo {author} {\bibfnamefont {L.}~\bibnamefont {Wang}}, \bibinfo {author}
  {\bibfnamefont {T.}~\bibnamefont {Wolf}}, \bibinfo {author} {\bibfnamefont
  {P.}~\bibnamefont {Schweiss}}, \ and\ \bibinfo {author} {\bibfnamefont
  {C.}~\bibnamefont {Meingast}},\ }\bibfield  {title} {\enquote {\bibinfo
  {title} {{Superconductivity-induced re-entrance of the orthorhombic
  distortion in Ba$_{1-x}$K$_x$Fe$_2$As$_2$}},}\ }\href {\doibase
  10.1038/ncomms8911} {\bibfield  {journal} {\bibinfo  {journal} {Nat.
  Commun.}\ }\textbf {\bibinfo {volume} {6}},\ \bibinfo {pages} {7911}
  (\bibinfo {year} {2015})}\BibitemShut {NoStop}%
\bibitem [{\citenamefont {Wang}\ \emph {et~al.}(2016)\citenamefont {Wang},
  \citenamefont {Hardy}, \citenamefont {B\"ohmer}, \citenamefont {Wolf},
  \citenamefont {Schweiss},\ and\ \citenamefont {Meingast}}]{Wang2016}%
  \BibitemOpen
  \bibfield  {author} {\bibinfo {author} {\bibfnamefont {L.}~\bibnamefont
  {Wang}}, \bibinfo {author} {\bibfnamefont {F.}~\bibnamefont {Hardy}},
  \bibinfo {author} {\bibfnamefont {A.~E.}\ \bibnamefont {B\"ohmer}}, \bibinfo
  {author} {\bibfnamefont {T.}~\bibnamefont {Wolf}}, \bibinfo {author}
  {\bibfnamefont {P.}~\bibnamefont {Schweiss}}, \ and\ \bibinfo {author}
  {\bibfnamefont {C.}~\bibnamefont {Meingast}},\ }\bibfield  {title} {\enquote
  {\bibinfo {title} {Complex phase diagram of
  {Ba}$_{1-x}${Na}$_{x}${Fe}$_{2}${As}$_{2}$: A multitude of phases striving
  for the electronic entropy},}\ }\href {\doibase 10.1103/PhysRevB.93.014514}
  {\bibfield  {journal} {\bibinfo  {journal} {Phys. Rev. B}\ }\textbf {\bibinfo
  {volume} {93}},\ \bibinfo {pages} {014514} (\bibinfo {year}
  {2016})}\BibitemShut {NoStop}%
\bibitem [{\citenamefont {Taddei}\ \emph {et~al.}(2016)\citenamefont {Taddei},
  \citenamefont {Allred}, \citenamefont {Bugaris}, \citenamefont {Lapidus},
  \citenamefont {Krogstad}, \citenamefont {Stadel}, \citenamefont {Claus},
  \citenamefont {Chung}, \citenamefont {Kanatzidis}, \citenamefont
  {Rosenkranz}, \citenamefont {Osborn},\ and\ \citenamefont
  {Chmaissem}}]{Taddei2016}%
  \BibitemOpen
  \bibfield  {author} {\bibinfo {author} {\bibfnamefont {K.~M.}\ \bibnamefont
  {Taddei}}, \bibinfo {author} {\bibfnamefont {J.~M.}\ \bibnamefont {Allred}},
  \bibinfo {author} {\bibfnamefont {D.~E.}\ \bibnamefont {Bugaris}}, \bibinfo
  {author} {\bibfnamefont {S.}~\bibnamefont {Lapidus}}, \bibinfo {author}
  {\bibfnamefont {M.~J.}\ \bibnamefont {Krogstad}}, \bibinfo {author}
  {\bibfnamefont {R.}~\bibnamefont {Stadel}}, \bibinfo {author} {\bibfnamefont
  {H.}~\bibnamefont {Claus}}, \bibinfo {author} {\bibfnamefont {D.~Y.}\
  \bibnamefont {Chung}}, \bibinfo {author} {\bibfnamefont {M.~G.}\ \bibnamefont
  {Kanatzidis}}, \bibinfo {author} {\bibfnamefont {S.}~\bibnamefont
  {Rosenkranz}}, \bibinfo {author} {\bibfnamefont {R.}~\bibnamefont {Osborn}},
  \ and\ \bibinfo {author} {\bibfnamefont {O.}~\bibnamefont {Chmaissem}},\
  }\bibfield  {title} {\enquote {\bibinfo {title} {Detailed magnetic and
  structural analysis mapping a robust magnetic ${C}_{4}$ dome in
  {Sr}$_{1-x}${Na}$_{x}${Fe}$_{2}${As}$_{2}$},}\ }\href {\doibase
  10.1103/PhysRevB.93.134510} {\bibfield  {journal} {\bibinfo  {journal} {Phys.
  Rev. B}\ }\textbf {\bibinfo {volume} {93}},\ \bibinfo {pages} {134510}
  (\bibinfo {year} {2016})}\BibitemShut {NoStop}%
\bibitem [{\citenamefont {Wang}\ \emph {et~al.}(2019)\citenamefont {Wang},
  \citenamefont {He}, \citenamefont {Scherer}, \citenamefont {Hardy},
  \citenamefont {Schweiss}, \citenamefont {Wolf}, \citenamefont {Merz},
  \citenamefont {Andersen},\ and\ \citenamefont {Meingast}}]{Wang2019}%
  \BibitemOpen
  \bibfield  {author} {\bibinfo {author} {\bibfnamefont {Liran}\ \bibnamefont
  {Wang}}, \bibinfo {author} {\bibfnamefont {Mingquan}\ \bibnamefont {He}},
  \bibinfo {author} {\bibfnamefont {Daniel~D.}\ \bibnamefont {Scherer}},
  \bibinfo {author} {\bibfnamefont {Fr\'{e}d\'{e}ric}\ \bibnamefont {Hardy}},
  \bibinfo {author} {\bibfnamefont {Peter}\ \bibnamefont {Schweiss}}, \bibinfo
  {author} {\bibfnamefont {Thomas}\ \bibnamefont {Wolf}}, \bibinfo {author}
  {\bibfnamefont {Michael}\ \bibnamefont {Merz}}, \bibinfo {author}
  {\bibfnamefont {Brian~M.}\ \bibnamefont {Andersen}}, \ and\ \bibinfo {author}
  {\bibfnamefont {Christoph}\ \bibnamefont {Meingast}},\ }\bibfield  {title}
  {\enquote {\bibinfo {title} {{Competing Electronic Phases near the Onset of
  Superconductivity in Hole-doped SrFe$_2$As$_2$}},}\ }\href {\doibase
  10.7566/JPSJ.88.104710} {\bibfield  {journal} {\bibinfo  {journal} {J. Phys.
  Soc. Jpn.}\ }\textbf {\bibinfo {volume} {88}},\ \bibinfo {pages} {104710}
  (\bibinfo {year} {2019})}\BibitemShut {NoStop}%
\bibitem [{\citenamefont {Hassinger}\ \emph {et~al.}(2016)\citenamefont
  {Hassinger}, \citenamefont {Gredat}, \citenamefont {Valade}, \citenamefont
  {de~Cotret}, \citenamefont {Cyr-Choini\`ere}, \citenamefont {Juneau-Fecteau},
  \citenamefont {Reid}, \citenamefont {Kim}, \citenamefont {Tanatar},
  \citenamefont {Prozorov}, \citenamefont {Shen}, \citenamefont {Wen},
  \citenamefont {Doiron-Leyraud},\ and\ \citenamefont
  {Taillefer}}]{Hassinger2016}%
  \BibitemOpen
  \bibfield  {author} {\bibinfo {author} {\bibfnamefont {E.}~\bibnamefont
  {Hassinger}}, \bibinfo {author} {\bibfnamefont {G.}~\bibnamefont {Gredat}},
  \bibinfo {author} {\bibfnamefont {F.}~\bibnamefont {Valade}}, \bibinfo
  {author} {\bibfnamefont {S.~Ren\'e}\ \bibnamefont {de~Cotret}}, \bibinfo
  {author} {\bibfnamefont {O.}~\bibnamefont {Cyr-Choini\`ere}}, \bibinfo
  {author} {\bibfnamefont {A.}~\bibnamefont {Juneau-Fecteau}}, \bibinfo
  {author} {\bibfnamefont {J.-Ph.}\ \bibnamefont {Reid}}, \bibinfo {author}
  {\bibfnamefont {H.}~\bibnamefont {Kim}}, \bibinfo {author} {\bibfnamefont
  {M.~A.}\ \bibnamefont {Tanatar}}, \bibinfo {author} {\bibfnamefont
  {R.}~\bibnamefont {Prozorov}}, \bibinfo {author} {\bibfnamefont
  {B.}~\bibnamefont {Shen}}, \bibinfo {author} {\bibfnamefont {H.-H.}\
  \bibnamefont {Wen}}, \bibinfo {author} {\bibfnamefont {N.}~\bibnamefont
  {Doiron-Leyraud}}, \ and\ \bibinfo {author} {\bibfnamefont {Louis}\
  \bibnamefont {Taillefer}},\ }\bibfield  {title} {\enquote {\bibinfo {title}
  {Expansion of the tetragonal magnetic phase with pressure in the iron
  arsenide superconductor {Ba}$_{1-x}${K}$_{x}${Fe}$_{2}${As}$_{2}$},}\ }\href
  {\doibase 10.1103/PhysRevB.93.144401} {\bibfield  {journal} {\bibinfo
  {journal} {Phys. Rev. B}\ }\textbf {\bibinfo {volume} {93}},\ \bibinfo
  {pages} {144401} (\bibinfo {year} {2016})}\BibitemShut {NoStop}%
\bibitem [{\citenamefont {Taddei}\ \emph {et~al.}(2017)\citenamefont {Taddei},
  \citenamefont {Allred}, \citenamefont {Bugaris}, \citenamefont {Lapidus},
  \citenamefont {Krogstad}, \citenamefont {Claus}, \citenamefont {Chung},
  \citenamefont {Kanatzidis}, \citenamefont {Osborn}, \citenamefont
  {Rosenkranz},\ and\ \citenamefont {Chmaissem}}]{Taddei2017}%
  \BibitemOpen
  \bibfield  {author} {\bibinfo {author} {\bibfnamefont {K.~M.}\ \bibnamefont
  {Taddei}}, \bibinfo {author} {\bibfnamefont {J.~M.}\ \bibnamefont {Allred}},
  \bibinfo {author} {\bibfnamefont {D.~E.}\ \bibnamefont {Bugaris}}, \bibinfo
  {author} {\bibfnamefont {S.~H.}\ \bibnamefont {Lapidus}}, \bibinfo {author}
  {\bibfnamefont {M.~J.}\ \bibnamefont {Krogstad}}, \bibinfo {author}
  {\bibfnamefont {H.}~\bibnamefont {Claus}}, \bibinfo {author} {\bibfnamefont
  {D.~Y.}\ \bibnamefont {Chung}}, \bibinfo {author} {\bibfnamefont {M.~G.}\
  \bibnamefont {Kanatzidis}}, \bibinfo {author} {\bibfnamefont
  {R.}~\bibnamefont {Osborn}}, \bibinfo {author} {\bibfnamefont
  {S.}~\bibnamefont {Rosenkranz}}, \ and\ \bibinfo {author} {\bibfnamefont
  {O.}~\bibnamefont {Chmaissem}},\ }\bibfield  {title} {\enquote {\bibinfo
  {title} {{Observation of the magnetic ${C}_{4}$ phase in
  {Ca}$_{1-x}${Na}$_{x}${Fe}$_{2}${As}$_{2}$ and its universality in the
  hole-doped 122 superconductors}},}\ }\href {\doibase
  10.1103/PhysRevB.95.064508} {\bibfield  {journal} {\bibinfo  {journal} {Phys.
  Rev. B}\ }\textbf {\bibinfo {volume} {95}},\ \bibinfo {pages} {064508}
  (\bibinfo {year} {2017})}\BibitemShut {NoStop}%
\bibitem [{\citenamefont {Yi}\ \emph {et~al.}(2018)\citenamefont {Yi},
  \citenamefont {Frano}, \citenamefont {Lu}, \citenamefont {He}, \citenamefont
  {Wang}, \citenamefont {Frandsen}, \citenamefont {Kemper}, \citenamefont {Yu},
  \citenamefont {Si}, \citenamefont {Wang}, \citenamefont {He}, \citenamefont
  {Hardy}, \citenamefont {Schweiss}, \citenamefont {Adelmann}, \citenamefont
  {Wolf}, \citenamefont {Hashimoto}, \citenamefont {Mo}, \citenamefont
  {Hussain}, \citenamefont {Le~Tacon}, \citenamefont {B\"ohmer}, \citenamefont
  {Lee}, \citenamefont {Shen}, \citenamefont {Meingast},\ and\ \citenamefont
  {Birgeneau}}]{Yi2018}%
  \BibitemOpen
  \bibfield  {author} {\bibinfo {author} {\bibfnamefont {M.}~\bibnamefont
  {Yi}}, \bibinfo {author} {\bibfnamefont {A.}~\bibnamefont {Frano}}, \bibinfo
  {author} {\bibfnamefont {D.~H.}\ \bibnamefont {Lu}}, \bibinfo {author}
  {\bibfnamefont {Y.}~\bibnamefont {He}}, \bibinfo {author} {\bibfnamefont
  {Meng}\ \bibnamefont {Wang}}, \bibinfo {author} {\bibfnamefont {B.~A.}\
  \bibnamefont {Frandsen}}, \bibinfo {author} {\bibfnamefont {A.~F.}\
  \bibnamefont {Kemper}}, \bibinfo {author} {\bibfnamefont {R.}~\bibnamefont
  {Yu}}, \bibinfo {author} {\bibfnamefont {Q.}~\bibnamefont {Si}}, \bibinfo
  {author} {\bibfnamefont {L.}~\bibnamefont {Wang}}, \bibinfo {author}
  {\bibfnamefont {M.}~\bibnamefont {He}}, \bibinfo {author} {\bibfnamefont
  {F.}~\bibnamefont {Hardy}}, \bibinfo {author} {\bibfnamefont
  {P.}~\bibnamefont {Schweiss}}, \bibinfo {author} {\bibfnamefont
  {P.}~\bibnamefont {Adelmann}}, \bibinfo {author} {\bibfnamefont
  {T.}~\bibnamefont {Wolf}}, \bibinfo {author} {\bibfnamefont {M.}~\bibnamefont
  {Hashimoto}}, \bibinfo {author} {\bibfnamefont {S.-K.}\ \bibnamefont {Mo}},
  \bibinfo {author} {\bibfnamefont {Z.}~\bibnamefont {Hussain}}, \bibinfo
  {author} {\bibfnamefont {M.}~\bibnamefont {Le~Tacon}}, \bibinfo {author}
  {\bibfnamefont {A.~E.}\ \bibnamefont {B\"ohmer}}, \bibinfo {author}
  {\bibfnamefont {D.-H.}\ \bibnamefont {Lee}}, \bibinfo {author} {\bibfnamefont
  {Z.-X.}\ \bibnamefont {Shen}}, \bibinfo {author} {\bibfnamefont
  {C.}~\bibnamefont {Meingast}}, \ and\ \bibinfo {author} {\bibfnamefont
  {R.~J.}\ \bibnamefont {Birgeneau}},\ }\bibfield  {title} {\enquote {\bibinfo
  {title} {{Spectral Evidence for Emergent Order in
  {Ba}$_{1-x}${Na}$_{x}${Fe}$_{2}${As}$_{2}$}},}\ }\href {\doibase
  10.1103/PhysRevLett.121.127001} {\bibfield  {journal} {\bibinfo  {journal}
  {Phys. Rev. Lett.}\ }\textbf {\bibinfo {volume} {121}},\ \bibinfo {pages}
  {127001} (\bibinfo {year} {2018})}\BibitemShut {NoStop}%
\bibitem [{\citenamefont {Avci}\ \emph {et~al.}(2014)\citenamefont {Avci},
  \citenamefont {Chmaissem}, \citenamefont {Allred}, \citenamefont
  {Rosenkranz}, \citenamefont {Eremin}, \citenamefont {Chubukov}, \citenamefont
  {Bugaris}, \citenamefont {Chung}, \citenamefont {Kanatzidis}, \citenamefont
  {Castellan}, \citenamefont {Schlueter}, \citenamefont {Claus}, \citenamefont
  {Khalyavin}, \citenamefont {Manuel}, \citenamefont {Daoud-Aladine},\ and\
  \citenamefont {Osborn}}]{Avci2014}%
  \BibitemOpen
  \bibfield  {author} {\bibinfo {author} {\bibfnamefont {S.}~\bibnamefont
  {Avci}}, \bibinfo {author} {\bibfnamefont {O.}~\bibnamefont {Chmaissem}},
  \bibinfo {author} {\bibfnamefont {J.~M.}\ \bibnamefont {Allred}}, \bibinfo
  {author} {\bibfnamefont {S.}~\bibnamefont {Rosenkranz}}, \bibinfo {author}
  {\bibfnamefont {I.}~\bibnamefont {Eremin}}, \bibinfo {author} {\bibfnamefont
  {A.~V.}\ \bibnamefont {Chubukov}}, \bibinfo {author} {\bibfnamefont {D.~E.}\
  \bibnamefont {Bugaris}}, \bibinfo {author} {\bibfnamefont {D.~Y.}\
  \bibnamefont {Chung}}, \bibinfo {author} {\bibfnamefont {M.~G.}\ \bibnamefont
  {Kanatzidis}}, \bibinfo {author} {\bibfnamefont {J.-P}\ \bibnamefont
  {Castellan}}, \bibinfo {author} {\bibfnamefont {J.~A.}\ \bibnamefont
  {Schlueter}}, \bibinfo {author} {\bibfnamefont {H.}~\bibnamefont {Claus}},
  \bibinfo {author} {\bibfnamefont {D.~D.}\ \bibnamefont {Khalyavin}}, \bibinfo
  {author} {\bibfnamefont {P.}~\bibnamefont {Manuel}}, \bibinfo {author}
  {\bibfnamefont {A.}~\bibnamefont {Daoud-Aladine}}, \ and\ \bibinfo {author}
  {\bibfnamefont {R.}~\bibnamefont {Osborn}},\ }\bibfield  {title} {\enquote
  {\bibinfo {title} {Magnetically driven suppression fo nematic order in an
  iron-based superconductor},}\ }\href {\doibase 10.1038/ncomms4845} {\bibfield
   {journal} {\bibinfo  {journal} {Nat. Commun.}\ }\textbf {\bibinfo {volume}
  {5}},\ \bibinfo {pages} {3845} (\bibinfo {year} {2014})}\BibitemShut
  {NoStop}%
\bibitem [{\citenamefont {Allred}\ \emph {et~al.}(2016)\citenamefont {Allred},
  \citenamefont {Taddei}, \citenamefont {Bugaris}, \citenamefont {Krogstad},
  \citenamefont {Lapidus}, \citenamefont {Chung}, \citenamefont {Claus},
  \citenamefont {Kanatzidis}, \citenamefont {Brown}, \citenamefont {Kang},
  \citenamefont {Fernandes}, \citenamefont {Eremin}, \citenamefont
  {Rosenkranz}, \citenamefont {Chmaissem},\ and\ \citenamefont
  {Osborn}}]{Allred2016}%
  \BibitemOpen
  \bibfield  {author} {\bibinfo {author} {\bibfnamefont {J.~M.}\ \bibnamefont
  {Allred}}, \bibinfo {author} {\bibfnamefont {K.~M.}\ \bibnamefont {Taddei}},
  \bibinfo {author} {\bibfnamefont {D.~E.}\ \bibnamefont {Bugaris}}, \bibinfo
  {author} {\bibfnamefont {M.~J.}\ \bibnamefont {Krogstad}}, \bibinfo {author}
  {\bibfnamefont {S.~H.}\ \bibnamefont {Lapidus}}, \bibinfo {author}
  {\bibfnamefont {D.~Y.}\ \bibnamefont {Chung}}, \bibinfo {author}
  {\bibfnamefont {H.}~\bibnamefont {Claus}}, \bibinfo {author} {\bibfnamefont
  {M.~G.}\ \bibnamefont {Kanatzidis}}, \bibinfo {author} {\bibfnamefont
  {D.~E.}\ \bibnamefont {Brown}}, \bibinfo {author} {\bibfnamefont
  {J.}~\bibnamefont {Kang}}, \bibinfo {author} {\bibfnamefont {R.~M.}\
  \bibnamefont {Fernandes}}, \bibinfo {author} {\bibfnamefont {I.}~\bibnamefont
  {Eremin}}, \bibinfo {author} {\bibfnamefont {S.}~\bibnamefont {Rosenkranz}},
  \bibinfo {author} {\bibfnamefont {O.}~\bibnamefont {Chmaissem}}, \ and\
  \bibinfo {author} {\bibfnamefont {R.}~\bibnamefont {Osborn}},\ }\bibfield
  {title} {\enquote {\bibinfo {title} {{Double-Q spin-density wave in iron
  arsenide superconductors}},}\ }\href {\doibase 10.1038/nphys3629} {\bibfield
  {journal} {\bibinfo  {journal} {Nat. Phys.}\ }\textbf {\bibinfo {volume}
  {12}},\ \bibinfo {pages} {493} (\bibinfo {year} {2016})}\BibitemShut
  {NoStop}%
\bibitem [{\citenamefont {Wa\ss{}er}\ \emph {et~al.}(2015)\citenamefont
  {Wa\ss{}er}, \citenamefont {Schneidewind}, \citenamefont {Sidis},
  \citenamefont {Wurmehl}, \citenamefont {Aswartham}, \citenamefont
  {B\"uchner},\ and\ \citenamefont {Braden}}]{Wasser2015}%
  \BibitemOpen
  \bibfield  {author} {\bibinfo {author} {\bibfnamefont {F.}~\bibnamefont
  {Wa\ss{}er}}, \bibinfo {author} {\bibfnamefont {A.}~\bibnamefont
  {Schneidewind}}, \bibinfo {author} {\bibfnamefont {Y.}~\bibnamefont {Sidis}},
  \bibinfo {author} {\bibfnamefont {S.}~\bibnamefont {Wurmehl}}, \bibinfo
  {author} {\bibfnamefont {S.}~\bibnamefont {Aswartham}}, \bibinfo {author}
  {\bibfnamefont {B.}~\bibnamefont {B\"uchner}}, \ and\ \bibinfo {author}
  {\bibfnamefont {M.}~\bibnamefont {Braden}},\ }\bibfield  {title} {\enquote
  {\bibinfo {title} {{Spin reorientation in
  {Ba}$_{0.65}${Na}$_{0.35}${Fe}$_{2}${As}$_{2}$ studied by single-crystal
  neutron diffraction}},}\ }\href {\doibase 10.1103/PhysRevB.91.060505}
  {\bibfield  {journal} {\bibinfo  {journal} {Phys. Rev. B}\ }\textbf {\bibinfo
  {volume} {91}},\ \bibinfo {pages} {060505(R)} (\bibinfo {year}
  {2015})}\BibitemShut {NoStop}%
\bibitem [{\citenamefont {Mallett}\ \emph
  {et~al.}(2015{\natexlab{a}})\citenamefont {Mallett}, \citenamefont
  {Pashkevich}, \citenamefont {Gusev}, \citenamefont {Wolf},\ and\
  \citenamefont {Bernhard}}]{Mallett2015a}%
  \BibitemOpen
  \bibfield  {author} {\bibinfo {author} {\bibfnamefont {B.~P.~P.}\
  \bibnamefont {Mallett}}, \bibinfo {author} {\bibfnamefont {Yu.~G.}\
  \bibnamefont {Pashkevich}}, \bibinfo {author} {\bibfnamefont
  {A.}~\bibnamefont {Gusev}}, \bibinfo {author} {\bibfnamefont {Th.}\
  \bibnamefont {Wolf}}, \ and\ \bibinfo {author} {\bibfnamefont
  {C.}~\bibnamefont {Bernhard}},\ }\bibfield  {title} {\enquote {\bibinfo
  {title} {{Muon spin rotation study of the magnetic structure in the
  tetragonal antiferromagnetic state of weakly underdoped
  Ba$_{1-x}$K$_x$Fe$_2$As$_2$}},}\ }\href {\doibase
  10.1209/0295-5075/111/57001} {\bibfield  {journal} {\bibinfo  {journal}
  {{EPL} (Europhysics Letters)}\ }\textbf {\bibinfo {volume} {111}},\ \bibinfo
  {pages} {57001} (\bibinfo {year} {2015}{\natexlab{a}})}\BibitemShut {NoStop}%
\bibitem [{\citenamefont {Hoyer}\ \emph {et~al.}(2016)\citenamefont {Hoyer},
  \citenamefont {Fernandes}, \citenamefont {Levchenko},\ and\ \citenamefont
  {Schmalian}}]{Hoyer2016}%
  \BibitemOpen
  \bibfield  {author} {\bibinfo {author} {\bibfnamefont {Mareike}\ \bibnamefont
  {Hoyer}}, \bibinfo {author} {\bibfnamefont {Rafael~M.}\ \bibnamefont
  {Fernandes}}, \bibinfo {author} {\bibfnamefont {Alex}\ \bibnamefont
  {Levchenko}}, \ and\ \bibinfo {author} {\bibfnamefont {J\"org}\ \bibnamefont
  {Schmalian}},\ }\bibfield  {title} {\enquote {\bibinfo {title}
  {{Disorder-promoted ${C}_{4}$-symmetric magnetic order in iron-based
  superconductors}},}\ }\href {\doibase 10.1103/PhysRevB.93.144414} {\bibfield
  {journal} {\bibinfo  {journal} {Phys. Rev. B}\ }\textbf {\bibinfo {volume}
  {93}},\ \bibinfo {pages} {144414} (\bibinfo {year} {2016})}\BibitemShut
  {NoStop}%
\bibitem [{\citenamefont {Guo}\ \emph {et~al.}(2019)\citenamefont {Guo},
  \citenamefont {Yue}, \citenamefont {Iida}, \citenamefont {Kamazawa},
  \citenamefont {Chen}, \citenamefont {Han}, \citenamefont {Zhang},\ and\
  \citenamefont {Li}}]{Guo2019}%
  \BibitemOpen
  \bibfield  {author} {\bibinfo {author} {\bibfnamefont {Jianqing}\
  \bibnamefont {Guo}}, \bibinfo {author} {\bibfnamefont {Li}~\bibnamefont
  {Yue}}, \bibinfo {author} {\bibfnamefont {Kazuki}\ \bibnamefont {Iida}},
  \bibinfo {author} {\bibfnamefont {Kazuya}\ \bibnamefont {Kamazawa}}, \bibinfo
  {author} {\bibfnamefont {Lei}\ \bibnamefont {Chen}}, \bibinfo {author}
  {\bibfnamefont {Tingting}\ \bibnamefont {Han}}, \bibinfo {author}
  {\bibfnamefont {Yan}\ \bibnamefont {Zhang}}, \ and\ \bibinfo {author}
  {\bibfnamefont {Yuan}\ \bibnamefont {Li}},\ }\bibfield  {title} {\enquote
  {\bibinfo {title} {Preferred magnetic excitations in the iron-based
  {Sr}$_{1-x}${Na}$_{x}${Fe}$_{2}${As}$_{2}$ superconductor},}\ }\href
  {\doibase 10.1103/PhysRevLett.122.017001} {\bibfield  {journal} {\bibinfo
  {journal} {Phys. Rev. Lett.}\ }\textbf {\bibinfo {volume} {122}},\ \bibinfo
  {pages} {017001} (\bibinfo {year} {2019})}\BibitemShut {NoStop}%
\bibitem [{\citenamefont {Homes}\ \emph {et~al.}(1993)\citenamefont {Homes},
  \citenamefont {Reedyk}, \citenamefont {Crandles},\ and\ \citenamefont
  {Timusk}}]{Homes1993}%
  \BibitemOpen
  \bibfield  {author} {\bibinfo {author} {\bibfnamefont {Christopher~C.}\
  \bibnamefont {Homes}}, \bibinfo {author} {\bibfnamefont {M.}~\bibnamefont
  {Reedyk}}, \bibinfo {author} {\bibfnamefont {D.~A.}\ \bibnamefont
  {Crandles}}, \ and\ \bibinfo {author} {\bibfnamefont {T.}~\bibnamefont
  {Timusk}},\ }\bibfield  {title} {\enquote {\bibinfo {title} {{Technique for
  measuring the reflectance of irregular, submillimeter-sized samples}},}\
  }\href {\doibase 10.1364/AO.32.002976} {\bibfield  {journal} {\bibinfo
  {journal} {Appl. Opt.}\ }\textbf {\bibinfo {volume} {32}},\ \bibinfo {pages}
  {2976--2983} (\bibinfo {year} {1993})}\BibitemShut {NoStop}%
\bibitem [{Sup()}]{Suplmt}%
  \BibitemOpen
  \href@noop {} {}\bibinfo {note} {See Supplemental Material at [URL will be
  inserted by publisher] for details of the experimental reflectivity and
  Kramers-Kronig analysis, which includes
  Refs.~\onlinecite{Wooten,Dressel-Book,Singh1994,Singh1991,Wien2k}.}\BibitemShut
  {Stop}%
\bibitem [{\citenamefont {Dai}\ \emph {et~al.}(2016)\citenamefont {Dai},
  \citenamefont {Akrap}, \citenamefont {Bud'ko}, \citenamefont {Canfield},\
  and\ \citenamefont {Homes}}]{Homes2016}%
  \BibitemOpen
  \bibfield  {author} {\bibinfo {author} {\bibfnamefont {Y.~M.}\ \bibnamefont
  {Dai}}, \bibinfo {author} {\bibfnamefont {Ana}\ \bibnamefont {Akrap}},
  \bibinfo {author} {\bibfnamefont {S.~L.}\ \bibnamefont {Bud'ko}}, \bibinfo
  {author} {\bibfnamefont {P.~C.}\ \bibnamefont {Canfield}}, \ and\ \bibinfo
  {author} {\bibfnamefont {C.~C.}\ \bibnamefont {Homes}},\ }\bibfield  {title}
  {\enquote {\bibinfo {title} {{Optical properties of $A${Fe}$_{2}${As}$_{2}$
  ($A=\,$Ca, Sr, and Ba) single crystals}},}\ }\href {\doibase
  10.1103/PhysRevB.94.195142} {\bibfield  {journal} {\bibinfo  {journal} {Phys.
  Rev. B}\ }\textbf {\bibinfo {volume} {94}},\ \bibinfo {pages} {195142}
  (\bibinfo {year} {2016})}\BibitemShut {NoStop}%
\bibitem [{\citenamefont {Homes}\ \emph {et~al.}(2018)\citenamefont {Homes},
  \citenamefont {Dai}, \citenamefont {Akrap}, \citenamefont {Bud'ko},\ and\
  \citenamefont {Canfield}}]{Homes2018}%
  \BibitemOpen
  \bibfield  {author} {\bibinfo {author} {\bibfnamefont {C.~C.}\ \bibnamefont
  {Homes}}, \bibinfo {author} {\bibfnamefont {Y.~M.}\ \bibnamefont {Dai}},
  \bibinfo {author} {\bibfnamefont {Ana}\ \bibnamefont {Akrap}}, \bibinfo
  {author} {\bibfnamefont {S.~L.}\ \bibnamefont {Bud'ko}}, \ and\ \bibinfo
  {author} {\bibfnamefont {P.~C.}\ \bibnamefont {Canfield}},\ }\bibfield
  {title} {\enquote {\bibinfo {title} {{Vibrational anomalies in
  $A${Fe}$_{2}${As}$_{2}$ ($A=\,$Ca, Sr, and Ba) single crystals}},}\ }\href
  {\doibase 10.1103/PhysRevB.98.035103} {\bibfield  {journal} {\bibinfo
  {journal} {Phys. Rev. B}\ }\textbf {\bibinfo {volume} {98}},\ \bibinfo
  {pages} {035103} (\bibinfo {year} {2018})}\BibitemShut {NoStop}%
\bibitem [{\citenamefont {Mallett}\ \emph
  {et~al.}(2015{\natexlab{b}})\citenamefont {Mallett}, \citenamefont {Marsik},
  \citenamefont {Yazdi-Rizi}, \citenamefont {Wolf}, \citenamefont {B\"ohmer},
  \citenamefont {Hardy}, \citenamefont {Meingast}, \citenamefont {Munzar},\
  and\ \citenamefont {Bernhard}}]{Mallett2015b}%
  \BibitemOpen
  \bibfield  {author} {\bibinfo {author} {\bibfnamefont {B.~P.~P.}\
  \bibnamefont {Mallett}}, \bibinfo {author} {\bibfnamefont {P.}~\bibnamefont
  {Marsik}}, \bibinfo {author} {\bibfnamefont {M.}~\bibnamefont {Yazdi-Rizi}},
  \bibinfo {author} {\bibfnamefont {Th.}\ \bibnamefont {Wolf}}, \bibinfo
  {author} {\bibfnamefont {A.~E.}\ \bibnamefont {B\"ohmer}}, \bibinfo {author}
  {\bibfnamefont {F.}~\bibnamefont {Hardy}}, \bibinfo {author} {\bibfnamefont
  {C.}~\bibnamefont {Meingast}}, \bibinfo {author} {\bibfnamefont
  {D.}~\bibnamefont {Munzar}}, \ and\ \bibinfo {author} {\bibfnamefont
  {C.}~\bibnamefont {Bernhard}},\ }\bibfield  {title} {\enquote {\bibinfo
  {title} {{Infrared Study of the Spin Reorientation Transition and Its
  Reversal in the Superconducting State in Underdoped
  Ba$_{1-x}$K$_x$Fe$_2$As$_2$}},}\ }\href {\doibase
  10.1103/PhysRevLett.115.027003} {\bibfield  {journal} {\bibinfo  {journal}
  {Phys. Rev. Lett.}\ }\textbf {\bibinfo {volume} {115}},\ \bibinfo {pages}
  {027003} (\bibinfo {year} {2015}{\natexlab{b}})}\BibitemShut {NoStop}%
\bibitem [{\citenamefont {Singh}(2008)}]{Singh2008}%
  \BibitemOpen
  \bibfield  {author} {\bibinfo {author} {\bibfnamefont {D.~J.}\ \bibnamefont
  {Singh}},\ }\bibfield  {title} {\enquote {\bibinfo {title} {{Electronic
  structure and doping in BaFe$_2$As$_2$ and LiFeAs: Density functional
  calculations}},}\ }\href {\doibase 10.1103/PhysRevB.78.094511} {\bibfield
  {journal} {\bibinfo  {journal} {Phys. Rev. B}\ }\textbf {\bibinfo {volume}
  {78}},\ \bibinfo {pages} {094511} (\bibinfo {year} {2008})}\BibitemShut
  {NoStop}%
\bibitem [{\citenamefont {Fink}\ \emph {et~al.}(2009)\citenamefont {Fink},
  \citenamefont {Thirupathaiah}, \citenamefont {Ovsyannikov}, \citenamefont
  {D{\"u}rr}, \citenamefont {Follath}, \citenamefont {Huang}, \citenamefont
  {de~Jong}, \citenamefont {Golden}, \citenamefont {Zhang}, \citenamefont
  {Jeschke}, \citenamefont {Valent\'{i}}, \citenamefont {Felser}, \citenamefont
  {Dastjani~Farahani}, \citenamefont {Rotter},\ and\ \citenamefont
  {Johrendt}}]{Fink2009}%
  \BibitemOpen
  \bibfield  {author} {\bibinfo {author} {\bibfnamefont {J.}~\bibnamefont
  {Fink}}, \bibinfo {author} {\bibfnamefont {S.}~\bibnamefont {Thirupathaiah}},
  \bibinfo {author} {\bibfnamefont {R.}~\bibnamefont {Ovsyannikov}}, \bibinfo
  {author} {\bibfnamefont {H.~A.}\ \bibnamefont {D{\"u}rr}}, \bibinfo {author}
  {\bibfnamefont {R.}~\bibnamefont {Follath}}, \bibinfo {author} {\bibfnamefont
  {Y.}~\bibnamefont {Huang}}, \bibinfo {author} {\bibfnamefont
  {S.}~\bibnamefont {de~Jong}}, \bibinfo {author} {\bibfnamefont {M.~S.}\
  \bibnamefont {Golden}}, \bibinfo {author} {\bibfnamefont {Yu-Zhong}\
  \bibnamefont {Zhang}}, \bibinfo {author} {\bibfnamefont {H.~O.}\ \bibnamefont
  {Jeschke}}, \bibinfo {author} {\bibfnamefont {R.}~\bibnamefont
  {Valent\'{i}}}, \bibinfo {author} {\bibfnamefont {C.}~\bibnamefont {Felser}},
  \bibinfo {author} {\bibfnamefont {S.}~\bibnamefont {Dastjani~Farahani}},
  \bibinfo {author} {\bibfnamefont {M.}~\bibnamefont {Rotter}}, \ and\ \bibinfo
  {author} {\bibfnamefont {D.}~\bibnamefont {Johrendt}},\ }\bibfield  {title}
  {\enquote {\bibinfo {title} {Electronic structure studies of
  {BaFe}$_2${As}$_2$ by angle-resolved photoemission spectroscopy},}\ }\href
  {\doibase 10.1103/PhysRevB.79.155118} {\bibfield  {journal} {\bibinfo
  {journal} {Phys. Rev. B}\ }\textbf {\bibinfo {volume} {79}},\ \bibinfo
  {pages} {155118} (\bibinfo {year} {2009})}\BibitemShut {NoStop}%
\bibitem [{\citenamefont {Wu}\ \emph {et~al.}(2010)\citenamefont {Wu},
  \citenamefont {Bari{\v{s}}i{\'{c}}}, \citenamefont {Kallina}, \citenamefont
  {Faridian}, \citenamefont {Gorshunov}, \citenamefont {Drichko}, \citenamefont
  {Li}, \citenamefont {Lin}, \citenamefont {Cao}, \citenamefont {Xu},
  \citenamefont {Wang},\ and\ \citenamefont {Dressel}}]{Wu2010}%
  \BibitemOpen
  \bibfield  {author} {\bibinfo {author} {\bibfnamefont {D.}~\bibnamefont
  {Wu}}, \bibinfo {author} {\bibfnamefont {N.}~\bibnamefont
  {Bari{\v{s}}i{\'{c}}}}, \bibinfo {author} {\bibfnamefont {P.}~\bibnamefont
  {Kallina}}, \bibinfo {author} {\bibfnamefont {A.}~\bibnamefont {Faridian}},
  \bibinfo {author} {\bibfnamefont {B.}~\bibnamefont {Gorshunov}}, \bibinfo
  {author} {\bibfnamefont {N.}~\bibnamefont {Drichko}}, \bibinfo {author}
  {\bibfnamefont {L.~J.}\ \bibnamefont {Li}}, \bibinfo {author} {\bibfnamefont
  {X.}~\bibnamefont {Lin}}, \bibinfo {author} {\bibfnamefont {G.~H.}\
  \bibnamefont {Cao}}, \bibinfo {author} {\bibfnamefont {Z.~A.}\ \bibnamefont
  {Xu}}, \bibinfo {author} {\bibfnamefont {N.~L.}\ \bibnamefont {Wang}}, \ and\
  \bibinfo {author} {\bibfnamefont {M.}~\bibnamefont {Dressel}},\ }\bibfield
  {title} {\enquote {\bibinfo {title} {{Optical investigations of the normal
  and superconducting states reveal two electronic subsystems in iron
  pnictides}},}\ }\href {\doibase 10.1103/PhysRevB.81.100512} {\bibfield
  {journal} {\bibinfo  {journal} {Phys. Rev. B}\ }\textbf {\bibinfo {volume}
  {81}},\ \bibinfo {pages} {100512(R)} (\bibinfo {year} {2010})}\BibitemShut
  {NoStop}%
\bibitem [{\citenamefont {Yin}\ \emph {et~al.}(2011)\citenamefont {Yin},
  \citenamefont {Haule},\ and\ \citenamefont {Kotliar}}]{Yin2011}%
  \BibitemOpen
  \bibfield  {author} {\bibinfo {author} {\bibfnamefont {Z.~P.}\ \bibnamefont
  {Yin}}, \bibinfo {author} {\bibfnamefont {K.}~\bibnamefont {Haule}}, \ and\
  \bibinfo {author} {\bibfnamefont {G.}~\bibnamefont {Kotliar}},\ }\bibfield
  {title} {\enquote {\bibinfo {title} {Magnetism and charge dynamics in iron
  pnictides},}\ }\href {\doibase 10.1038/nphys1923} {\bibfield  {journal}
  {\bibinfo  {journal} {Nat. Phys.}\ }\textbf {\bibinfo {volume} {7}},\
  \bibinfo {pages} {294--297} (\bibinfo {year} {2011})}\BibitemShut {NoStop}%
\bibitem [{\citenamefont {Ferrell}\ and\ \citenamefont
  {Glover}(1958)}]{Ferrell1958}%
  \BibitemOpen
  \bibfield  {author} {\bibinfo {author} {\bibfnamefont {Richard~A.}\
  \bibnamefont {Ferrell}}\ and\ \bibinfo {author} {\bibfnamefont {Rolfe~E.}\
  \bibnamefont {Glover}},\ }\bibfield  {title} {\enquote {\bibinfo {title}
  {{Conductivity of Superconducting Films: A Sum Rule}},}\ }\href {\doibase
  10.1103/PhysRev.109.1398} {\bibfield  {journal} {\bibinfo  {journal} {Phys.
  Rev.}\ }\textbf {\bibinfo {volume} {109}},\ \bibinfo {pages} {1398--1399}
  (\bibinfo {year} {1958})}\BibitemShut {NoStop}%
\bibitem [{\citenamefont {Tinkham}\ and\ \citenamefont
  {Ferrell}(1959)}]{Tinkham1959}%
  \BibitemOpen
  \bibfield  {author} {\bibinfo {author} {\bibfnamefont {M.}~\bibnamefont
  {Tinkham}}\ and\ \bibinfo {author} {\bibfnamefont {R.~A.}\ \bibnamefont
  {Ferrell}},\ }\bibfield  {title} {\enquote {\bibinfo {title} {{Determination
  of the Superconducting Skin Depth from the Energy Gap and Sum Rule}},}\
  }\href {\doibase 10.1103/PhysRevLett.2.331} {\bibfield  {journal} {\bibinfo
  {journal} {Phys. Rev. Lett.}\ }\textbf {\bibinfo {volume} {2}},\ \bibinfo
  {pages} {331--333} (\bibinfo {year} {1959})}\BibitemShut {NoStop}%
\bibitem [{\citenamefont {Homes}\ \emph {et~al.}(2015)\citenamefont {Homes},
  \citenamefont {Dai}, \citenamefont {Wen}, \citenamefont {Xu},\ and\
  \citenamefont {Gu}}]{Homes2015}%
  \BibitemOpen
  \bibfield  {author} {\bibinfo {author} {\bibfnamefont {C.~C.}\ \bibnamefont
  {Homes}}, \bibinfo {author} {\bibfnamefont {Y.~M.}\ \bibnamefont {Dai}},
  \bibinfo {author} {\bibfnamefont {J.~S.}\ \bibnamefont {Wen}}, \bibinfo
  {author} {\bibfnamefont {Z.~J.}\ \bibnamefont {Xu}}, \ and\ \bibinfo {author}
  {\bibfnamefont {G.~D.}\ \bibnamefont {Gu}},\ }\bibfield  {title} {\enquote
  {\bibinfo {title} {{FeTe}$_{0.55}${Se}$_{0.45}$: A multiband superconductor
  in the clean and dirty limit},}\ }\href {\doibase 10.1103/PhysRevB.91.144503}
  {\bibfield  {journal} {\bibinfo  {journal} {Phys. Rev. B}\ }\textbf {\bibinfo
  {volume} {91}},\ \bibinfo {pages} {144503} (\bibinfo {year}
  {2015})}\BibitemShut {NoStop}%
\bibitem [{\citenamefont {Homes}\ \emph {et~al.}(2004)\citenamefont {Homes},
  \citenamefont {Dordevic}, \citenamefont {Strongin}, \citenamefont {Bonn},
  \citenamefont {Liang}, \citenamefont {Hardy}, \citenamefont {Komiya},
  \citenamefont {Ando}, \citenamefont {Yu}, \citenamefont {Kaneko},
  \citenamefont {Zhao}, \citenamefont {Greven}, \citenamefont {Basov},\ and\
  \citenamefont {Timusk}}]{Homes2004}%
  \BibitemOpen
  \bibfield  {author} {\bibinfo {author} {\bibfnamefont {C.~C.}\ \bibnamefont
  {Homes}}, \bibinfo {author} {\bibfnamefont {S.~V.}\ \bibnamefont {Dordevic}},
  \bibinfo {author} {\bibfnamefont {M.}~\bibnamefont {Strongin}}, \bibinfo
  {author} {\bibfnamefont {D.~A.}\ \bibnamefont {Bonn}}, \bibinfo {author}
  {\bibfnamefont {Ruixing}\ \bibnamefont {Liang}}, \bibinfo {author}
  {\bibfnamefont {W.~N.}\ \bibnamefont {Hardy}}, \bibinfo {author}
  {\bibfnamefont {Seiki}\ \bibnamefont {Komiya}}, \bibinfo {author}
  {\bibfnamefont {Yoichi}\ \bibnamefont {Ando}}, \bibinfo {author}
  {\bibfnamefont {G.}~\bibnamefont {Yu}}, \bibinfo {author} {\bibfnamefont
  {N.}~\bibnamefont {Kaneko}}, \bibinfo {author} {\bibfnamefont
  {X.}~\bibnamefont {Zhao}}, \bibinfo {author} {\bibfnamefont {M.}~\bibnamefont
  {Greven}}, \bibinfo {author} {\bibfnamefont {D.~N.}\ \bibnamefont {Basov}}, \
  and\ \bibinfo {author} {\bibfnamefont {T.}~\bibnamefont {Timusk}},\
  }\bibfield  {title} {\enquote {\bibinfo {title} {{Universal scaling relation
  in high-temperature superconductors}},}\ }\href {\doibase
  10.1038/nature02673} {\bibfield  {journal} {\bibinfo  {journal} {Nature
  (London)}\ }\textbf {\bibinfo {volume} {430}},\ \bibinfo {pages} {539}
  (\bibinfo {year} {2004})}\BibitemShut {NoStop}%
\bibitem [{\citenamefont {Homes}\ \emph
  {et~al.}(2005{\natexlab{a}})\citenamefont {Homes}, \citenamefont {Dordevic},
  \citenamefont {Bonn}, \citenamefont {Liang}, \citenamefont {Hardy},\ and\
  \citenamefont {Timusk}}]{Homes2005a}%
  \BibitemOpen
  \bibfield  {author} {\bibinfo {author} {\bibfnamefont {C.~C.}\ \bibnamefont
  {Homes}}, \bibinfo {author} {\bibfnamefont {S.~V.}\ \bibnamefont {Dordevic}},
  \bibinfo {author} {\bibfnamefont {D.~A.}\ \bibnamefont {Bonn}}, \bibinfo
  {author} {\bibfnamefont {Ruixing}\ \bibnamefont {Liang}}, \bibinfo {author}
  {\bibfnamefont {W.~N.}\ \bibnamefont {Hardy}}, \ and\ \bibinfo {author}
  {\bibfnamefont {T.}~\bibnamefont {Timusk}},\ }\bibfield  {title} {\enquote
  {\bibinfo {title} {Coherence, incoherence, and scaling along the $c$ axis of
  {Y}{Ba}$_{2}${Cu}$_{3}${O}$_{6+x}$},}\ }\href {\doibase
  10.1103/PhysRevB.71.184515} {\bibfield  {journal} {\bibinfo  {journal} {Phys.
  Rev. B}\ }\textbf {\bibinfo {volume} {71}},\ \bibinfo {pages} {184515}
  (\bibinfo {year} {2005}{\natexlab{a}})}\BibitemShut {NoStop}%
\bibitem [{\citenamefont {Homes}\ \emph
  {et~al.}(2005{\natexlab{b}})\citenamefont {Homes}, \citenamefont {Dordevic},
  \citenamefont {Valla},\ and\ \citenamefont {Strongin}}]{Homes2005b}%
  \BibitemOpen
  \bibfield  {author} {\bibinfo {author} {\bibfnamefont {C.~C.}\ \bibnamefont
  {Homes}}, \bibinfo {author} {\bibfnamefont {S.~V.}\ \bibnamefont {Dordevic}},
  \bibinfo {author} {\bibfnamefont {T.}~\bibnamefont {Valla}}, \ and\ \bibinfo
  {author} {\bibfnamefont {M.}~\bibnamefont {Strongin}},\ }\bibfield  {title}
  {\enquote {\bibinfo {title} {{Scaling of the superfluid density in
  high-temperature superconductors}},}\ }\href {\doibase
  10.1103/PhysRevB.72.134517} {\bibfield  {journal} {\bibinfo  {journal} {Phys.
  Rev. B}\ }\textbf {\bibinfo {volume} {72}},\ \bibinfo {pages} {134517}
  (\bibinfo {year} {2005}{\natexlab{b}})}\BibitemShut {NoStop}%
\bibitem [{\citenamefont {Tu}\ \emph {et~al.}(2010)\citenamefont {Tu},
  \citenamefont {Li}, \citenamefont {Liu}, \citenamefont {Punnoose},
  \citenamefont {Gong}, \citenamefont {Ren}, \citenamefont {Li}, \citenamefont
  {Cao}, \citenamefont {Xu},\ and\ \citenamefont {Homes}}]{Tu2010}%
  \BibitemOpen
  \bibfield  {author} {\bibinfo {author} {\bibfnamefont {J.~J.}\ \bibnamefont
  {Tu}}, \bibinfo {author} {\bibfnamefont {J.}~\bibnamefont {Li}}, \bibinfo
  {author} {\bibfnamefont {W.}~\bibnamefont {Liu}}, \bibinfo {author}
  {\bibfnamefont {A.}~\bibnamefont {Punnoose}}, \bibinfo {author}
  {\bibfnamefont {Y.}~\bibnamefont {Gong}}, \bibinfo {author} {\bibfnamefont
  {Y.~H.}\ \bibnamefont {Ren}}, \bibinfo {author} {\bibfnamefont {L.~J.}\
  \bibnamefont {Li}}, \bibinfo {author} {\bibfnamefont {G.~H.}\ \bibnamefont
  {Cao}}, \bibinfo {author} {\bibfnamefont {Z.~A.}\ \bibnamefont {Xu}}, \ and\
  \bibinfo {author} {\bibfnamefont {C.~C.}\ \bibnamefont {Homes}},\ }\bibfield
  {title} {\enquote {\bibinfo {title} {Optical properties of the iron arsenic
  superconductor {BaFe}$_{1.85}${Co}$_{0.15}${As}$_{2}$},}\ }\href {\doibase
  10.1103/PhysRevB.82.174509} {\bibfield  {journal} {\bibinfo  {journal} {Phys.
  Rev. B}\ }\textbf {\bibinfo {volume} {82}},\ \bibinfo {pages} {174509}
  (\bibinfo {year} {2010})}\BibitemShut {NoStop}%
\bibitem [{\citenamefont {Yang}\ \emph {et~al.}(2019)\citenamefont {Yang},
  \citenamefont {Dai}, \citenamefont {Yu}, \citenamefont {Sui}, \citenamefont
  {Cai}, \citenamefont {Ren}, \citenamefont {Hwang}, \citenamefont {Xiao},
  \citenamefont {Zhou}, \citenamefont {Qiu},\ and\ \citenamefont
  {Homes}}]{Yang2019}%
  \BibitemOpen
  \bibfield  {author} {\bibinfo {author} {\bibfnamefont {Run}\ \bibnamefont
  {Yang}}, \bibinfo {author} {\bibfnamefont {Yaomin}\ \bibnamefont {Dai}},
  \bibinfo {author} {\bibfnamefont {Jia}\ \bibnamefont {Yu}}, \bibinfo {author}
  {\bibfnamefont {Qiangtao}\ \bibnamefont {Sui}}, \bibinfo {author}
  {\bibfnamefont {Yongqing}\ \bibnamefont {Cai}}, \bibinfo {author}
  {\bibfnamefont {Zhian}\ \bibnamefont {Ren}}, \bibinfo {author} {\bibfnamefont
  {Jungseek}\ \bibnamefont {Hwang}}, \bibinfo {author} {\bibfnamefont {Hong}\
  \bibnamefont {Xiao}}, \bibinfo {author} {\bibfnamefont {Xingjiang}\
  \bibnamefont {Zhou}}, \bibinfo {author} {\bibfnamefont {Xianggang}\
  \bibnamefont {Qiu}}, \ and\ \bibinfo {author} {\bibfnamefont
  {Christopher~C.}\ \bibnamefont {Homes}},\ }\bibfield  {title} {\enquote
  {\bibinfo {title} {{Unravelling the mechanism of the semiconducting-like
  behavior and its relation to superconductivity in
  $({\mathrm{CaFe}}_{1\ensuremath{-}x}{\mathrm{Pt}}_{x}\mathrm{As}{)}_{10}{\mathrm{Pt}}_{3}{\mathrm{As}}_{8}$}},}\
  }\href {\doibase 10.1103/PhysRevB.99.144520} {\bibfield  {journal} {\bibinfo
  {journal} {Phys. Rev. B}\ }\textbf {\bibinfo {volume} {99}},\ \bibinfo
  {pages} {144520} (\bibinfo {year} {2019})}\BibitemShut {NoStop}%
\bibitem [{\citenamefont {Smith}(2001)}]{Smith2001}%
  \BibitemOpen
  \bibfield  {author} {\bibinfo {author} {\bibfnamefont {N.~V.}\ \bibnamefont
  {Smith}},\ }\bibfield  {title} {\enquote {\bibinfo {title} {{Classical
  generalization of the Drude formula for the optical conductivity}},}\ }\href
  {\doibase 10.1103/PhysRevB.64.155106} {\bibfield  {journal} {\bibinfo
  {journal} {Phys. Rev. B}\ }\textbf {\bibinfo {volume} {64}},\ \bibinfo
  {pages} {155106} (\bibinfo {year} {2001})}\BibitemShut {NoStop}%
\bibitem [{Note1()}]{Note1}%
  \BibitemOpen
  \bibinfo {note} {Replacing the broad Drude term with the expression in
  Eq.~(\ref {eq:smith}) and fitting to the real and imaginary parts of the
  optical conductivity using a non-linear least-squares technique yields
  $\omega _p \simeq 5350$~cm$^{-1}$, $1/\tau \simeq 146$~cm$^{-1}$, and
  $c=-0.7$. The plasma frequency is larger because it now describes both the
  localized as well as free carriers; $\omega _{p}^2 \simeq \omega _{p,D2}^2 +
  \Omega _0^2$.}\BibitemShut {Stop}%
\bibitem [{\citenamefont {Dai}\ \emph {et~al.}(2012)\citenamefont {Dai},
  \citenamefont {Xu}, \citenamefont {Shen}, \citenamefont {Wen}, \citenamefont
  {Hu}, \citenamefont {Qiu},\ and\ \citenamefont {Lobo}}]{Dai2012}%
  \BibitemOpen
  \bibfield  {author} {\bibinfo {author} {\bibfnamefont {Y.~M.}\ \bibnamefont
  {Dai}}, \bibinfo {author} {\bibfnamefont {B.}~\bibnamefont {Xu}}, \bibinfo
  {author} {\bibfnamefont {B.}~\bibnamefont {Shen}}, \bibinfo {author}
  {\bibfnamefont {H.~H.}\ \bibnamefont {Wen}}, \bibinfo {author} {\bibfnamefont
  {J.~P.}\ \bibnamefont {Hu}}, \bibinfo {author} {\bibfnamefont {X.~G.}\
  \bibnamefont {Qiu}}, \ and\ \bibinfo {author} {\bibfnamefont {R.~P. S.~M.}\
  \bibnamefont {Lobo}},\ }\bibfield  {title} {\enquote {\bibinfo {title}
  {{Pseudogap in underdoped Ba$_{1-x}$K$_{x}$Fe$_{2}$As$_{2}$ as seen via
  optical conductivity}},}\ }\href {\doibase 10.1103/PhysRevB.86.100501}
  {\bibfield  {journal} {\bibinfo  {journal} {Phys. Rev. B}\ }\textbf {\bibinfo
  {volume} {86}},\ \bibinfo {pages} {100501(R)} (\bibinfo {year}
  {2012})}\BibitemShut {NoStop}%
\bibitem [{\citenamefont {Zabolotnyy}\ \emph {et~al.}(2009)\citenamefont
  {Zabolotnyy}, \citenamefont {Inosov}, \citenamefont {Evtushinsky},
  \citenamefont {Koitzsch}, \citenamefont {Kordyuk}, \citenamefont {Sun},
  \citenamefont {Park}, \citenamefont {Haug}, \citenamefont {Hinkov},
  \citenamefont {Boris}, \citenamefont {Lin}, \citenamefont {Knupfer},
  \citenamefont {Yaresko}, \citenamefont {B\"{u}chner}, \citenamefont
  {Varykhalov}, \citenamefont {Follath},\ and\ \citenamefont
  {Borisenko}}]{Zabolotnyy2009}%
  \BibitemOpen
  \bibfield  {author} {\bibinfo {author} {\bibfnamefont {V.~B.}\ \bibnamefont
  {Zabolotnyy}}, \bibinfo {author} {\bibfnamefont {D.~S.}\ \bibnamefont
  {Inosov}}, \bibinfo {author} {\bibfnamefont {D.~V.}\ \bibnamefont
  {Evtushinsky}}, \bibinfo {author} {\bibfnamefont {A.}~\bibnamefont
  {Koitzsch}}, \bibinfo {author} {\bibfnamefont {A.~A.}\ \bibnamefont
  {Kordyuk}}, \bibinfo {author} {\bibfnamefont {G.~L.}\ \bibnamefont {Sun}},
  \bibinfo {author} {\bibfnamefont {J.~T.}\ \bibnamefont {Park}}, \bibinfo
  {author} {\bibfnamefont {D.}~\bibnamefont {Haug}}, \bibinfo {author}
  {\bibfnamefont {V.}~\bibnamefont {Hinkov}}, \bibinfo {author} {\bibfnamefont
  {A.~V.}\ \bibnamefont {Boris}}, \bibinfo {author} {\bibfnamefont {C.~T.}\
  \bibnamefont {Lin}}, \bibinfo {author} {\bibfnamefont {M.}~\bibnamefont
  {Knupfer}}, \bibinfo {author} {\bibfnamefont {A.~N.}\ \bibnamefont
  {Yaresko}}, \bibinfo {author} {\bibfnamefont {B.}~\bibnamefont
  {B\"{u}chner}}, \bibinfo {author} {\bibfnamefont {A.}~\bibnamefont
  {Varykhalov}}, \bibinfo {author} {\bibfnamefont {R.}~\bibnamefont {Follath}},
  \ and\ \bibinfo {author} {\bibfnamefont {S.~V.}\ \bibnamefont {Borisenko}},\
  }\bibfield  {title} {\enquote {\bibinfo {title} {$(\pi,\pi)$ electronic order
  in iron arsenide superconductors},}\ }\href {\doibase 10.1038/nature07714}
  {\bibfield  {journal} {\bibinfo  {journal} {Nature}\ }\textbf {\bibinfo
  {volume} {457}},\ \bibinfo {pages} {569--572} (\bibinfo {year}
  {2009})}\BibitemShut {NoStop}%
\bibitem [{\citenamefont {Derondeau}\ \emph {et~al.}(2017)\citenamefont
  {Derondeau}, \citenamefont {Bisti}, \citenamefont {Kobayashi}, \citenamefont
  {Braun}, \citenamefont {Ebert}, \citenamefont {Rogalev}, \citenamefont {Shi},
  \citenamefont {Schmitt}, \citenamefont {Ma}, \citenamefont {Ding},
  \citenamefont {Strocov},\ and\ \citenamefont {Min\'{a}r}}]{Derondeau2017}%
  \BibitemOpen
  \bibfield  {author} {\bibinfo {author} {\bibfnamefont {Gerald}\ \bibnamefont
  {Derondeau}}, \bibinfo {author} {\bibfnamefont {Federico}\ \bibnamefont
  {Bisti}}, \bibinfo {author} {\bibfnamefont {Masaki}\ \bibnamefont
  {Kobayashi}}, \bibinfo {author} {\bibfnamefont {J\"{u}rgen}\ \bibnamefont
  {Braun}}, \bibinfo {author} {\bibfnamefont {Hubert}\ \bibnamefont {Ebert}},
  \bibinfo {author} {\bibfnamefont {Victor~A.}\ \bibnamefont {Rogalev}},
  \bibinfo {author} {\bibfnamefont {Ming}\ \bibnamefont {Shi}}, \bibinfo
  {author} {\bibfnamefont {Thorsten}\ \bibnamefont {Schmitt}}, \bibinfo
  {author} {\bibfnamefont {Junzhang}\ \bibnamefont {Ma}}, \bibinfo {author}
  {\bibfnamefont {Hong}\ \bibnamefont {Ding}}, \bibinfo {author} {\bibfnamefont
  {Vladimir~N.}\ \bibnamefont {Strocov}}, \ and\ \bibinfo {author}
  {\bibfnamefont {J\'{a}n}\ \bibnamefont {Min\'{a}r}},\ }\bibfield  {title}
  {\enquote {\bibinfo {title} {{Fermi surface and effective masses in
  photoemission response of the (Ba$_{1-x}$K$_x$)Fe$_2$As$_2$
  superconductor}},}\ }\href {\doibase 10.1038/s41598-017-09480-y} {\bibfield
  {journal} {\bibinfo  {journal} {Sci. Rep.}\ }\textbf {\bibinfo {volume}
  {7}},\ \bibinfo {pages} {8787} (\bibinfo {year} {2017})}\BibitemShut
  {NoStop}%
\bibitem [{\citenamefont {Yi}\ \emph {et~al.}(2009)\citenamefont {Yi},
  \citenamefont {Lu}, \citenamefont {Analytis}, \citenamefont {Chu},
  \citenamefont {Mo}, \citenamefont {He}, \citenamefont {Hashimoto},
  \citenamefont {Moore}, \citenamefont {Mazin}, \citenamefont {Singh},
  \citenamefont {Hussain}, \citenamefont {Fisher},\ and\ \citenamefont
  {Shen}}]{Yi2009}%
  \BibitemOpen
  \bibfield  {author} {\bibinfo {author} {\bibfnamefont {M.}~\bibnamefont
  {Yi}}, \bibinfo {author} {\bibfnamefont {D.~H.}\ \bibnamefont {Lu}}, \bibinfo
  {author} {\bibfnamefont {J.~G.}\ \bibnamefont {Analytis}}, \bibinfo {author}
  {\bibfnamefont {J.-H.}\ \bibnamefont {Chu}}, \bibinfo {author} {\bibfnamefont
  {S.-K.}\ \bibnamefont {Mo}}, \bibinfo {author} {\bibfnamefont {R.-H.}\
  \bibnamefont {He}}, \bibinfo {author} {\bibfnamefont {M.}~\bibnamefont
  {Hashimoto}}, \bibinfo {author} {\bibfnamefont {R.~G.}\ \bibnamefont
  {Moore}}, \bibinfo {author} {\bibfnamefont {I.~I.}\ \bibnamefont {Mazin}},
  \bibinfo {author} {\bibfnamefont {D.~J.}\ \bibnamefont {Singh}}, \bibinfo
  {author} {\bibfnamefont {Z.}~\bibnamefont {Hussain}}, \bibinfo {author}
  {\bibfnamefont {I.~R.}\ \bibnamefont {Fisher}}, \ and\ \bibinfo {author}
  {\bibfnamefont {Z.-X.}\ \bibnamefont {Shen}},\ }\bibfield  {title} {\enquote
  {\bibinfo {title} {Unconventional electronic reconstruction in undoped
  ({Ba},{Sr}){Fe}$_{2}${As}$_{2}$ across the spin density wave transition},}\
  }\href {\doibase 10.1103/PhysRevB.80.174510} {\bibfield  {journal} {\bibinfo
  {journal} {Phys. Rev. B}\ }\textbf {\bibinfo {volume} {80}},\ \bibinfo
  {pages} {174510} (\bibinfo {year} {2009})}\BibitemShut {NoStop}%
\bibitem [{\citenamefont {Wooten}(1972)}]{Wooten}%
  \BibitemOpen
  \bibfield  {author} {\bibinfo {author} {\bibfnamefont {F.}~\bibnamefont
  {Wooten}},\ }\href@noop {} {\emph {\bibinfo {title} {{Optical Properties of
  Solids}}}}\ (\bibinfo  {publisher} {Academic Press},\ \bibinfo {address} {New
  York},\ \bibinfo {year} {1972})\ pp.\ \bibinfo {pages} {244--250}\BibitemShut
  {NoStop}%
\bibitem [{\citenamefont {Dressel}\ and\ \citenamefont
  {Gr{\"u}ner}(2001)}]{Dressel-Book}%
  \BibitemOpen
  \bibfield  {author} {\bibinfo {author} {\bibfnamefont {M.}~\bibnamefont
  {Dressel}}\ and\ \bibinfo {author} {\bibfnamefont {G.}~\bibnamefont
  {Gr{\"u}ner}},\ }\href@noop {} {\emph {\bibinfo {title} {{Electrodynamics of
  Solids}}}}\ (\bibinfo  {publisher} {Cambridge University Press},\ \bibinfo
  {address} {Cambridge},\ \bibinfo {year} {2001})\BibitemShut {NoStop}%
\bibitem [{\citenamefont {Singh}(1994)}]{Singh1994}%
  \BibitemOpen
  \bibfield  {author} {\bibinfo {author} {\bibfnamefont {D.~J.}\ \bibnamefont
  {Singh}},\ }\href@noop {} {\emph {\bibinfo {title} {Planewaves,
  Pseudopotentials and the LAPW method}}}\ (\bibinfo  {publisher} {Kluwer
  Adademic},\ \bibinfo {address} {Boston},\ \bibinfo {year} {1994})\BibitemShut
  {NoStop}%
\bibitem [{\citenamefont {Singh}(1991)}]{Singh1991}%
  \BibitemOpen
  \bibfield  {author} {\bibinfo {author} {\bibfnamefont {David}\ \bibnamefont
  {Singh}},\ }\bibfield  {title} {\enquote {\bibinfo {title} {{Ground-state
  properties of lanthanum: Treatment of extended-core states}},}\ }\href
  {\doibase 10.1103/PhysRevB.43.6388} {\bibfield  {journal} {\bibinfo
  {journal} {Phys. Rev. B}\ }\textbf {\bibinfo {volume} {43}},\ \bibinfo
  {pages} {6388--6392} (\bibinfo {year} {1991})}\BibitemShut {NoStop}%
\bibitem [{Wie()}]{Wien2k}%
  \BibitemOpen
  \href@noop {} {}\bibinfo {note} {P. Blaha, K. Schwarz, G.~K.~H. Madsen, D.
  Kvasnicka and J. Luitz, WIEN2k, {\it An augmented plane wave plus local
  orbitals program for calculating crystal properties} (Techn.
  Universit{\"{a}}t Wien, Austria, 2001).}\BibitemShut {Stop}%
\end{thebibliography}

%
% Supplementary material
%

\clearpage
\newpage

\newpage
\vspace*{-2.1cm}
\hspace*{-2.5cm}
{
  \centering
  \includegraphics[width=1.2\textwidth]{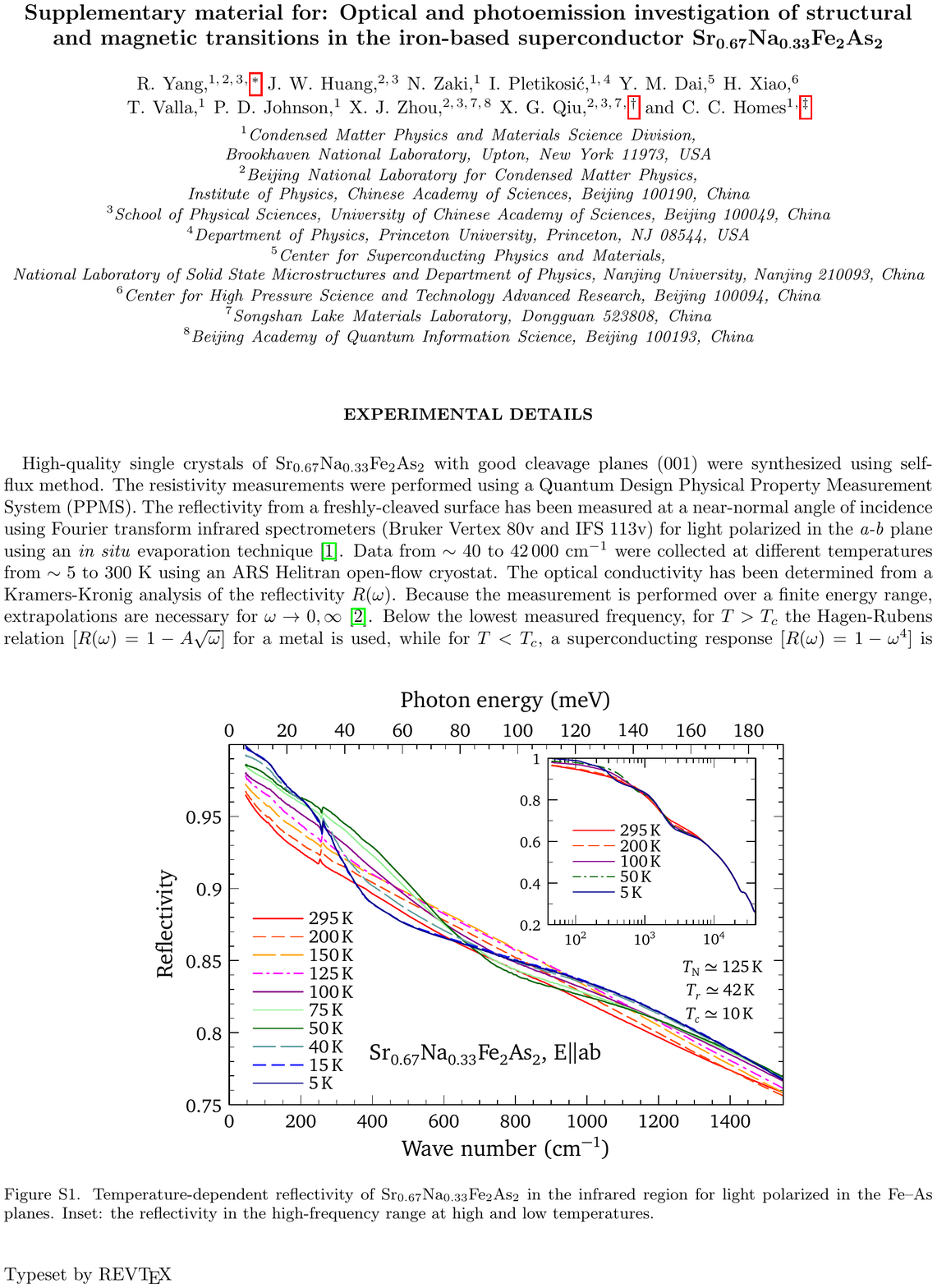} \\
  \ \\
}

\newpage
\vspace*{-2.1cm}
\hspace*{-2.5cm}
{
  \centering
  \includegraphics[width=1.2\textwidth]{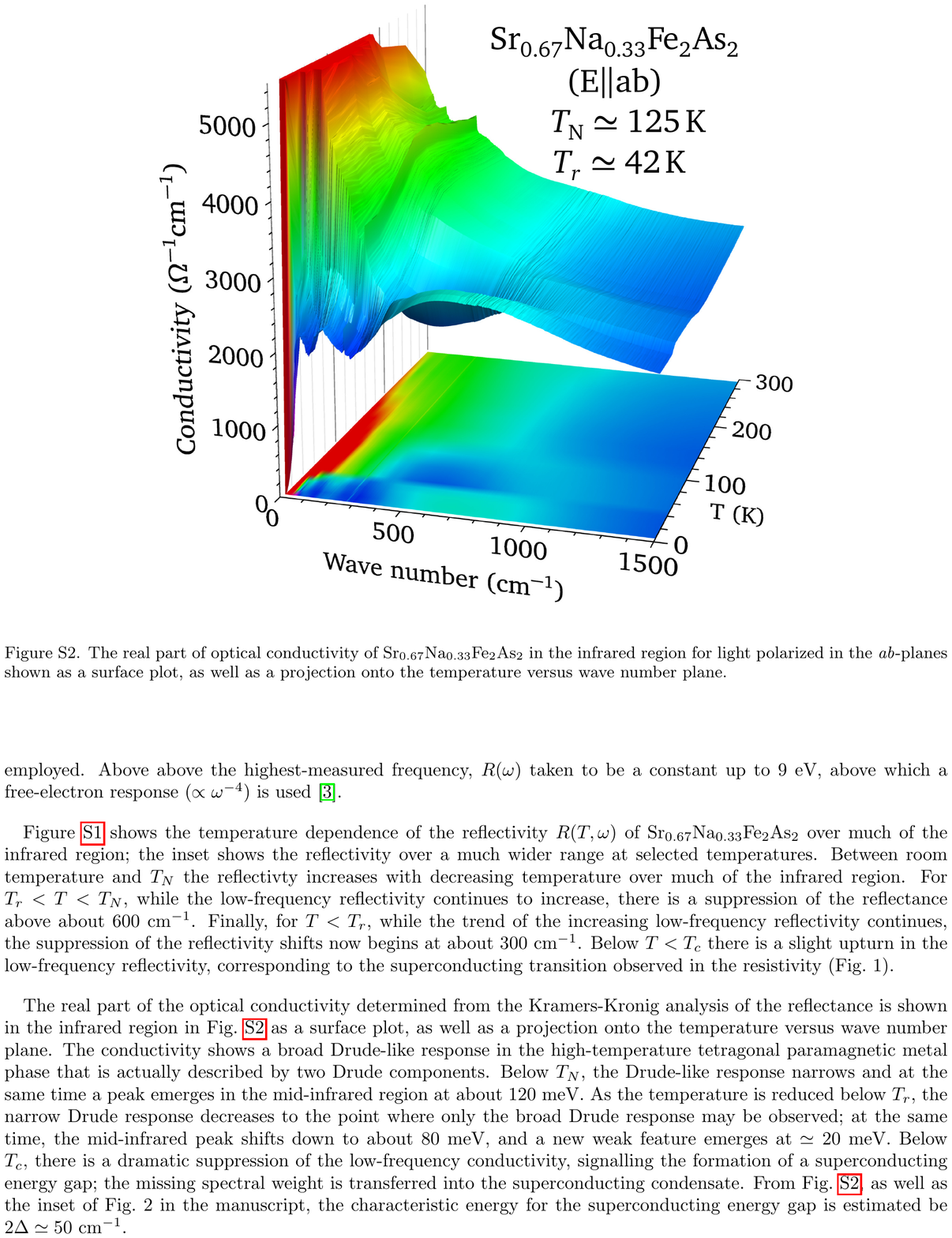} \\
  \ \\
}

\newpage
\vspace*{-2.1cm}
\hspace*{-2.5cm}
{
  \centering
  \includegraphics[width=1.2\textwidth]{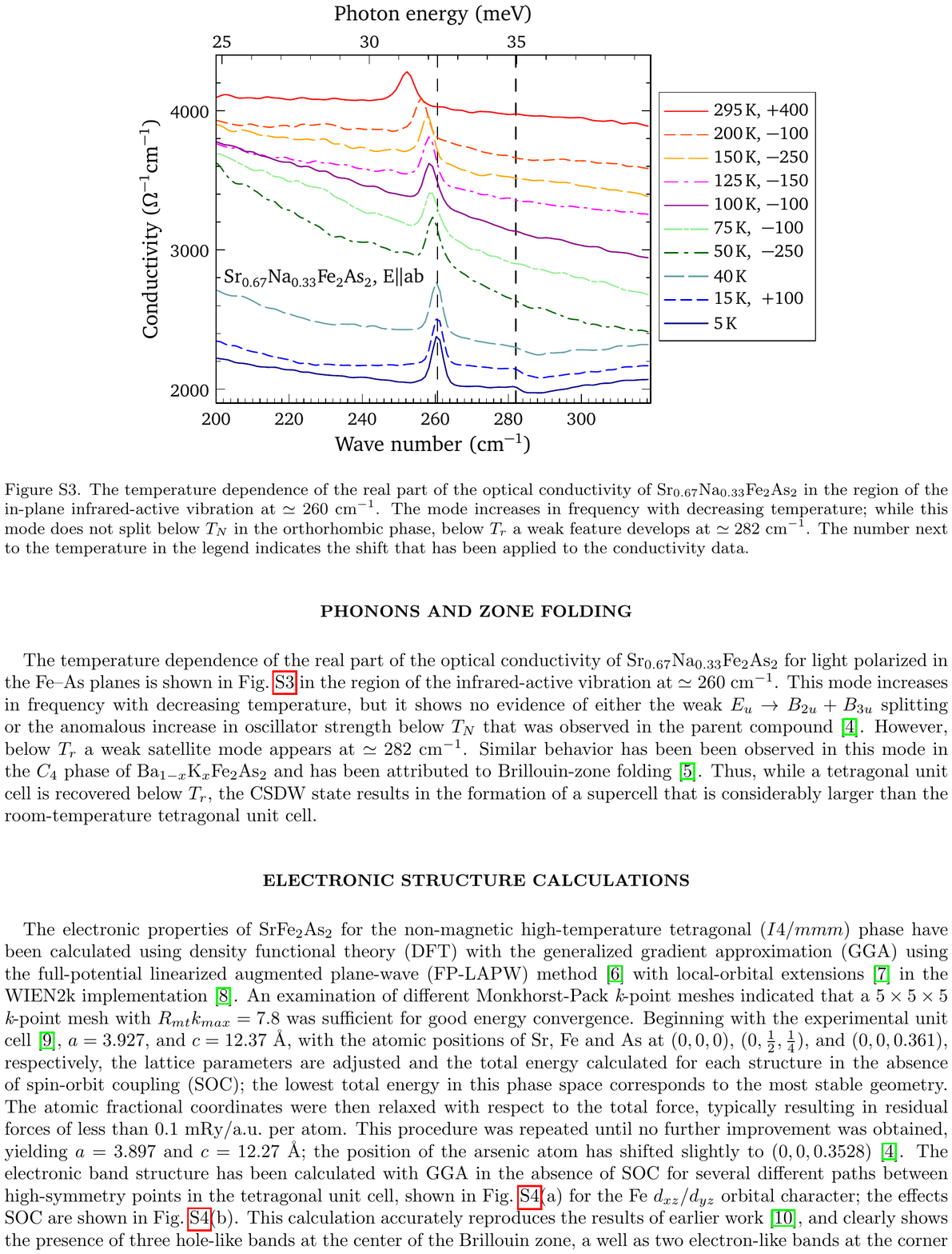} \\
  \ \\
}

\newpage
\vspace*{-2.1cm}
\hspace*{-2.5cm}
{
  \centering
  \includegraphics[width=1.2\textwidth]{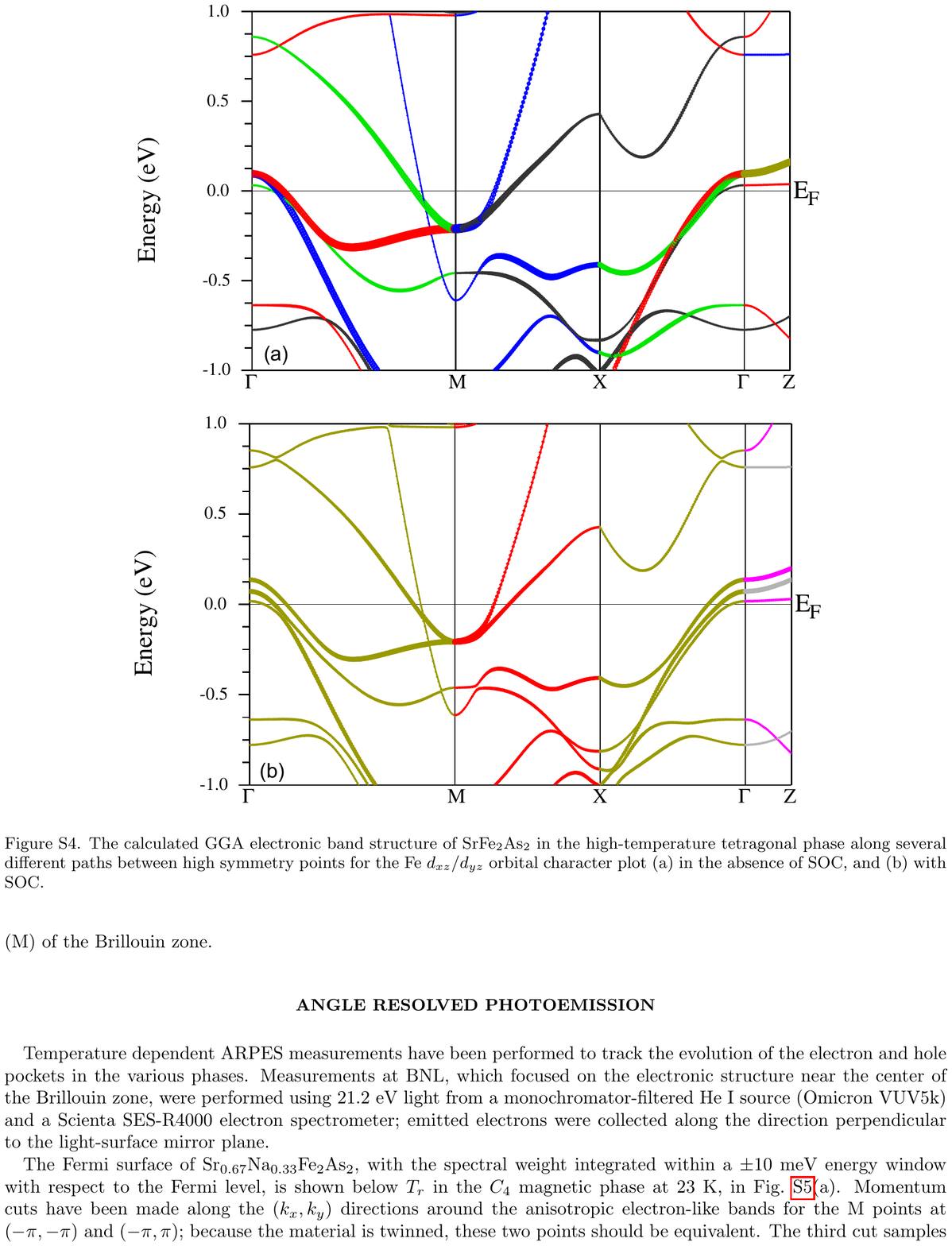} \\
  \ \\
}

\newpage
\vspace*{-2.1cm}
\hspace*{-2.5cm}
{
  \centering
  \includegraphics[width=1.2\textwidth]{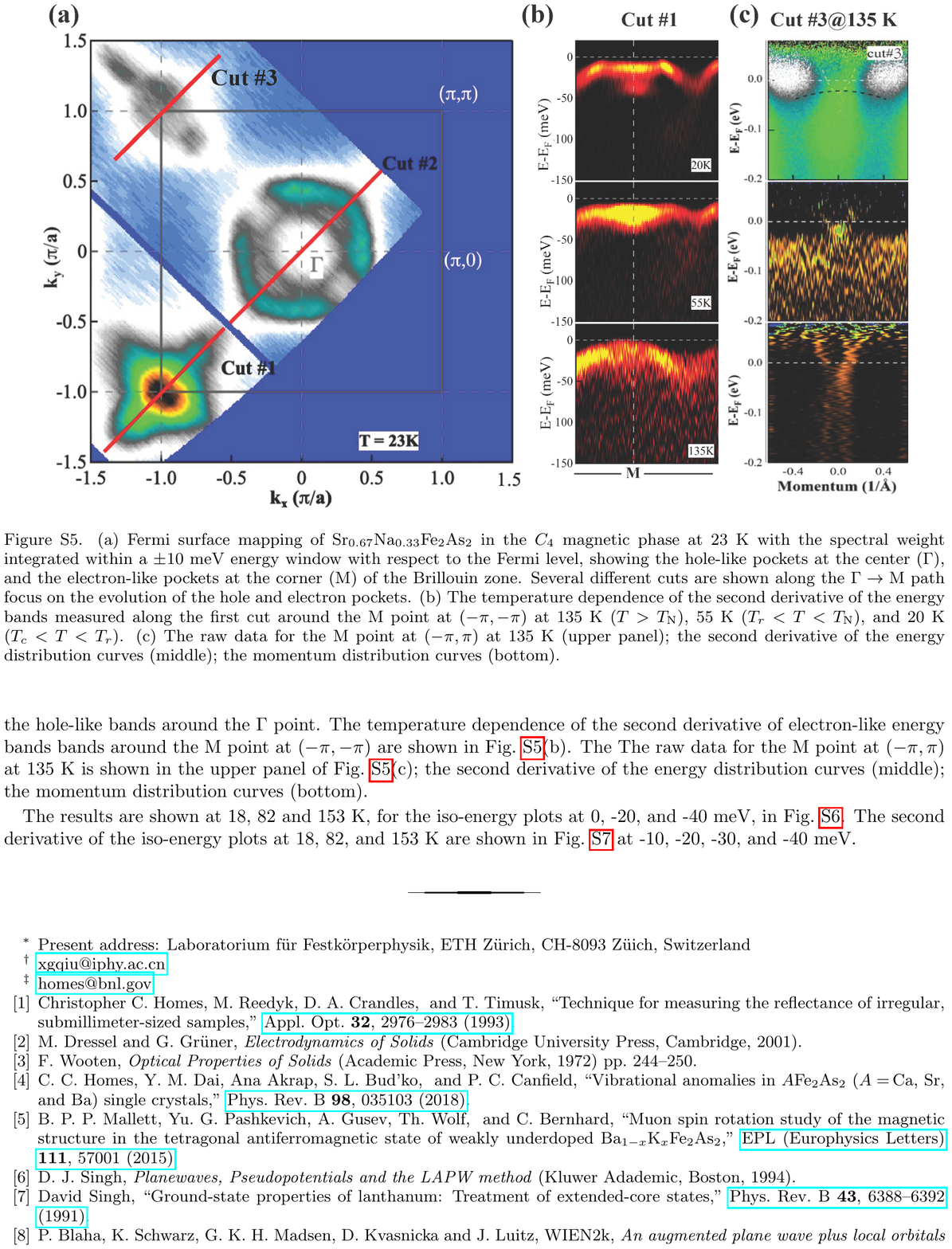} \\
  \ \\
}

\newpage
\vspace*{-2.1cm}
\hspace*{-2.5cm}
{
  \centering
  \includegraphics[width=1.2\textwidth]{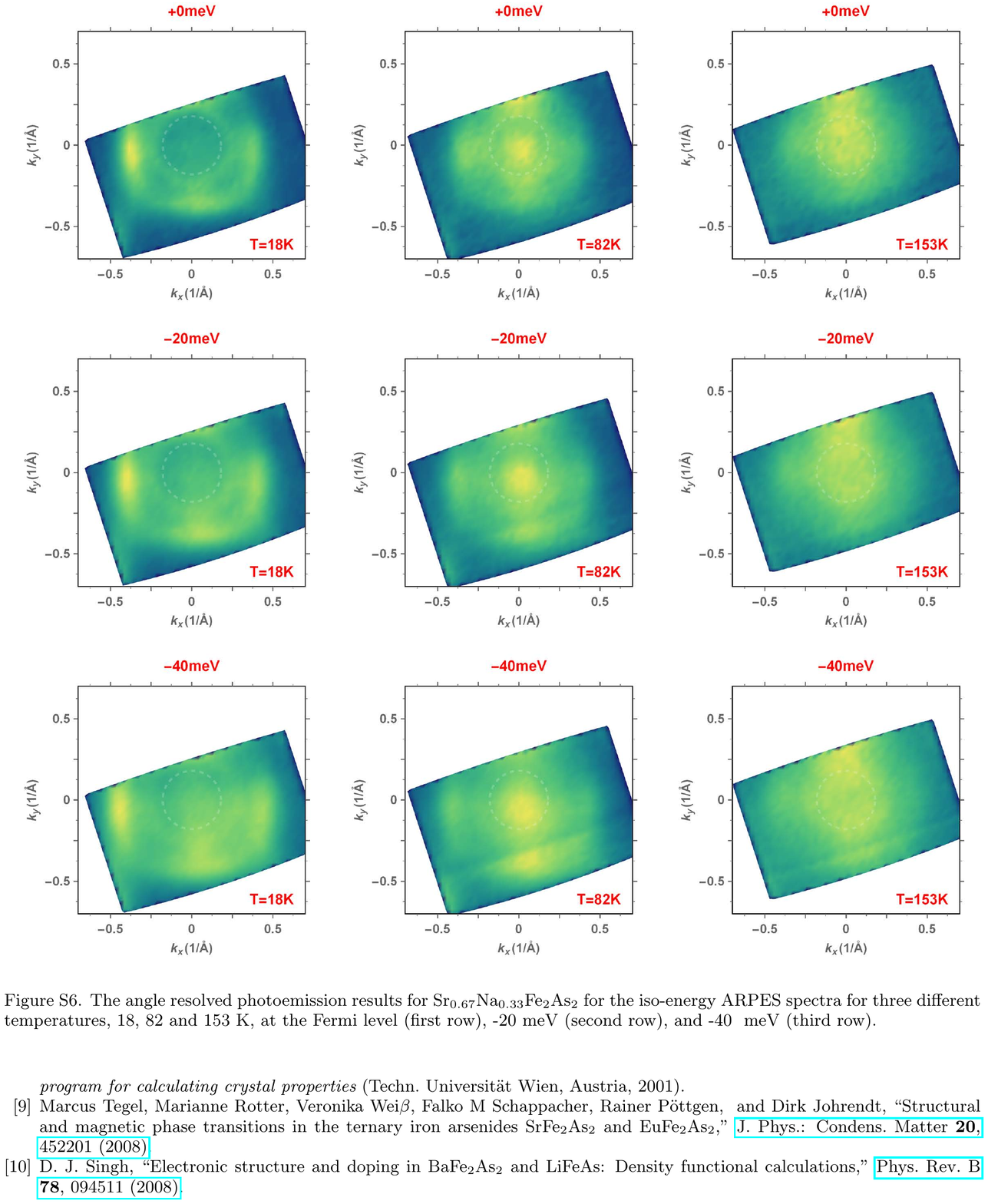} \\
  \ \\
}

\newpage
\vspace*{-2.1cm}
\hspace*{-2.5cm}
{
  \centering
  \includegraphics[width=1.2\textwidth]{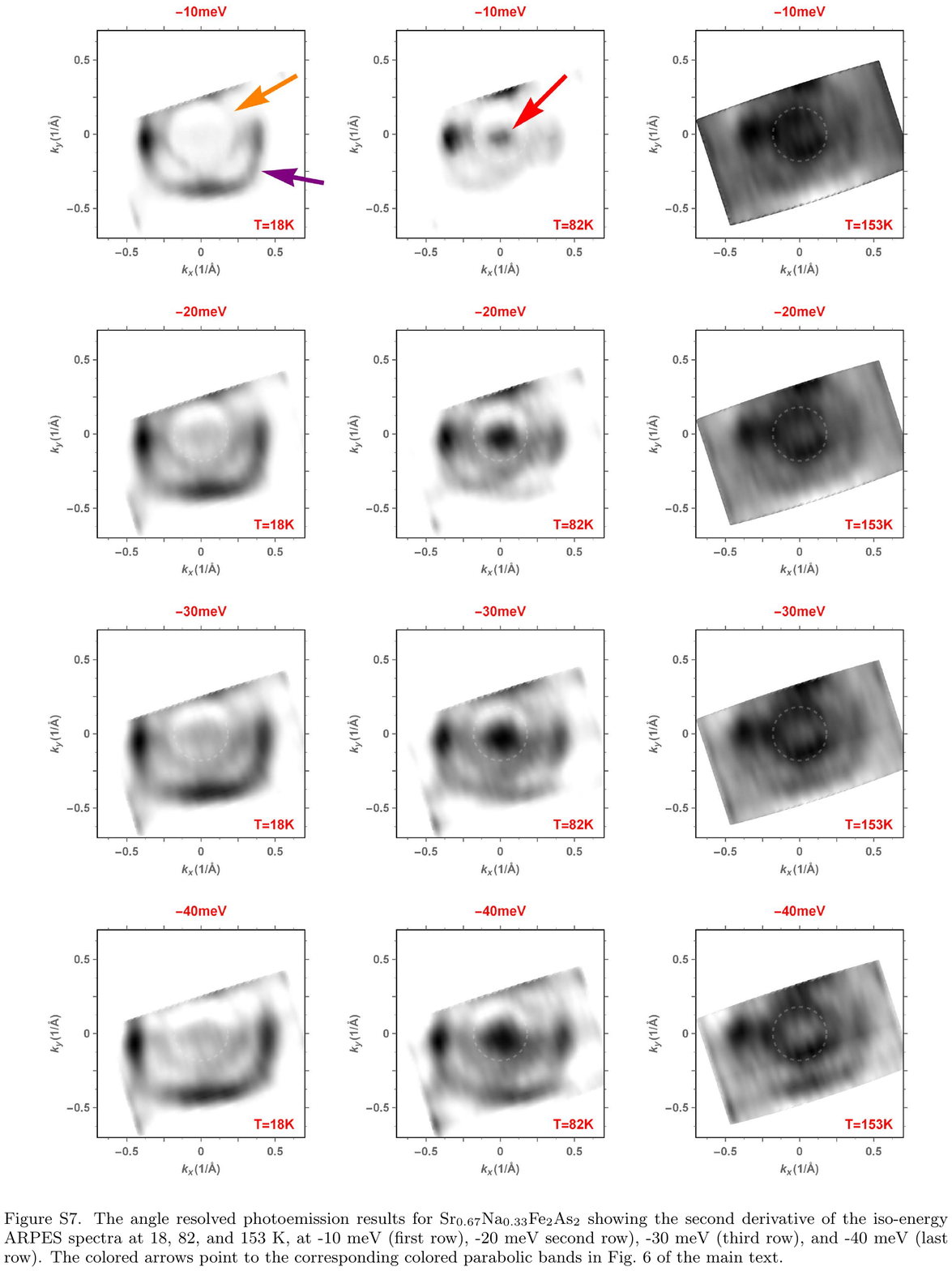} \\
  \ \\
}

\end{document}